\documentclass[twocolumn]{aastex631}

\let\tablenum\relax
\usepackage{hyperref}

\usepackage[separate-uncertainty,multi-part-units = single,group-digits=false]{siunitx}
\usepackage{bm}

\usepackage{chemformula}
\usepackage{amsmath}
\usepackage{float}
\DeclareMathOperator*{\argminA}{arg\,min}

\setcitestyle{notesep={ }}
\shorttitle{Discovery and Characterization of HIP~21152~B}
\shortauthors{Franson et al.}

\graphicspath{{./}{figures/}}

\begin{document}

\title{Astrometric Accelerations as Dynamical Beacons: Discovery and Characterization of HIP 21152 B, the First T-Dwarf Companion in the Hyades\footnote{Based in part on data collected at Subaru Telescope, which is operated by the National Astronomical Observatory of Japan.}}

\author[0000-0003-4557-414X]{Kyle Franson}
\altaffiliation{NSF Graduate Research Fellow} \affiliation{Department of Astronomy, The University of Texas at Austin, Austin, TX 78712, USA}

\author[0000-0003-2649-2288]{Brendan P. Bowler}
\affiliation{Department of Astronomy, The University of Texas at Austin, Austin, TX 78712, USA}

\author[0000-0002-7520-8389]{Mariangela Bonavita}
\affiliation{School of Physical Sciences, The Open University, Walton Hall, Milton Keynes, MK7 6AA, UK}

\author{Timothy D. Brandt}
\affiliation{Department of Physics, University of California, Santa Barbara, Santa Barbara, CA 93106, USA}

\author[0000-0001-8892-4045]{Minghan Chen}
\affiliation{Department of Physics, University of California, Santa Barbara, Santa Barbara, CA 93106, USA}

\author[0000-0001-9992-4067]{Matthias Samland}
\affiliation{Max-Planck-Institut f\"ur Astronomie, K\"onigstuhl 17, 69117 Heidelberg, Germany}

\author[0000-0002-3726-4881]{Zhoujian Zhang}
\affiliation{Department of Astronomy, The University of Texas at Austin, Austin, TX 78712, USA}

\author[0000-0001-6960-0256]{Anna Lueber}
\affiliation{Center for Space and Habitability, University of Bern, Gesellschaftsstrasse 6, 3012 Bern, Switzerland}
\affiliation{Ludwig Maximilian University, University Observatory Munich, Scheinerstr. 1, Munich D-81679, Germany}

\author[0000-0003-1907-5910]{Kevin Heng}
\affiliation{Center for Space and Habitability, University of Bern, Gesellschaftsstrasse 6, 3012 Bern, Switzerland}
\affiliation{University of Warwick, Department of Physics, Astronomy \& Astrophysics Group, Coventry CV4 7AL, UK}
\affiliation{Ludwig Maximilian University, University Observatory Munich, Scheinerstr. 1, Munich D-81679, Germany}

\author[0000-0003-4269-3311]{Daniel Kitzmann}
\affiliation{Center for Space and Habitability, University of Bern, Gesellschaftsstrasse 6, 3012 Bern, Switzerland}

\author{Trevor Wolf}
\affiliation{Department of Aerospace Engineering and Engineering Mechanics, The University of Texas at Austin, Austin, TX 78712, USA}

\author{Brandon A. Jones}
\affiliation{Department of Aerospace Engineering and Engineering Mechanics, The University of Texas at Austin, Austin, TX 78712, USA}

\author[0000-0001-6532-6755]{Quang H. Tran}
\affiliation{Department of Astronomy, The University of Texas at Austin, Austin, TX 78712, USA}

\author[0000-0001-8170-7072]{Daniella C. Bardalez Gagliuffi}
\affiliation{American Museum of Natural History, 200 Central Park West, New York, NY 10024, USA}

\author[0000-0003-4614-7035]{Beth Biller}
\affiliation{SUPA, Institute for Astronomy, The University of Edinburgh, Royal Observatory, Blackford Hill, Edinburgh, EH9 3HJ, UK}

\author{Jeffrey Chilcote}
\affiliation{Department of Physics, University of Notre Dame, 225 Nieuwland Science Hall, Notre Dame, IN 46556, USA}

\author[0000-0003-0800-0593]{Justin R. Crepp}
\affiliation{Department of Physics, University of Notre Dame, 225 Nieuwland Science Hall, Notre Dame, IN 46556, USA}

\author[0000-0001-9823-1445]{Trent J. Dupuy}
\affiliation{Institute for Astronomy, University of Edinburgh, Royal Observatory, Blackford Hill, Edinburgh, EH9 3HJ, UK}

\author[0000-0001-6251-0573]{Jacqueline Faherty}
\affiliation{American Museum of Natural History, 200 Central Park West, New York, NY 10024, USA}

\author[0000-0002-2428-9932]{Cl{\'e}mence Fontanive}
\affiliation{Center for Space and Habitability, University of Bern, Gesellschaftsstrasse 6, 3012 Bern, Switzerland}

\author{Tyler D. Groff}
\affiliation{NASA-Goddard Space Flight Center, Greenbelt, MD, USA}

\author[0000-0003-2195-6805]{Raffaele Gratton}
\affiliation{INAF Osservatorio Astronomico di Padova, Vicolo dell'Osservatorio 5,  35121 Padova, Italy}

\author[0000-0002-1097-9908]{Olivier Guyon}
\affiliation{Subaru Telescope, National Astronomical Observatory of Japan, 
650 North A`oh$\bar{o}$k$\bar{u}$ Place, Hilo, HI 96720, USA}
\affiliation{Steward Observatory, University of Arizona, Tucson, AZ 85721, USA}
\affiliation{Astrobiology Center of NINS, 2-21-1 Osawa, Mitaka, Tokyo 181-8588, Japan}

\author[0000-0003-0054-2953]{Rebecca Jensen-Clem}
\affiliation{Astronomy \& Astrophysics Department, University of California, Santa Cruz, CA 95064, USA}

\author[0000-0001-5213-6207]{Nemanja Jovanovic}
\affiliation{Department of Astronomy, California Institute of Technology, 1200 East California Boulevard, Pasadena, CA 91125, USA}

\author{N. Jeremy Kasdin}
\affiliation{Department of Mechanical Engineering, Princeton University, Princeton, NJ, USA}

\author[0000-0002-3047-1845]{Julien Lozi}
\affiliation{Subaru Telescope, National Astronomical Observatory of Japan, 
650 North A`oh$\bar{o}$k$\bar{u}$ Place, Hilo, HI 96720, USA}

\author[0000-0002-7965-2815]{Eugene A. Magnier}
\affiliation{Institute for Astronomy, University of Hawai`i, 2680 Woodlawn Drive, Honolulu, HI 96822, USA}

\author[0000-0002-7989-2595]{Koraljka Mu{\v z}i{\'c}}
\affiliation{CENTRA, Faculdade de Ci\^{e}ncias, Universidade de Lisboa, Ed. C8, Campo Grande, P-1749-016 Lisboa, Portugal}

\author[0000-0002-1838-4757]{Aniket Sanghi}
\affiliation{Department of Astronomy, The University of Texas at Austin, Austin, TX 78712, USA}

\author[0000-0002-9807-5435]{Christopher A. Theissen}
\altaffiliation{NASA Sagan Fellow}
\affiliation{Center for Astrophysics and Space Sciences, University of California, San Diego, 9500 Gilman Drive, La Jolla, CA 92093, USA}

\begin{abstract}
Benchmark brown dwarf companions with well-determined ages and model-independent masses are powerful tools to test substellar evolutionary models and probe the formation of giant planets and brown dwarfs. Here, we report the independent discovery of HIP~21152~B, the first imaged brown dwarf companion in the Hyades, and conduct a comprehensive orbital and atmospheric characterization of the system. HIP~21152 was targeted in an ongoing high-contrast imaging campaign of stars exhibiting proper motion changes between Hipparcos and Gaia, and was also recently identified by \citet{bonavitaResultsCopainsPilot_2022} and \citet{kuzuharaDirectImagingDiscovery_2022}. Our Keck/NIRC2 and SCExAO/CHARIS imaging of HIP~21152 revealed a comoving companion at a separation of $0\farcs37$ (16 au). We perform a joint orbit fit of all available relative astrometry and radial velocities together with the Hipparcos-Gaia proper motions, yielding a dynamical mass of $24^{+6}_{-4}\,\mathrm{M_{Jup}}$, which is $1{-}2{\sigma}$ lower than evolutionary model predictions. Hybrid grids that include the evolution of cloud properties best reproduce the dynamical mass. We also identify a comoving wide-separation ($1837\arcsec$ or $\SI{7.9e4}{au}$) early-L dwarf with an inferred mass near the hydrogen-burning limit. Finally, we analyze the spectra and photometry of HIP~21152~B using the \citet{saumonEvolutionDwarfsColor_2008} atmospheric models and a suite of retrievals. The best-fit grid-based models have $f_{\mathrm{sed}}=2$, indicating the presence of clouds, $T_{\mathrm{eff}}=\SI{1400}{K}$, and $\log{g}=\SI{4.5}{dex}$. These results are consistent with the object's spectral type of $\mathrm{T0\pm1}$. As the first benchmark brown dwarf companion in the Hyades, HIP~21152~B joins the small but growing number of substellar companions with well-determined ages and dynamical masses.

\end{abstract}


\section{Introduction \label{sec:intro}}

Brown dwarfs are objects that predominantly form like stars but fail to reach sufficient masses (${\sim }70{-}\SI{80}{M_{Jup}}$; \citealt{dupuyIndividualDynamicalMasses_2017, fernandesEvolutionaryModelsUltracool_2019}) to sustain hydrogen fusion, instead cooling and fading over their lifetimes \citep{kumarStructureStarsVery_1963}. As these objects radiate away their internal energy, their colors and spectra change dramatically as a rich variety of chemical species form and condense in their atmospheres. During this process, they pass through a series of associated spectral transitions spanning the M, L, T, and Y spectral classes \citep{kirkpatrickNewSpectralTypes_2005, cushingDiscoveryDwarfsUsing_2011}. 

Brown dwarfs are an important bridge population between gas giants and low-mass stars. They share much of the same atmospheric chemistry as self-luminous giant planets, but are significantly brighter and easier to observe. While over one thousand field brown dwarfs have been identified, only ${\sim} 200$ brown dwarf companions have been discovered via imaging, most of which are on wide orbits (${>}\SI{100}{au}$; \citealt{bestwilliamm.j.UltracoolsheetPhotometryAstrometry_2020}). These systems serve as important benchmarks for testing atmospheric models, as their host stars enable constraints to be placed on their ages and compositions \citep[e.g.,][]{dupuyDynamicalMassSubstellar_2009,dupuyNewEvidenceSubstellar_2014,brandtPreciseDynamicalMasses_2021,zhangUniformForwardmodelingAnalysis_2021a}. They additionally comprise an excellent comparison population to imaged planets to delineate the upper boundary of planet formation (e.g., \citealt{nielsenGeminiPlanetImager_2019, bowlerPopulationlevelEccentricityDistributions_2020}).

The gold standard for benchmark systems are objects with well-constrained ages and independent mass measurements. Since substellar objects follow mass-luminosity-\emph{age} relations instead of the mass-luminosity relations of main-sequence stars, the masses of directly imaged planets and brown dwarfs are typically inferred via low-temperature cooling models (e.g., \citealt{burrowsNongrayTheoryExtrasolar_1997,baraffeEvolutionaryModelsCool_2003, saumonEvolutionDwarfsColor_2008,marleauConstrainingInitialEntropy_2014, phillipsNewSetAtmosphere_2020,marleySonoraBrownDwarf_2021}). Independent mass measurements are critical to empirically calibrate and test evolutionary models. These models encapsulate assumptions about the origin, interior structure, and atmospheres of substellar objects. This is especially true in the planetary regime, where different formation channels may significantly alter the luminosities of young objects \citep{fortneySyntheticSpectraColors_2008,spiegelSpectralPhotometricDiagnostics_2012}, with core-accretion scenarios \citep{pollackFormationGiantPlanets_1996} leaving planets with lower initial entropies (``cold-start" models) than gravitational instability \citep{bossGiantPlanetFormation_1997} or turbulent fragmentation \citep{bateDependenceStarFormation_2009} routes (the ``hot-start" pathway). The burning of deuterium for brown dwarfs (and lithium at high masses; \citealt{gharib-nezhadFollowingLithiumTracing_2021}) impacts their evolution by slowing their cooling \citep{spiegelDeuteriumburningMassLimit_2011}. Their atmospheres further act to regulate thermal evolution; the presence and properties of clouds, different chemical species, and non-equilibrium processes all affect the emergent spectra and the resultant luminosity evolution \citep{burrowsTheoryBrownDwarfs_2001}. By testing evolutionary models and their input physics, benchmark systems therefore provide a direct window into the formation, thermal evolution, and interior structure of substellar objects.

Direct masses of substellar companions can be obtained through measurements of the gravitational reflex motion they exert on their host stars. Because most imaged substellar companions are on wide orbits, orbital motion is challenging to observe for the majority of systems. To date, there have been less than 20 precisely measured dynamical masses of substellar objects with well-constrained ages and luminosities \citep[see recent compilation in][]{fransonDynamicalMassYoung_2022}. The majority of these studies couple observations of the relative orbital motion of the companion with absolute astrometry of the host star \citep[e.g.,][]{maireOrbitalSpectralCharacterization_2020,brandtPreciseDynamicalMasses_2019,brandtImprovedDynamicalMasses_2021,fransonDynamicalMassYoung_2022}, usually from small proper motion changes between Hipparcos and Gaia, and, in some cases, long-term radial velocity trends \citep[e.g.,][]{creppDynamicalMassThreedimensional_2012,cheethamDirectImagingUltracool_2018, bowlerOrbitDynamicalMass_2018,rickmanSpectralAtmosphericCharacterisation_2020,bowlerMcdonaldAcceleratingStars_2021}. 

This growing collection of benchmark systems has presented a conflicting story about the reliability of widely used evolutionary models. Though the majority of dynamical masses are consistent with model predictions to within ${\sim}10{-}15$\% of predicted masses, several companions are significantly less massive  \citep{dupuyDynamicalMassSubstellar_2009,beattySignificantOverluminosityTransiting_2018,rickmanSpectralAtmosphericCharacterisation_2020} and more massive \citep{cheethamDirectImagingUltracool_2018, brandtFirstDynamicalMass_2021, bowlerMcdonaldAcceleratingStars_2021} than the predicted masses given their luminosities and ages. While over-massive companions can be explained by unresolved binarity, convincing theoretical explanations for the under-massive cases have remained elusive. There is a pressing need for new benchmark systems to test evolutionary models across a wide range of masses, ages, and luminosities.

\begin{figure}
    \centering
    \includegraphics[width=\linewidth]{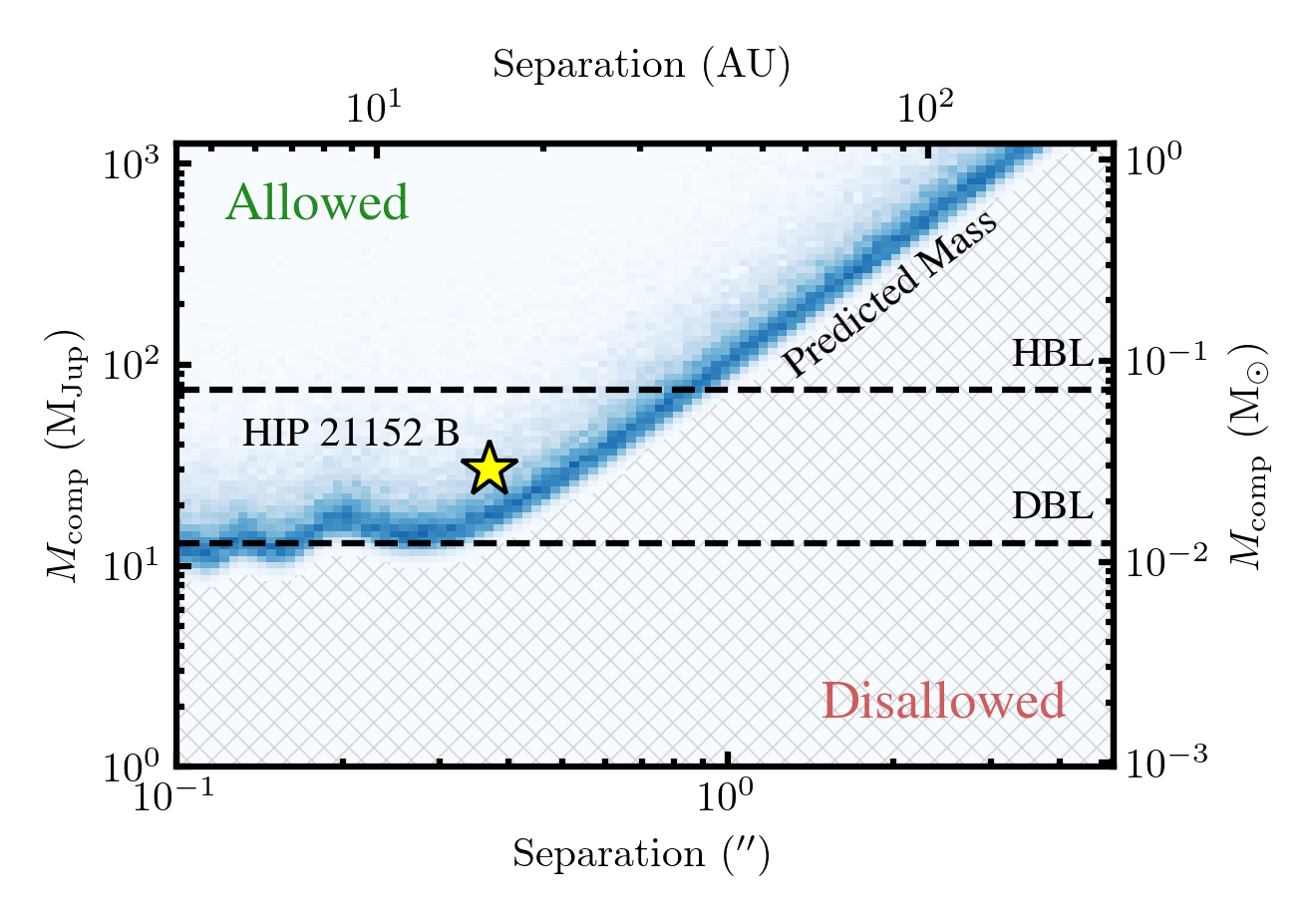}
    \caption{Predicted mass as a function of projected separation for HIP~21152~B based on the $\SI{8.4 \pm 0.8}{m.s^{-1}yr^{-1}}$ Hipparcos-Gaia acceleration of its host star in the HGCA. Colors indicate the relative probability of agreement; companions with separations and masses below the blue curve cannot produce the observed change in proper motion. HIP~21152's astrometric acceleration indicates the presence of a substellar companion within 1 arcsec or a star at wider separations. The dotted lines denote the hydrogen-burning limit (HBL; ${\sim} \SI{75}{M_{Jup}}$) and the deuterium-burning limit (DBL; ${\sim} \SI{13}{M_{Jup}}$). The yellow star on the plot shows the dynamical mass ($24^{+6}_{-4} \, \mathrm{M_{Jup}}$) and separation ($0\farcs37$) of HIP~21152~B, which agrees with the prediction.
    \label{fig:predmass}}
\end{figure}

HIP~21152 is an F5 star in the ${\sim}\SI{650}{Myr}$ Hyades cluster \citep{perrymanHyadesDistanceStructure_1998,lodieuAgeHyadesLithium_2020}. Here, we report the discovery and atmospheric characterization of HIP~21152~B, a $24^{+6}_{-4} \, \mathrm{M_{Jup}}$ brown dwarf orbiting at a separation of ${\sim}\SI{370}{mas}$ (${\sim} \SI{16}{au}$) and the first directly imaged T-dwarf companion in the Hyades. Due to the precise age and metallicity constraints from the cluster membership, HIP~21152~B offers an excellent opportunity to robustly test substellar atmospheric and evolutionary models. In parallel to our independent detection, \citet{kuzuharaDirectImagingDiscovery_2022} and \citet{bonavitaResultsCopainsPilot_2022} identified this companion with SCExAO/CHARIS, Keck/NIRC2, and VLT/SPHERE. Our paper presents the synthesis of all available data on the system. We combine our new Keck/NIRC2 $L'$-band photometry and SCExAO/CHARIS ($1.15-\SI{2.39}{\mu m}$) spectrum, together with the VLT/SPHERE ($0.95-\SI{1.65}{\mu m}$) spectrum from \citet{bonavitaResultsCopainsPilot_2022}, RVs from \citet{kuzuharaDirectImagingDiscovery_2022}, and all available astrometry to carry out a comprehensive orbital and atmospheric characterization of the system.

\begin{deluxetable}{lcc}
\tabletypesize{\footnotesize}
\tablecaption{\label{tab:prop}Properties of HIP 21152 AB}
\tablehead{\colhead{Property} & \colhead{Value} & \colhead{Refs}}
\startdata
\multicolumn{3}{c}{HIP 21152 A} \\
\hline
$\alpha_{2000.0}$ & 04:32:04.81 & 1\\
$\delta_{2000.0}$ & $+05$:24:36.2 & 1\\
$\pi$ (mas) & $23.109 \pm 0.028$ & 1\\
Distance (pc) & $43.27 \pm 0.05$ & 1\\
$\mu_{\alpha,\mathrm{EDR3}}$\tablenotemark{a} ($\si{mas.yr^{-1}}$) & $112.17 \pm 0.05$ & 2\\
$\mu_{\delta,\mathrm{EDR3}}$ ($\si{mas.yr^{-1}}$) & $7.76 \pm 0.05$ & 2\\
$a$\tablenotemark{b} ($\si{mas.yr^{-2}}$) & $0.041 \pm 0.004$ & 3\\
$a$\tablenotemark{b} ($\si{m.s^{-1}yr^{-1}}$) & $8.4 \pm 0.8$ & 3\\
SpT & F5V & 4\\
Mass\tablenotemark{c} ($\mathrm{M_\odot}$) & $1.40 \pm 0.05$ & 3\\
Age (Myr) & $650 \pm 100$ & 5, 6, 7\\
$T_{\mathrm{eff}}$ (K) & $6655 \pm 125$ & 8\\
$\log g$ (dex) & $4.3 \pm 0.2$ & 8\\
$\mathrm{[Fe/H]}$ (dex) & $0.13 \pm 0.05$ & 9\\
RV ($\si{km.s^{-1}}$) & $40.6 \pm 1.5$ & 3\\
$v \sin i$ ($\si{km.s^{-1}}$) & $45.6 \pm 1.8$ & 3\\
$\mathrm{RUWE_{EDR3}}$ & 0.950 & 1\\
$V$ (mag) & $6.352 \pm 0.002$ & 10\\
Gaia $G$ (mag) & $6.2678 \pm 0.0028$ & 1\\
$J$ (mag) & $5.593 \pm 0.024$ & 11\\
$H$ (mag) & $5.385 \pm 0.020$ & 11\\
$K_s$ (mag) & $5.333 \pm 0.021$ & 11\\
$W1$ (mag) & $5.348 \pm 0.053$ & 12\\
\hline
\multicolumn{3}{c}{HIP 21152 B} \\
\hline
Mass ($\mathrm{M_{Jup}}$) & $24^{+6}_{-4}$ & 3\\
SpT & $\mathrm{T0 \pm 1}$ & 3\\
$\log(L_{\mathrm{bol}}/\mathrm{L_\odot})$ (dex) & $-4.57 \pm 0.07$ & 3\\
$T_{\mathrm{eff}}$\tablenotemark{d} (K) & $1300 \pm 50$ & 3\\
\enddata
\tablenotetext{a}{Proper motion in R.A. includes a factor of $\cos \delta$.}
\tablenotetext{b}{Calculated from proper motion difference between Hipparcos-Gaia joint proper motion and Gaia EDR3 proper motion in \citet{brandtHipparcosgaiaCatalogAccelerations_2021}.}
\tablenotetext{c}{Determined by taking the mean and standard deviation of masses for HIP~21152 from \citet{vansadersFastStarSlow_2013}, \citet{douglasFactoryBeehiveIi_2014}, \citet{davidAgesEarlytypeStars_2015}, \citet{reinersRadialVelocityPhoton_2020}, \citet{allendeprietoFundamentalParametersNearby_1999}, \citet{roserDeepAllskyCensus_2011}, \citet{paceLithium67Main_2012}, \citet{kopytovaSingleStarsHyades_2016}, \citet{lodieu3dViewHyades_2019}, and \citet{bochanskiFundamentalPropertiesComoving_2018}.}
\tablenotetext{d}{Calculated using the companion's bolometric luminosity and its model-inferred radius of \SI{0.997 \pm 0.023}{R_{Jup}} (see Section \ref{sec:atm_model_comp}). The best fit atmospheric model had $T_{\mathrm{eff}} = \SI{1400}{K}$, while atmospheric retrievals produced $T_{\mathrm{eff}} \sim 1400 {-} \SI{1600}{K}$.}
\tablerefs{(1) \citet{gaiacollaborationGaiaEarlyData_2021}; (2) \citet{brandtHipparcosgaiaCatalogAccelerations_2021}; (3) This work; (4) \citet{oblakEstimationPhotometricSpectral_1981}; (5) \citet{gossageAgeDeterminationsHyades_2018}; (6) \citet{degennaroInvertingColormagnitudeDiagrams_2009}; (7) \citet{lodieuAgeHyadesLithium_2020}; (8) \citet{gebranChemicalCompositionDwarfs_2010}; (9) \citet{boesgaardLithiumHyadesHyades_1988}; (10) \citet{jonerHomogeneousPhotometryHyades_2006}; (11) \citet{skrutskieTwoMicronAll_2006}; (12) \citet{maroccoCatwise2020Catalog_2021}.}
\end{deluxetable}

\begin{figure*}[hbt!]
    \centering
    \includegraphics[width=\textwidth]{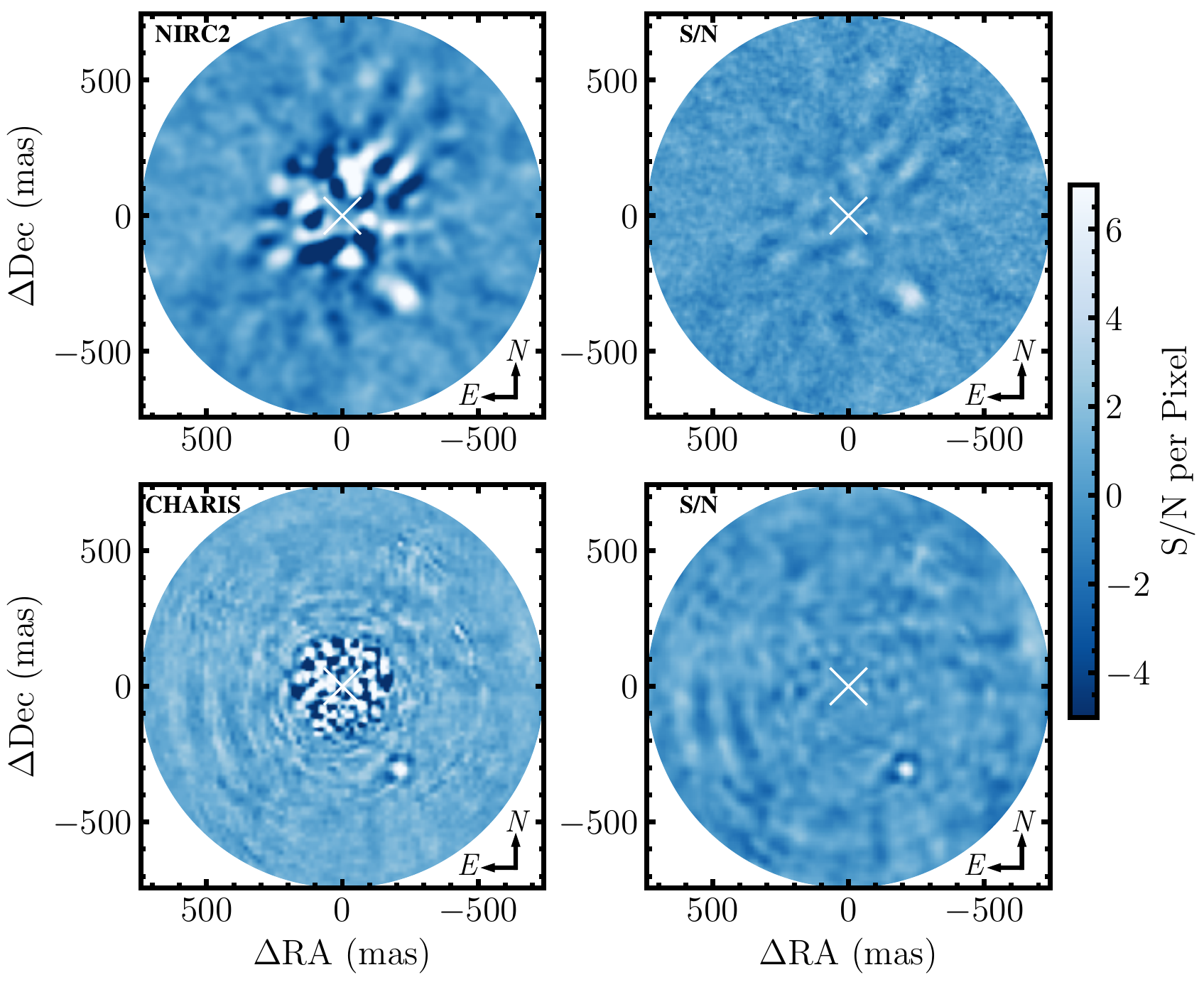}
    \caption{Keck/NIRC2 $L^\prime$ (top) and SCExAO/CHARIS $1.15{-}\SI{2.39}{\mu m}$ (bottom) imaging of HIP~21152~B. The left panels show the PSF-subtracted images for each dataset, while the right panels shows the corresponding S/N maps. Each frame is oriented so that north is up and east is to the left. The S/N maps are generated by measuring the flux in 0.5 FWHM-radius circular apertures at different separations and comparing the flux to noise estimated through additional non-overlapping apertures at the same separation. We apply a Gaussian filter with a standard deviation of 1.5 pixels to the Keck/NIRC2 $L^{\prime}$ imaging to enhance features at the scale of the recovered FWHM of the companion.
    \label{fig:red_img}}
\end{figure*}

\begin{deluxetable*}{ccccccc}
\tablecaption{\label{tab:rel_astrometry}Relative Astrometry of HIP 21152 B}
\tablehead{\colhead{Date} & \colhead{Epoch} & \colhead{Filter} & \colhead{Separation} & \colhead{PA} & \colhead{Instrument} & \colhead{Reference}\\ \colhead{(UT)} & \colhead{(UT)} & \colhead{ } & \colhead{(mas)} & \colhead{(\si{\degree})} & \colhead{ } & \colhead{ }}
\startdata
2019 Oct 26 & 2019.818 & $J$\tablenotemark{a} & $422.4 \pm 1.5$ & $217.06 \pm 0.20$ & SPHERE/IFS & \citet{bonavitaResultsCopainsPilot_2022} \\
2020 Oct 07 & 2020.767 & $1.15{-}\SI{2.39}{\mu m}$ & $408 \pm 4$ & $217.4 \pm 0.7$ & CHARIS & \citet{kuzuharaDirectImagingDiscovery_2022} \\
2020 Dec 04 & 2020.925 & $1.15{-}\SI{2.39}{\mu m}$ & $401 \pm 4$ & $216.7 \pm 0.7$ & CHARIS & \citet{kuzuharaDirectImagingDiscovery_2022} \\
2020 Dec 25 & 2020.982 & $L'$ & $406 \pm 6$ & $216.4 \pm 0.8$ & NIRC2 & \citet{kuzuharaDirectImagingDiscovery_2022} \\
2021 Oct 14 & 2021.785 & $1.15{-}\SI{2.39}{\mu m}$ & $379 \pm 5$ & $216.9 \pm 0.8$ & CHARIS & \citet{kuzuharaDirectImagingDiscovery_2022} \\
2021 Dec 21 & 2021.971 & $L'$ & $371 \pm 6$ & $217.8 \pm 0.8$ & NIRC2 & This Work \\
2022 Feb 27 & 2022.156 & $1.15{-}\SI{2.39}{\mu m}$ & $373 \pm 4$ & $215.4 \pm 0.5$ & CHARIS & This Work
\enddata
\tablenotetext{a}{Astrometry from $J$-band portion of SPHERE/IFS spectrum ($0.95{-}\SI{1.65}{\mu m}$).}
\end{deluxetable*}

\section{The Astrometric Accelerations as Dynamical Beacons Survey}
HIP~21152 was observed as part of the Astrometric Accelerations as Dynamical Beacons survey---an ongoing high-contrast imaging campaign targeting stars with small proper motion differences between Hipparcos and Gaia. The goal of our program is to image new long-period planets and brown dwarfs orbiting young stars. We aim to improve the efficiency of discoveries by observing stars with astrometric accelerations consistent with being caused by wide-separation substellar companions.

HIP~21152 exhibits a significant\footnote{$\chi^2 = 175$, which corresponds to $13\sigma$ with two degrees of freedom.} proper motion difference between Hipparcos and Gaia EDR3 in the Hipparcos-Gaia Catalog of Accelerations \citep[HGCA;][]{brandtHipparcosGaiaCatalog_2018, brandtHipparcosgaiaCatalogAccelerations_2021}. The HGCA provides three proper motion values: the Hipparcos proper motion, the Gaia EDR3 proper motion, and a joint Hipparcos-Gaia measurement from the difference in sky-positions between the two missions. The latter two measurements produce an average tangential acceleration of \SI{0.041 \pm 0.004}{mas.yr^{-2}}, which corresponds to a physical acceleration of \SI{8.4 \pm 0.8}{m.s^{-1}yr^{-1}} in the plane of the sky. 

Our strategy to prioritize potential targets is to compute joint probability maps in separation and companion mass for stars with significant low-amplitude HGCA accelerations. The procedure is summarized as follows. For each grid point in semi-major axis-companion mass space, 100 circular orbits\footnote{If we instead generate eccentric orbits, the mass predictions at a given semi-major axis increase slightly and take on a wider range of values, which has the effect of ``blurring out'' mass-separation predictions.} are generated with random orientations. The resulting acceleration distribution is then compared with the star's average acceleration and uncertainty through the K-S statistic \citep{kolmogorovSullaDeterminazioneEmpirica_1933, smirnovTableEstimatingGoodness_1948}. The mass prediction from HIP~21152's astrometric acceleration is shown in Figure \ref{fig:predmass}. This procedure implies that there is a brown dwarf within 1 arcsec or a stellar companion at wider separations. Our approach is similar to other analytical and numerical frameworks for predicting the nature of companions found with Hipparcos and Gaia \citep[e.g.,][]{kervellaStellarSubstellarCompanions_2019,kervellaStellarSubstellarCompanions_2022,derosaPossibleAstrometricSignature_2019}. By incorporating the sampling of the Hipparcos and Gaia missions, orbital curvature and aliasing is taken into account, and at wide separations the predictions mirror the power law relation between companion mass and separation from \citet{brandtPreciseDynamicalMasses_2019}.

\section{Host Star Properties \label{sec:props}}
HIP~21152 ($=$HD~28736, HR~1436, BD+05~674) is a bright ($V{=}\SI{6.352}{mag}$; \citealt{jonerHomogeneousPhotometryHyades_2006}) F5V star \citep{oblakEstimationPhotometricSpectral_1981} with a long history of being recognized as a reliable member of the Hyades cluster\footnote{Gaia EDR3 proper motions of $\mu_{\alpha} = \SI{112.17 \pm 0.05}{mas.yr^{-1}}$ $\mu_{\delta} = \SI{7.76 \pm 0.05}{mas.yr^{-1}}$ and our RV measurement of $\SI{40.6 \pm 1.5}{km.s^{-1}}$ from high-resolution spectroscopy (see Section \ref{sec:tull}) produce $UVW$ space motions of $U = \SI{-42.81 \pm 1.31}{km.s^{-1}}$, $V = \SI{-20.25 \pm 0.23}{km.s^{-1}}$, and $W = \SI{-0.19 \pm 0.70}{km.s^{-1}}$. \citet{gagneBanyanXiBanyan_2018} lists similar average space motions for the Hyades of $(U,V,W) = (-42.27,-18.79, -1.47) \ \si{km.s^{-1}}$. BANYAN-$\Sigma$ \citep{gagneBanyanXiBanyan_2018} gives a 99.5\% membership probabiliy in the Hyades for the EDR3 proper motions and our RV measurement.} \citep[e.g.,][]{vanbuerenStructureHyadesCluster_1952,perrymanHyadesDistanceStructure_1998,lodieu3dViewHyades_2019,gaiacollaborationGaiaEarlyData_2021a}. HIP~21152 has an effective temperature of \SI{6655 \pm 125}{K} and surface gravity of \SI{4.3 \pm 0.2}{dex} \citep{gebranChemicalCompositionDwarfs_2010}. Its metallicity is $\mathrm{[Fe/H]} = \SI{0.13 \pm 0.05}{dex}$ \citep{boesgaardLithiumHyadesHyades_1988}. This star has a Renormalised Unit Weght Error (RUWE; \citealt{lindegrenRenormalisingAstrometricChisquare_2018}) statistic in Gaia EDR3 of 0.95. RUWE values characterize the goodness-of-fit of the 5-parameter astrometric solution; values above 1.4 can indicate the presence of unresolved binaries \citep{stassunParallaxSystematicsPhotocenter_2021}. Therefore, there is no evidence in Gaia EDR3 that HIP~21152 is an unresolved stellar binary. Long-term radial velocities of HIP 21152 reported in \citet{kuzuharaDirectImagingDiscovery_2022} (see Figure \ref{fig:orbit_panel}) have an RV jitter of \SI{39}{m/s}. Assuming coplanarity with HIP~21152~B ($i \approx 95{}^\circ$; Section \ref{sec:orbitfit}), a \SI{0.1}{M_\odot} binary companion would produce RV semi-amplitudes of \SI{7.7}{km/s} at \SI{0.1}{au}, \SI{3.4}{km/s} at \SI{0.5}{au}, and \SI{2.4}{km/s} at \SI{1}{au}. Variations at this level are not seen in HIP~21152's radial velocities.

The Hyades is the closest open cluster to the Sun \citep{lodieu3dViewHyades_2019}. The core radius is about \SI{3}{pc} and the tidal radius is about \SI{10}{pc} \citep{perrymanHyadesDistanceStructure_1998,roserDeepAllskyCensus_2011}. Stars within the tidal radius are generally bound, while stars beyond that radius are likely unbound due to tidal stripping from the Galaxy and are therefore less reliable to identify. \citet{lodieu3dViewHyades_2019} identified 710 candidate members within \SI{30}{pc} of the center of the cluster, corresponding to a total mass of \SI{343}{M_\odot}. Two tidal tails have been found using Gaia DR2 astrometry, extending out to distances of up to \SI{170}{pc} from the cluster center \citep{meingastExtendedStellarSystems_2019,roserHyadesTidalTails_2019}. HIP~21152's distance of \SI{9.75}{pc} from the center of the cluster \citep{lodieu3dViewHyades_2019} places it at the approximate tidal radius. Spectroscopic abundance measurements of Hyades members consistently point to a super-solar metallicity of the Hyades ($\mathrm{[Fe/H]} = 0.1{-}\SI{0.2}{dex}$; \citealt{branchIronAbundanceHyades_1980,boesgaardChemicalCompositionOpen_1990,cummingsWiynOpenCluster_2017,takedaSpectroscopicDeterminationStellar_2020}).

There is some debate about the age of the Hyades, but most modern estimates fall between $600{-}\SI{800}{Myr}$. Isochrone fits to the main-sequence turnoff produce ages of about 600--\SI{650}{Myr} \citep{perrymanHyadesDistanceStructure_1998,lebretonHeliumContentAge_2001}. Fits using evolutionary models with an updated treatment of stellar rotation by \citet{brandtBayesianAgesEarlytype_2015} and \citet{brandtAgeAgeSpread_2015} yielded a somewhat older age of $\sim \SI{800}{Myr}$, although \citet{gossageAgeDeterminationsHyades_2018} found younger values of $\sim \SI{680}{Myr}$ using models with a different implementation of rotation. \citet{degennaroInvertingColormagnitudeDiagrams_2009} determined a white-dwarf cooling age of \SI{648 \pm 45}{Myr}. Recent efforts to measure the cluster's age using the lithium depletion boundary have produced an age of \SI{650 \pm 70}{Myr} \citep{martinLithiumDepletionBoundary_2018,lodieuLithiumHyadesL5_2018,lodieuAgeHyadesLithium_2020}. For this work, we adopt a fiducial age of \SI{650 \pm 100}{Myr} based on these age estimates. The properties of the host star are summarized in Table \ref{tab:prop}.

\section{Observations \label{sec:obs}}
\subsection{Keck/NIRC2 Adaptive Optics Imaging \label{sec:nirc2_obs}}
We obtained high-contrast imaging of HIP~21152 on UT 2021 December 21 with the NIRC2 camera at W.M. Keck Observatory in $L'$-band (3.426--\SI{4.126}{\micron}) with the Vector Vortex Coronagraph \citep[VVC;][]{serabynKeckObservatoryInfrared_2017}. The Differential Image Motion Monitor (DIMM) seeing for the night averaged $0\farcs65$. The observations were carried out with natural guide star adaptive optics \citep{wizinowichAstronomicalScienceAdaptive_2013} and the visible-light Shack-Hartmann wavefront sensor. Images were taken in sequences of 20--30 science frames using the Quadrant Analysis of Coronagraphic Images for Tip-tilt Sensing (\texttt{QACITS}) algorithm, which centers the star behind the vortex mask by applying small tip-tilt corrections after each exposure \citep{hubyPostcoronagraphicTiptiltSensing_2015, hubyOnskyPerformanceQacits_2017}. Each sequence includes an off-axis unsaturated frame of the star for flux calibration and sky background frames for both the science images and the flux calibration image. The science frames consist of 90 coadds each with integration times of \SI{0.3}{s} using a subarray of the central $512\times512$ pixels for shorter readout times. Exposures were taken in pupil-tracking mode to facilitate Angular Differential Imaging \citep[ADI;][]{liuSubstructureCircumstellarDisk_2004, maroisAngularDifferentialImaging_2006}. Excluding short pointing optimization frames taken as part of the \texttt{QACITS} algorithm, we obtained a total of 86 exposures of HIP~21152, amounting to \SI{2322}{s} (\SI{38.7}{min}) of integration time and $\SI{55.9}{\degree}$\footnote{4.6 FWHM at the separation of HIP~21152~B.} of frame rotation.

Science frames are first flat fielded and dark-subtracted. Cosmic rays are removed using the \texttt{L.A.Cosmic} algorithm \citep{vandokkumCosmicRayRejection_2001} and geometric distortions are corrected by applying the solution from \citet{serviceNewDistortionSolution_2016} for the narrow-field mode of the NIRC2 camera. The sky background is modeled and subtracted with Principal Component Analysis (PCA) using the Vortex Image Processing (\texttt{VIP}) package \citep{gomezgonzalezVipVortexImage_2017}. Following \citet{xuanCharacterizingPerformanceNirc2_2018}, we fit four principal components to the sky background frames, which equals the number of sky exposures for the science images. These sky principal components are then subtracted from each frame. The sky background is estimated and subtracted from off-axis flux calibration frames in a similar fashion. Following sky subtraction, the science frames are co-registered through a cross-correlation approach developed by \citet{guizar-sicairosEfficientSubpixelImage_2008} which is implemented in the ``register\_translation" function of \texttt{scikit-image} and utilized by \texttt{VIP}. We co-register to a median-combined frame. The absolute centering is determined by fitting a negative 2D Gaussian to the vortex core.

PSF subtraction is carried out and astrometry is measured using the \texttt{VIP} package via a similar approach to that described in \citet{fransonDynamicalMassYoung_2022}. PCA is used to estimate the PSF for each image in the sequence \citep{amaraPynpointImageProcessing_2012, soummerDetectionCharacterizationExoplanets_2012}. Then, the PSFs are subtracted and science frames are derotated and coadded. To select the total number of PCA components $n_{\mathrm{comp}}$, we run PSF subtractions from $n_{\mathrm{comp}} = 1$ to $n_{\mathrm{comp}} = 30$, measuring the resultant companion S/N in the reduced images. S/N is determined via the method of \citet{mawetFundamentalLimitationsHigh_2014}, which imposes a penalty at small separations to account for the small number of resolution elements. The highest S/N is produced for the reduction using 10 components, so we adopt that for the measurement of our astrometry. The reduced image and S/N map are shown in Figure \ref{fig:red_img}. We detect HIP 21152 B with a S/N of 7.9. In Appendix \ref{sec:nirc2_paco}, we also present an independent reduction with a modified version of the PAtch COvariances (PACO) algorithm \citep{flasseurExoplanetDetectionAngular_2018}, which recovers the companion with a comparable S/N.

To minimize the introduction of systematics from the PSF subtraction method, we use the negative companion injection approach \citep[e.g.,][]{lagrangeGiantPlanetImaged_2010, maroisImagesFourthPlanet_2010} to measure astrometry. A PSF template is generated by median combining the four sky-subtracted off-axis calibration frames taken over the sequence. A negative version of this is then injected into the science frames at the approximate position and with the approximate flux of the companion. After PSF subtraction, the residuals at the position of the injected template indicate how well the parameters of the input PSF-template match the true values of the companion. The astrometry and photometry is first optimized through the \texttt{AMOEBA} downhill simplex algorithm \citep{nelderSimplexMethodFunction_1965}, using the sum of the $\chi^2$ residuals within a 1.5 FWHM aperture to assess how well the injected parameters match the companion. The parameter space is then finely explored using the \texttt{emcee} affine-invariant Markov-chain Monte Carlo (MCMC) ensemble sampler \citep{foreman-mackeyEmceeMcmcHammer_2013}. We use a total of 100 walkers over 251 steps per walker (25,100 total steps) and discard the first 30\% of each chain as burn-in. We assess convergence by both visually inspecting the chains and performing multiple runs of the routine, which yields identical astrometry.

The VVC has a transmission profile that extends well beyond the inner working angle (${\sim} \SI{125}{mas}$) of the coronagraph (see Figure 4 of \citealt{serabynKeckObservatoryInfrared_2017}), causing a small loss of light at the separation of HIP~21152~B (${\sim} \SI{370}{mas}$) and to a lesser extent at the separation of the off-axis PSF (${\sim} \SI{500}{mas}$). Thus, we correct for the coronagraph at two points in this procedure: the creation of the PSF template from the off-axis frames and each time a negative PSF is injected in the MCMC run. This is performed by computing the radial distance to the vortex center on a pixel-by-pixel basis across the image and interpolating a recently simulated PSF transmission profile of the VVC (G. Ruane, priv. communication, 2022) to correct for the throughput.

The MCMC run produces chains of separation, position angle, and flux ratio. We convert the flux ratio into an apparent $L'$-band magnitude by scaling the flux ratio to the $W1$ magnitude\footnote{Here we assume that $L' - W1 = \SI{0}{mag}$, because $W1$ and $L'$ are in the Rayleigh-Jeans tail of the F5 host star's spectral energy distribution.} of HIP~21152 in CatWISE2020 ($W1=\SI{5.348 \pm 0.053}{mag}$; \citealt{maroccoCatwise2020Catalog_2021}). Uncertainties are produced in a similar manner to \citet{fransonDynamicalMassYoung_2022}, incorporating the standard deviation of each parameter from the MCMC run and the uncertainty in the distortion solution, north alignment, and plate scale from \citet{serviceNewDistortionSolution_2016}. Following \citet{wangKeckNirc2Band_2020}, we also add a \SI{4.5}{mas} \texttt{QACITS} centering uncertainty \citep{hubyOnskyPerformanceQacits_2017} in quadrature to account for the average pointing accuracy provided by the \texttt{QACITS} controller. This centering uncertainty is divided by the separation measurement before being added in quadrature to the position angle uncertainty. Our astrometry is shown in Table \ref{tab:rel_astrometry}. We measure a separation of $\rho=\SI{371 \pm 6}{mas}$, position angle of $\theta=217\fdg8 \pm 0\fdg8$, and $L'$-band contrast of $\Delta L'=\SI{9.31 \pm 0.09}{mag}$, which corresponds to an apparent magnitude of $L' = \SI{14.66 \pm 0.11}{mag}$, absolute magnitude of $M_{L'} = \SI{11.48 \pm 0.11}{mag}$, and flux density of $f_{\lambda = \SI{3.8}{\mu m}}=\SI[{multi-part-units = brackets, separate-uncertainty = true}]{7.3 \pm 0.7e-17}{W.m^{-2}\mu m^{-1}}$. The flux density conversion uses the Mauna Kea Observatories $L'$ zeropoint of \SI{5.31e-11}{W.m^{-2}\mu m^{-1}} from \citet{tokunagaMaunaKeaObservatories_2005}. Our value of the $L'$ apparent magnitude of HIP~21152~B is consistent with the value from \citet{kuzuharaDirectImagingDiscovery_2022} of $L' = \SI{15.01 \pm 0.12}{mag}$ to within $2.3\sigma$.

\begin{deluxetable}{lcc} 
\tablecaption{\label{tab:charis_spectrum}2022 Feb. CHARIS Spectrum}
\tablehead{\colhead{$\lambda$} & \colhead{$f_\lambda \times 10^{-17}$} & \colhead{$\sigma_{f_\lambda} \times 10^{-17}$}\\
\colhead{($\mathrm{\mu m}$)} & \colhead{($\mathrm{erg / s / cm^2 / \text{\AA}}$)} & \colhead{($\mathrm{erg / s / cm^2 / \text{\AA}}$)}}
\startdata
1.160 & $2.0$ & $1.3$\\
1.200 & $1.2$ & $1.1$\\
1.241 & $2.1$ & $0.5$\\
1.284 & $2.6$ & $0.8$\\
1.329 & $1.3$ & $0.6$\\
1.375 & $0.4$ & $0.7$\\
1.422 & $1.1$ & $0.4$\\
1.471 & $1.4$ & $0.3$\\
1.522 & $1.6$ & $0.4$\\
1.575 & $2.5$ & $0.3$\\
1.630 & $2.3$ & $0.3$\\
1.686 & $2.32$ & $0.23$\\
1.744 & $1.40$ & $0.15$\\
1.805 & $0.79$ & $0.19$\\
1.867 & $1.22$ & $0.21$\\
1.932 & $0.80$ & $0.14$\\
1.999 & $1.2$ & $0.3$\\
2.068 & $1.44$ & $0.16$\\
2.139 & $1.39$ & $0.18$\\
2.213 & $1.04$ & $0.25$\\
2.290 & $1.04$ & $0.24$\\
2.369 & $0.5$ & $0.6$
\enddata
\end{deluxetable}

\begin{figure*}[hbt!]
    \centering
    \includegraphics[width=\textwidth]{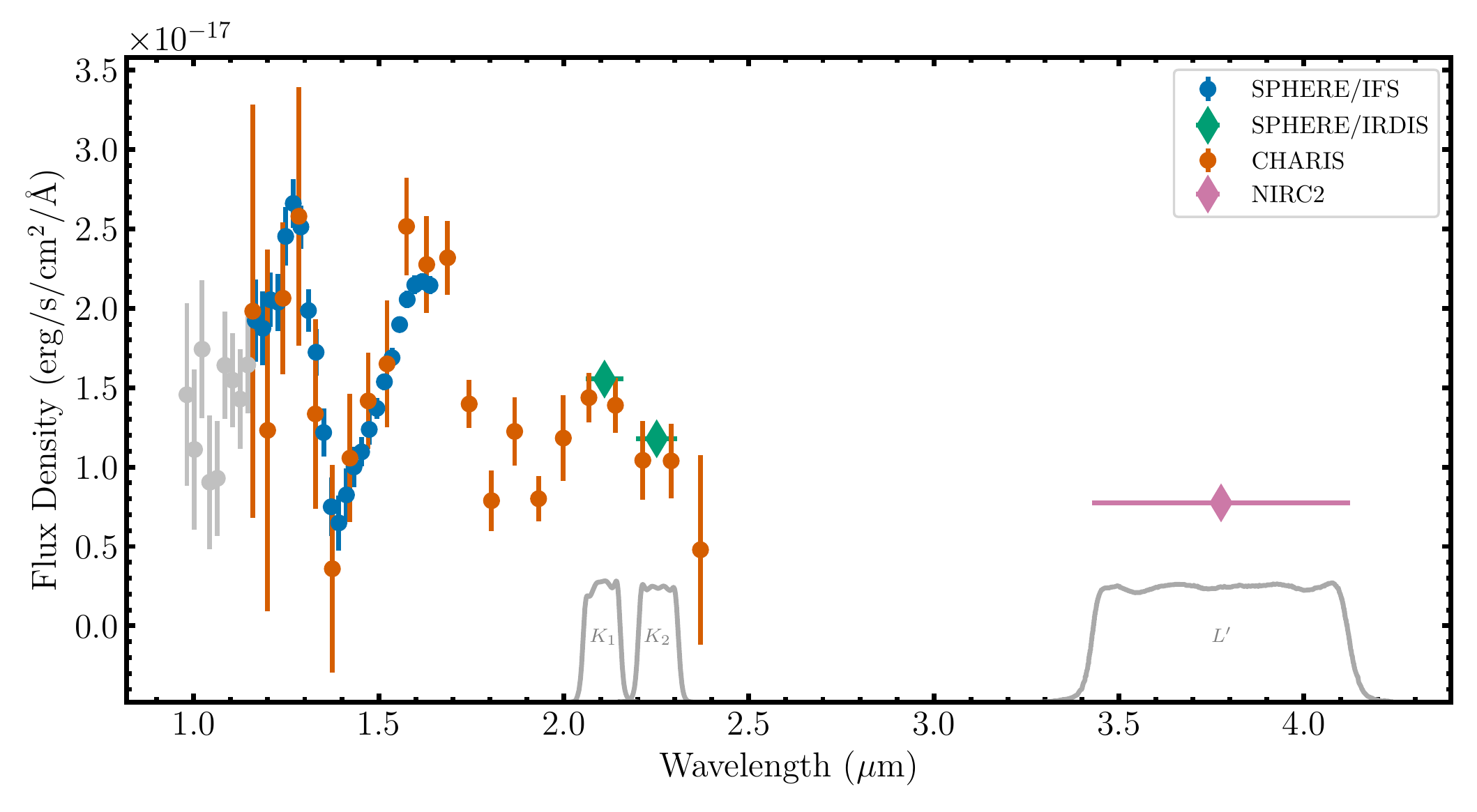}
    \caption{Spectra and photometry of HIP~21152~B. The blue points are the SPHERE/IFS spectrum from \citet{bonavitaResultsCopainsPilot_2022}, the green diamonds are the SPHERE/IRDIS $K_1$ and $K_2$ photometry from \citet{bonavitaResultsCopainsPilot_2022}, the orange points are our new SCExAO/CHARIS spectrum, and the pink diamond is our Keck/NIRC2 $L^\prime$ photometry. Due to the modest discrepancy between the two reductions of the SPHERE/IFS spectrum at short wavelengths (Section \ref{sec:second_reduction}), we exclude the $\lambda < \SI{1.15}{\mu m}$ region (shown in grey) from our analysis.
    \label{fig:spectrum}}
\end{figure*} 

\subsection{SCExAO/CHARIS Adaptive Optics Imaging \label{sec:charis_imaging}}
We obtained high-contrast imaging and spectroscopy of HIP~21152~B with the Coronagraphic High Angular Resolution Imaging Spectrograph \citep[CHARIS;][]{groffLaboratoryTestingPerformance_2016} on the Subaru Telescope in low-resolution ($R \sim 20$), broadband ($1.15-\SI{2.39}{\mu m}$) mode on UT 2022 February 28. Wavefront correction was provided by both the facility AO188 system, which removes lower-order abberations, and the Subaru Coronagraphic Extreme Adaptive Optics instrument (SCExAO; \citealt{jovanovicSubaruCoronagraphicExtreme_2015}), which provides high-order correction with a 2000-actuator deformable mirror (DM) to achieve typical Strehl ratios ${>}80\%$. The host star was occulted with the \SI{113}{mas} Lyot coronagraph. The DIMM seeing for the night averaged 0\farcs68. These data were taken in pupil-tracking mode to facilitate ADI in addition to Spectral Differential Imaging (SDI; \citealt{maroisEfficientSpeckleNoise_2000,sparksImagingSpectroscopyExtrasolar_2002}) enabled by the IFS. We obtained 101 exposures each with an integration time of \SI{30}{s}, amounting to $\SI{20.4}{\degree}$ of frame rotation over the sequence. Spectrophotometric calibration is carried out using satellite spots---fainter copies of the host star PSF. Four satellite spots are generated at a separation of 11.2 $\lambda / D$ (\SI{463}{mas} in $H$-band) by applying a modulation to the DM of two sine waves with amplitudes of \SI{50}{nm} \citep{jovanovicArtificialIncoherentSpeckles_2015}.

The CHARIS spectrum was extracted using the CHARIS raw data reduction pipeline \citep{brandtDataReductionPipeline_2017} and the CHARIS post-processing pipeline implemented in the \texttt{pyKLIP} package \citep{wangPyklipPsfSubtraction_2015}. The CHARIS raw data reduction pipeline constructs 3D data cubes from raw detector readouts using a $\chi^2$ spectral extraction algorithm detailed in \citet{brandtDataReductionPipeline_2017}. The extracted 3D data cubes consist of 2D images with wavelength as the third dimension. We then use the CHARIS post-processing tools in \texttt{pyKLIP} to perform PSF subtraction, measure astrometry, and extract the companion's spectrum. The host star position is determined in each 2D slice by averaging the positions of the four satellite spots. The spot positions are found through a global fit of the satellite spots over all wavelength bins and exposures. PSF subtraction is carried out with the Karhunen Lo\'eve Image Processing algorithm \citep{soummerDetectionCharacterizationExoplanets_2012}, which is a PCA-based algorithm that uses a Karhunen Lo\'eve transform \citep{karhunenUberLineareMethoden_1947,loeveFonctionsAleatoiresSecond_1948} to construct eigenimages. A total of 30 principal components (KL modes) are used in the PSF subtraction. Figure \ref{fig:red_img} shows the wavelength-collapsed reduced image and S/N map. We detect HIP~21152~B at a S/N of 9.6.

The companion's spectrum and astrometry are determined by forwarding modeling its PSF with the implementation of \texttt{KLIP-FM} \citep{pueyoDetectionCharacterizationExoplanets_2016} in \texttt{pyKLIP}. \texttt{KLIP-FM} uses perturbation-based forward modeling to account for the impact of over-subtraction and self-subtraction and reduce the number of false negatives produced by classical KLIP. Our PSF models for \texttt{KLIP-FM} are generated by subtracting the background from the satellite spots, extracting a 15-pixel square region about the spot positions, and averaging over the four spots in each image and the spots from all exposures in the sequence at a given wavelength. This provides one PSF model per wavelength bin.

To optimize the position of the companion, we vary the injected companion parameters (separation and position angle) using \texttt{emcee}. We use 100 walkers over 1000 steps per walker ($10^5$ total steps) to sample the parameter space. Astrometric calibration is provided by a short CHARIS sequence of the known binary HIP 55507 taken on the same night (see Appendix \ref{sec:hip55507}). A plate scale of \SI{16.15 \pm 0.1}{mas/spaxel} is adopted based on previous CHARIS calibration sequences (Chen et al., in prep.). Uncertainties in the astrometry include the MCMC-based measurement errors, the plate scale uncertainty, and for position angle, the uncertainty from the north alignment using the calibrator ($\pm 0\fdg3$, Appendix \ref{sec:hip55507}). These sources of error are added in quadrature and the resulting astrometry is reported in Table \ref{tab:rel_astrometry}.

Spectral extraction is then carried out with the \texttt{extractSpec} module in \texttt{pyKLIP}, which implements the \texttt{KLIP-FM} framework to determine the spectrum of a companion, given its astrometry. The spectrum (in units of contrast) is then calibrated by interpolating Castelli-Kurucz model atmospheres to the effective temperature ($T_{\mathrm{eff}} = \SI{6655 \pm 125}{K}$; \citealt{gebranChemicalCompositionDwarfs_2010}) and surface gravity ($\log g = \SI{4.3 \pm 0.2}{dex}$; \citealt{gebranChemicalCompositionDwarfs_2010}) of the host star. Finally, we scale our CHARIS spectrum to the SPHERE/IFS spectrum of the companion using Equation 2 of \citet{cushingAtmosphericParametersField_2008} to calculate the scale-factor that minimizes the $\chi^2$ value between the two spectra. The values of the extracted and rescaled 2022 CHARIS spectrum are listed in Table \ref{tab:charis_spectrum}. Figure \ref{fig:spectrum} shows this spectrum alongside the companion's SPHERE/IFS spectrum and additional photometry.

\begin{figure}
    \centering
    \includegraphics[width=\linewidth]{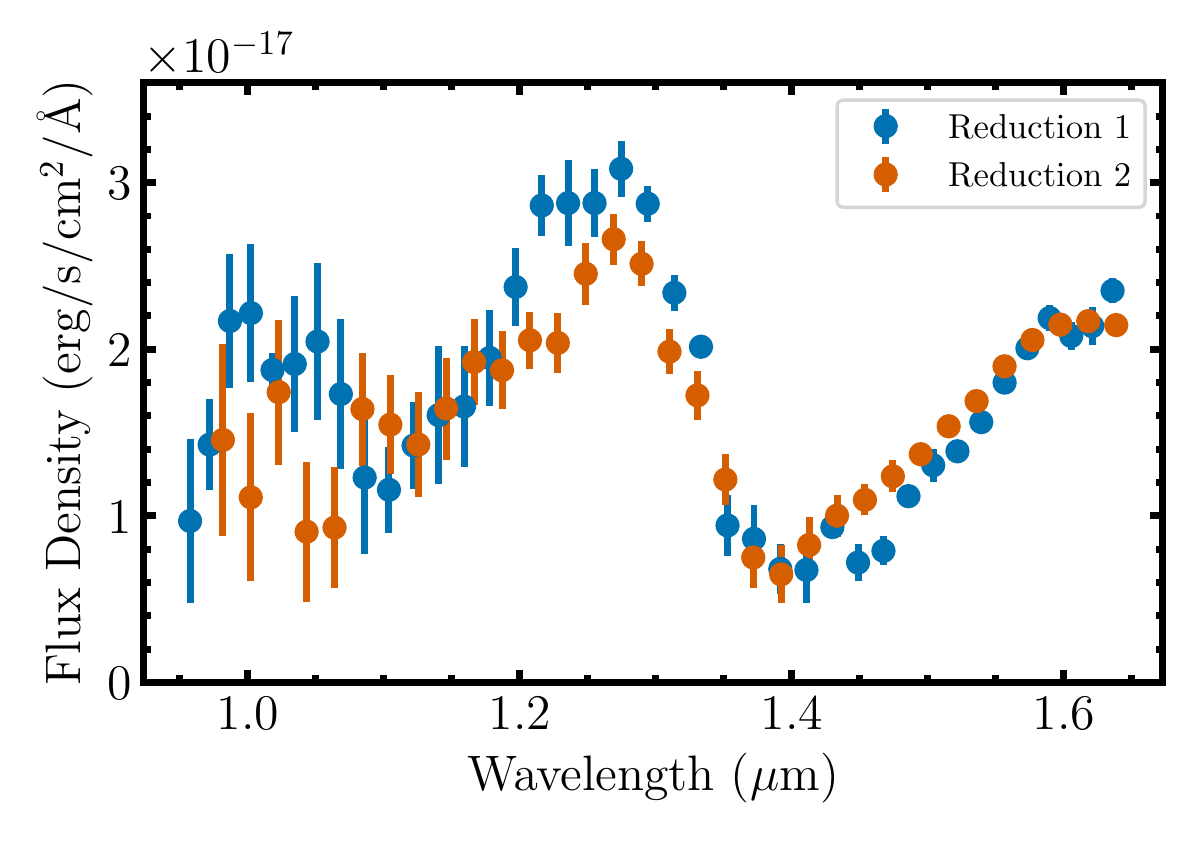}
    \caption{Comparison between two independent reductions of the SPHERE/IFS data. Reduction 1 uses the DRH pipeline complemented with additional steps described in \citet{langloisSphereInfraredSurvey_2021} and the SpeCal software to extract spectral cubes from the raw data, subtract the host star PSF, and obtain the companion's spectrum. Reduction 2 uses a modified version of the CHARIS pipeline and the TRAP post-processing pipeline to measure the companion's spectrum from the raw data. These two independent reductions produce similar results for $\lambda \gtrsim \SI{1.15}{\mu m}$. Reduction 2 tends to yield lower values for the companion flux blueward of $\SI{1.15}{\mu m}$ than Reduction 1. Due to this discrepancy, we exclude points with $\lambda < \SI{1.15}{\mu m}$ from this analysis. \label{fig:spec_comp}}
\end{figure}

\subsection{SPHERE/IFS Observations}
HIP~21152 was observed with the Spectro-Polarimetric High-contrast Exoplanet REsearch instrument \citep[SPHERE][]{beuzitSphereExoplanetImager_2019} on the Very Large Telescope (VLT) on UT 2019 November 26 as part of an independent high-contrast imaging survey of accelerating stars \citep{bonavitaResultsCopainsPilot_2022} selected using the Code for Orbital Parametrization of Astrometrically Inferred New Systems method \citep[COPAINS;][]{fontaniveNewMethodTarget_2019}.
The observations were carried out with the Integral Field Spectograph \citep[IFS;][]{claudiSphereIfsSpectro_2008} and Infra-Red Dual-beam Imaging and Spectroscopy \citep[IRDIS]{dohlenInfraredDualImaging_2008} instruments observing in parallel. The data were taken in \texttt{IRDIFS-EXT} mode, which routes $Y{-}H$-band light to the IFS and $K$-band to IRDIS. $K1$ and $K2$-band filters were used for the IRDIS dual-band imaging \citep[DBI;][]{viganPhotometricCharacterizationExoplanets_2010}.

The observing sequence consists of two flux-calibration sub-sequences in which the host star is offset from the coronagraph. In addition, two sub-sequences with four symmetric satellite spots for centering were obtained, a short sky observing sequence was taken for fine correction of the hot pixel variation during the night, and the science observations were acquired with the host star behind the coronagraph. The IRDIS and IFS data sets were reduced using the SPHERE Data Reduction and Handling (DRH) automated pipeline \citep{pavlovSphereDataReduction_2008} complemented with additional steps implemented via the SPHERE Data Center \citep{mesaPerformanceVltPlanet_2015,delormeSphereDataCenter_2017} for an improved wavelength calibration, and bad pixel and cross-talk correction \citep[see][for details]{langloisSphereInfraredSurvey_2021}. Images of the astrometric reference field 47 Tuc observed with SPHERE on UT 2019 September 07 and UT 2019 November 28 were used for calibration. The plate scale and true north values used are based on the long-term analysis of the GTO astrometric calibration described by \cite{maireSphereIrdisIfs_2016}. 

The SpeCal software \citep{galicherAstrometricPhotometricAccuracies_2018} was used to subtract the host star's PSF using Principal Component Analysis and Angular and Spectral Differential Imaging \citep[PCA-ASDI;][]{mesaPerformanceVltPlanet_2015} for the IFS and the Template Locally Optimised Combination of Images \citep[TLOCI;][]{maroisGpiPsfSubtraction_2014,galicherAstrometricPhotometricAccuracies_2018} approach for IRDIS. Using ADI in combination with negative companion injection, SpeCal provides the brightness and position of the companion, and the associated uncertainties, in the IFS $J$ and $H$ channels and in IRDIS $K1$ and $K2$ bands. For our orbit fit, we use the $J$-band value reported in \citet{bonavitaResultsCopainsPilot_2022}.

Spectral extraction of the IFS spectrum of HIP~21152~B is performed by injecting negative PSF templates generated from the flux calibration frames at the average companion position for a given wavelength slice and adjusting the flux density to minimize the root mean square of the residuals in a $9\times 9$-pixel square about the companion's mean position. Uncertainties for each point are estimated by repeating this process for five noise estimation points at the same separation as HIP~21152~B but offset in PA by a multiple of $\SI{60}{\degree}$. This produces the spectrum shown in Figure \ref{fig:spec_comp}. The values of this reduction of the spectrum are reported in Table \ref{tab:ifs_spectrum}, denoted by ``Reduction 1.''

\subsubsection{Independent Reduction of SPHERE/IFS Observations\label{sec:second_reduction}}
We carried out a second independent reduction of the IFS spectrum using the CHARIS pipeline \citep{brandtDataReductionPipeline_2017} to extract the spectral data cubes from the raw data and the TRAP post-processing pipeline \citep{samlandTrapTemporalSystematics_2021} to fit the planet signal and detrend the systematic noise. The CHARIS pipeline has been recently adapted to reduce SPHERE/IFS data \citep{samlandSpectralCubeExtraction_2022} which offers several advantages over the DRH pipeline: 1) it implements improved spectral extraction methods such as `optimal extraction' and `least-square fitting' using an instrument model, 2) it provides a correct wavelength solution natively without re-interpolating the data, 3) it uses the instrument model for cross-talk correction, and 4) it significantly reduces artifacts in low SNR channels. We use this pipeline to extract 3D data cubes from each raw detector readout using the optimal extraction approach \citep{brandtDataReductionPipeline_2017}. Analogous to the DRH pipeline, the position of the star is measured based on the satellite spots before and after the coronagraphic sequence. However, the unsaturated PSF images (flux frames) taken before and after the ADI sequence are relatively under-exposed in some wavelengths. We therefore use the satellite spots to obtain a higher-SNR unsaturated PSF model. The closest coronagraphic raw frames were first scaled and subtracted from the star center frame before extracting the images using the CHARIS pipeline. This effectively removes the static speckle background. The four satellite spots were then combined and averaged across all available frames. We determined a single star-to-spot flux ratio between the calibrated flux frames and the satellite spot frames as measured in a circular aperture ($R=3$ pixel) and averaged over all wavelengths to scale the satellite spot PSF to the flux-calibrated unsaturated stellar PSF frames. The first two unsaturated flux frames after the coronagraphic sequence were excluded as they showed signs of persistence.

The TRAP post-processing was performed using 20\% of the available principal components to detrend the temporal systematics and otherwise adopted default parameters. The reduction was done channel-by-channel without using SDI, such that there is no overfitting or bias from training data on other spectral channels. The spectrum was then extracted at the best-fit position determined from the wavelength-combined detection image. We calibrate this spectrum using a PHOENIX-Gaia model spectrum appropriate for the temperature and surface gravity of the host star, with the absolute calibration provided by using Equation 2 of \citep{cushingAtmosphericParametersField_2008} to calculate the appropriate scale factor to anchor this reduction to Reduction 1. The values of this reduction of the spectrum is reported in Table \ref{tab:ifs_spectrum} under ``Reduction 2."

Figure \ref{fig:spec_comp} displays the two reductions of the SPHERE/IFS data. While they are similar for $\lambda \gtrsim \SI{1.15}{\mu m}$, they are discrepant at shorter wavelengths. We thus exclude flux densities at $\lambda < \SI{1.15}{\mu m}$ from our analysis. The peak flux density in $J$-band is lower for Reduction 2 than Reduction 1, with the two reductions having different spectral shapes from $1.2{-}\SI{1.25}{\mu m}$. Additionally, Reduction 2 produces a slightly higher flux from ${\sim}1.4{-}\SI{1.55}{\mu m}$. Based on a comparison of atmospheric retrievals using both versions of the spectrum, we ultimately adopt Reduction 2 for our spectral analysis, model comparison, and retrieval work in Section \ref{sec:spec_analysis}.\footnote{Atmospheric retrievals using Reduction 1 (Appendix \ref{sec:retrieval_red1}) produce significantly higher effective temperatures of $\sim 1800{-}\SI{1900}{K}$ than retrievals restricted to the CHARIS spectrum ($T_{\mathrm{eff}} \sim \SI{1400}{K}$) or retrievals that use Reduction 2 ($T_{\mathrm{eff}} \sim 1400{-}\SI{1600}{K}$; Section \ref{sec:retrieval}), likely owing to the increased flux in the $J$-band. This higher effective temperature is inconsistent with the effective temperature from the Stefan-Boltzmann law using the companion's bolometric luminosity and a model-inferred radius ($T_{\mathrm{eff}}{=}\SI{1300 \pm 50}{K}$; Section \ref{sec:atm_model_comp}).} A comparison with atmospheric models and a suite of retrievals using Reduction 1 are presented in Appendices \ref{sec:atm_model_comp_red1} and \ref{sec:retrieval_red1}, respectively.
 
\subsection{Tull Coud\'e Spectra \label{sec:tull}}

We are uniformly acquiring high-resolution optical spectra of accelerating stars to further vet for close binaries and characterize targets in our survey. As part of this effort, we observed HIP~21152 with the Tull Coud\'e spectrograph \citep{tullHighresolutionCrossdispersedEchelle_1995} at McDonald Observatory's 2.7-m Harlan J. Smith telescope on UT 2021 October 15.  Two spectra were obtained at an airmass of 1.11 with integration times of \SI{50}{s} and \SI{100}{s}. The 1$\farcs$2 slit was used, resulting in a resolving power of $R$=60,000 in a setup covering 56 orders (with gaps between orders) from \SI{3870}{\angstrom}--\SI{10450}{\angstrom}. Several slowly rotating RV standards spanning a range of spectral types were targeted in the same setup throughout the night.

Spectra are extracted and wavelength calibrated using a custom pipeline. 2-dimensional curved spectral traces are fit with a polynomial and remapped to a 1-dimensional horizontal trace, then each order is optimally extracted following \citet{horneOptimalExtractionAlgorithm_1986}.
Spectral orders are continuum normalized by dividing a fifth-order polynomial fit to each extracted spectrum.
Wavelength solutions for each order are determined using a ThAr emission lamp spectrum taken on the same night
with an identical setup as the science target.  A ThAr emission line list is compiled from \citet{lovisNewListThorium_2007} and \citet{murphySelectionTharLines_2007},
and a third-order polynomial solution is derived for each order to map pixels to wavelengths.

The radial velocity and projected rotational velocity of HIP~21152 are derived by cross correlating each order with observations
of the stable F2 star HD 207978.  
Because some orders also contain strong telluric features, especially at red-optical wavelengths,
only 42 orders are used for the RV and $v \sin i$ measurements.
Each order of the standard is individually broadened by convolving it with a broadening kernel assuming
a linear limb darkening law with a limb-darkening coefficient ($\epsilon$) of 0.6 \citep{grayObservationAnalysisStellar_2005a}.
A fine grid of $v \sin i$ values is sampled; the $v \sin i$ and RV values that produce 
the highest cross-correlation function peak are adopted for each order.
RVs relative to HD 207978 are corrected for barycentric motion and shifted to an absolute scale using the
measured radial velocity of \SI{18.92 \pm 0.02}{km.s^{-1}} from \citet{soubiranGaiaDataRelease_2018}.
Projected rotational velocities are computed as the quadrature sum of the intrinsic broadening of HD 20797---\SI{6.2}{km.s^{-1}} based on
four measurements from \citet{glebockiVizierOnlineData_2005}---and the additional
broadening applied to the standard star spectrum.

After removing outlier measurements following a biweight estimator (Equation~9 in \citealt{beersMeasuresLocationScale_1990}),
we determine an RV of \SI{40.6 \pm 2.2}{km.s^{-1}} and a  $v \sin i$ value of \SI{45.8 \pm 2.5}{km.s^{-1}} for the 
50-s integration observation.  For the 100~s observation we find an RV of \SI{40.6 \pm 2.0}{km.s^{-1}} and a 
$v \sin i$ value of \SI{45.3 \pm 2.5}{km.s^{-1}}.
We adopt the weighted mean and weighted standard deviation of these two observations for our final measurements: 
RV = \SI{40.6 \pm 1.5}{km.s^{-1}} and $v \sin i=\SI{45.6 \pm 1.8}{km.s^{-1}}$.
These are in good agreement with values from the literature.  For example, the RV from $Gaia$ DR2 
\citep{gaiacollaborationGaiaDataRelease_2018} is 41.5 $\pm$ 0.5 km s$^{-1}$
and the projected rotational velocity of HIP~21152 from \citet{glebockiVizierOnlineData_2005} is \SI{42}{km.s^{-1}} based on three measurements.

\begin{figure*}
    \centering
    \includegraphics[width=0.65\textwidth]{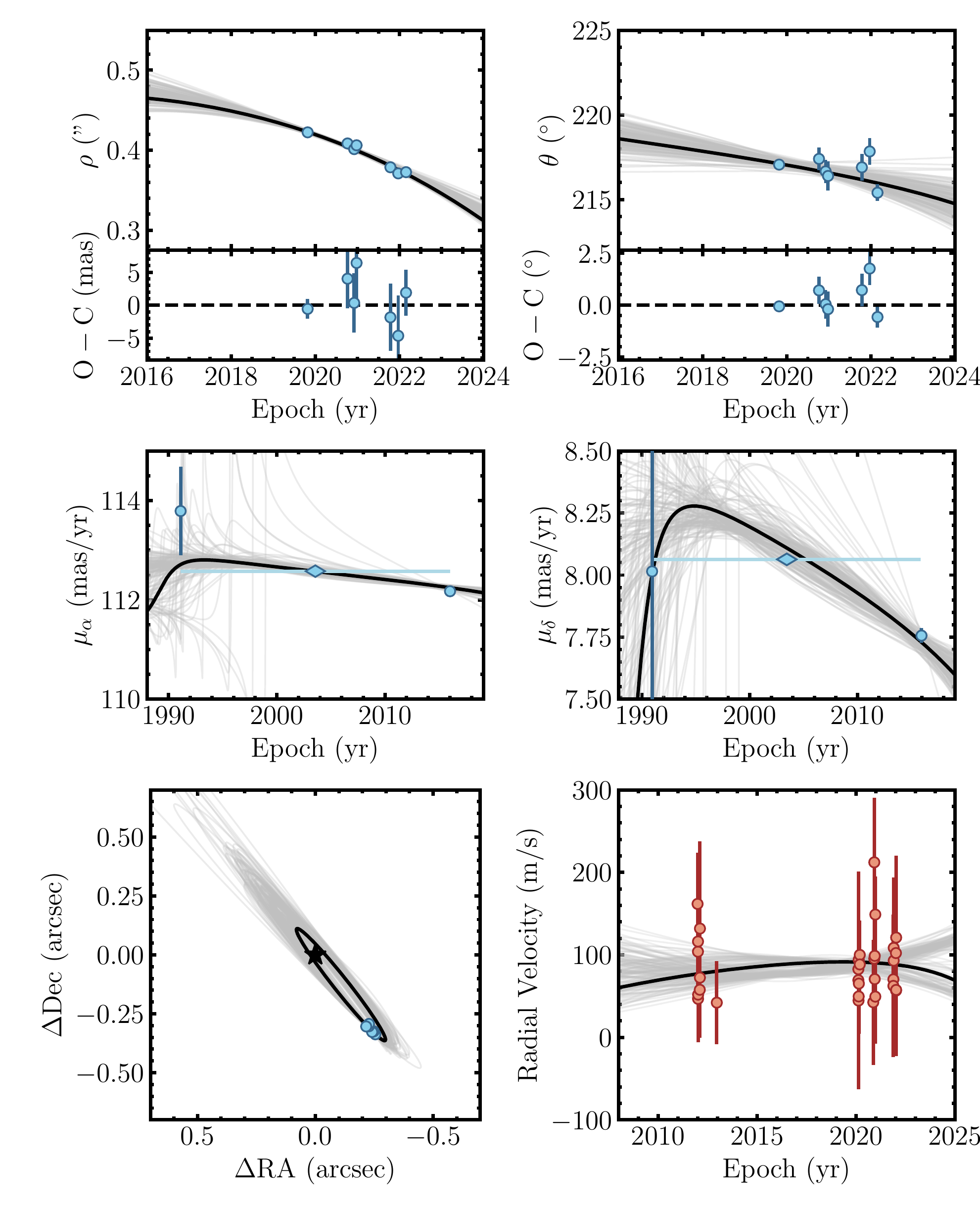}
    \caption{Orbit fit of HIP~21152~B compared with relative astrometry (top panels), HGCA proper motions (middle panels), sky-projected orbit (bottom left panel), and radial velocities (bottom right panel). The gray curves show 150 randomly drawn orbits from the MCMC chains. The maximum-likelihood orbit is highlighted in black. The middle proper motion (diamond) in the proper motion panels is a joint proper motion from the difference in sky-position between Hipparcos and Gaia EDR3. It is an average proper motion that reflects the reflex motion over the entire 25 years between the missions. 
    \label{fig:orbit_panel}}
\end{figure*}

\begin{figure*}
    \centering
    \includegraphics[width=0.8\textwidth]{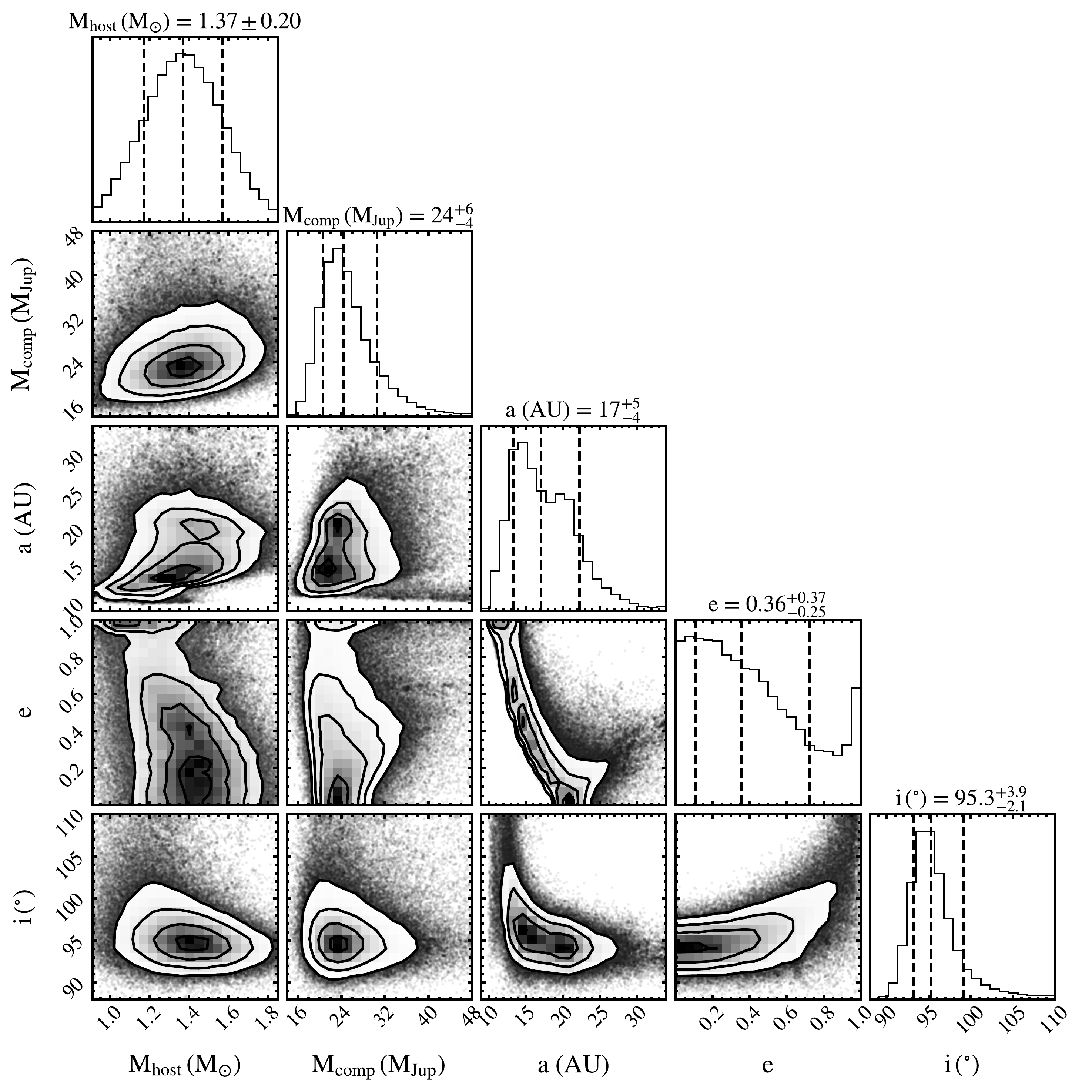}
    \caption{Joint posterior distributions for the primary mass ($M_{\mathrm{host}}$), companion mass ($M_{\mathrm{comp}}$), semi-major axis ($a$), eccentricity ($e$), and inclination ($i$) from the orbit fit of HIP~21152~B. Diagonal panels show the marginalized posterior for each parameter. We measure a dynamical mass of $24^{+6}_{-4} \, \mathrm{M_{Jup}}$ for HIP~21152~B.
    \label{fig:orbit_corner}}
\end{figure*}

\begin{deluxetable*}{lccc} 
\tablecaption{\label{tab:elements}HIP 21152 B Orbit Fit Results}
\tablehead{\colhead{Parameter} & \colhead{Median $\pm 1\sigma$} & \colhead{95.4\% C.I.} & \colhead{Prior}}
\startdata
\multicolumn{4}{c}{Fitted Parameters} \\
\hline
$M_{\mathrm{comp}}$ $(\mathrm{M_{Jup}})$ & ${24}_{-4}^{+6}$ & (18, 46) & $1/M_{\mathrm{comp}}$ (log-flat)\\
$M_{\mathrm{host}}$ $(\mathrm{M_\odot})$ & $1.37 \pm 0.20$ & (0.98, 1.77) & $\SI{1.4 \pm 0.2}{M_\odot}$ (Gaussian)\\
$a$ $(\mathrm{AU})$ & ${17}_{-4}^{+5}$ & (11, 36) & $1/a$ (log-flat)\\
$i$ $(\si{\degree})$ & ${95.3}_{-2.1}^{+3.9}$ & (91.3, 125.0) & $\sin (i)$, $\SI{0}{\degree} < i < \SI{180}{\degree}$\\
$\sqrt{e} \sin{\omega}$ & ${0.1}_{-0.4}^{+0.3}$ & (-0.5, 0.6) & Uniform\\
$\sqrt{e} \cos{\omega}$ & ${0.0}_{-0.7}^{+0.6}$ & (-0.9, 0.9) & Uniform\\
$\Omega$ $(\si{\degree})$ & ${39.3}_{-1.5}^{+2.6}$\tablenotemark{a} & (36.1, 50.0)\tablenotemark{a} & Uniform\\
$\lambda_{\mathrm{ref}}$ $(\si{\degree})$\tablenotemark{b} & ${180}_{-30}^{+160}$ & (10, 360) & Uniform\\
Parallax $(\si{mas})$ & $23.109 \pm 0.030$ & (23.049, 23.169) & $\SI{23.109 \pm 0.028}{mas}$ (Gaussian)\\
$\mu_\alpha$ ($\si{mas.yr^{-1}}$) & ${112.37}_{-0.05}^{+0.06}$ & (112.28, 112.55) & Uniform\\
$\mu_\delta$ ($\si{mas.yr^{-1}}$) & ${7.80}_{-0.06}^{+0.07}$ & (7.68, 7.99) & Uniform\\
RV Jitter $\sigma_{\mathrm{RV}}$ ($\si{m.s^{-1}}$) & ${<}5.293$\tablenotemark{c} & . . . & $1/\sigma_{\mathrm{RV}}$ (log-flat), $\sigma_{\mathrm{RV}} \in (0, 1000\si{m.s^{-1}}]$\\
\hline
\multicolumn{4}{c}{Derived Parameters} \\
\hline
$P$ (yr) & ${59}_{-16}^{+31}$ & (34, 183) & . . .\\
$e$ & ${0.36}_{-0.25}^{+0.37}$ & (0.02, 0.98) & . . .\\
$\omega$ $(\si{\degree})$ & ${160}_{-130}^{+160}$ & (10, 350) & . . .\\
$T_0$ $(\mathrm{JD})$ & ${2449000}_{-4000}^{+15000}$ & (2431000, 2468000) & . . .\\
$q$ $(=M_{\mathrm{comp}}/M_{\mathrm{host}})$ & ${0.017}_{-0.003}^{+0.005}$ & (0.012, 0.035) & . . .
\enddata
\tablenotetext{a}{The $\Omega$ posterior consists of two distinct peaks separated by \SI{180}{\degree}. The values shown in the table correspond to the higher peak. The other peak is located at ${218.4}_{-1.6}^{+2.3}{}^\circ$ with a 95.4\% confidence interval of (205.9, 227.6).}\tablenotetext{b}{Mean longitude at the reference epoch of 2010.0.}
\tablenotetext{c}{$2\sigma$ upper limit. The median value is \SI{0}{m.s^{-1}}.}

\end{deluxetable*}
%file: ../../../results/HIP21152_chain016_10e6.fits
%burnin: 100000

\section{3D Orbit and Dynamical Mass \label{sec:orbitfit}}
Our orbit fit incorporates all available relative astrometry and radial velocities from this work, \citet{kuzuharaDirectImagingDiscovery_2022}, and \citet{bonavitaResultsCopainsPilot_2022}. It also includes the astrometric acceleration between Hipparcos and Gaia-EDR3, which enables the measurement of a precise dynamical mass of HIP~21152~B. The fit is performed with \texttt{orvara} \citep{brandtOrvaraEfficientCode_2021}, which uses the parallel-tempered MCMC (PT-MCMC) ensemble sampler in \texttt{emcee} to sample the orbital parameter posteriors. In PT-MCMC, individual chains span a range of ``temperatures." Low-temperature chains more accurately sample the neighborhood of a $\chi^2$ minimum, while higher-temperature chains are capable of escaping local minima and accessing the entire parameter space. Periodically, chains swap positions, which enables low-temperature chains to be sensitive to additional peaks in the posterior distribution accessed by the higher-temperature ones. This strategy is well-suited for orbit fitting of imaged companions, as their typically short orbit arcs can produce complex, multi-modal posteriors. 

For HIP 21152, we use 100 walkers, 20 temperatures, and $10^7$ total steps to sample the parameter space. The coldest chain is adopted as the posterior distribution. \texttt{orvara} fits nine quantities: the host star mass ($M_{\mathrm{host}}$), the companion mass ($M_{\mathrm{comp}}$), semi-major axis ($a$), inclination ($i$), eccentricity ($e$), argument of periastron ($\omega$), longitude of ascending node ($\Omega$), the longitude at the reference epoch of 2010.0 ($\lambda_{\mathrm{ref}}$), and an RV jitter term ($\sigma_{\mathrm{RV}}$). Eccentricity and argument of periastron are fit as $\sqrt{e} \sin \omega$ and $\sqrt{e} \cos \omega$ to avoid the Lucy-Sweeney bias against circular orbits \citep{lucySpectroscopicBinariesCircular_1971}. The code analytically marginalizes over instrumental RV zeropoints, parallax, and barycentric proper motion. 

We adopt uninformative priors for all quantities excluding primary mass. For the host-star mass, we use a broad Gaussian prior of $1.4 \pm 0.2 \, \mathrm{M_\odot}$\footnote{To verify that the host-star mass prior minimally impacts the resulting orbital elements, we also performed joint orbit fits with a narrow prior of \SI{1.40 \pm 0.10}{M_{Jup}} and a wide prior of \SI{1.4 
\pm 0.5}{M_{Jup}}. Both runs produced consistent orbit elements within $1\sigma$, with the narrow prior yielding $M_{\mathrm{comp}} = 24^{+6}_{-4} \, \mathrm{M_{Jup}}$, $a = 17^{+5}_{-3} \, \mathrm{au}$, $e = 0.35^{+0.31}_{-0.24}$, and $i = 95.2^{+3.0}_{-2.0} {}^\circ$ and the wide prior producing $M_{\mathrm{comp}} = 25^{+7}_{-5} \, \mathrm{M_{Jup}}$, $a = 17^{+5}_{-4} \, \mathrm{au}$, $e = 0.4^{+0.4}_{-0.3}$, and $i = 95^{+6}_{-2}{}^\circ$.}, which encompasses typical mass estimates of HIP~21152 in the literature (see Table \ref{tab:prop}) but also allows for potential systematic errors that may be larger than the dispersion of quoted values. Log-flat priors are adopted for the companion mass, semi-major axis, and RV jitter. An isotropic, $\sin i$ prior is used for inclination. All other quantities are assigned uniform priors. The first 50\% of each walker is discarded as burn-in (\num{5e6} steps). We assess convergence by checking that different portions of the walkers and starting positions produce the same posteriors.

Results of the orbit fit are listed in Table \ref{tab:elements}. Figure \ref{fig:orbit_panel} compares the relative astrometry, HGCA proper motions, and radial velocities against a sample of the orbit solutions. The black curves highlight the maximum likelihood orbit. Note that the central point on the proper motion panels is a joint Hipparcos-Gaia proper motion from the difference in sky position between the two epochs. Unlike the Hipparcos and Gaia proper motions, this point is an average proper motion over the 25-year time baseline between the missions. Figure \ref{fig:orbit_corner} shows the posterior distributions for a subset of the orbital elements. We measure a semi-major axis of $a = 17^{+5}_{-4} \, \mathrm{au}$, a nearly edge-on inclination of $95.3^{+3.9}_{-2.1} {}^{\circ}$, and an orbital period of $P = 59^{+31}_{-16} \, \mathrm{yr}$. The companion is moving towards its host star at a rate of ${\sim}\SI{25}{mas/yr}$. The orbit fit yields a dynamical mass for HIP~21152~B of $M_{\mathrm{comp}} = 24^{+6}_{-4} \, \mathrm{M_{Jup}}$, which corresponds to a mass ratio of $q = 0.017^{+0.005}_{-0.003}$.

Our dynamical mass measurement is consistent with the value from \citet{bonavitaResultsCopainsPilot_2022} of \SI{22 \pm 7}{M_{Jup}} which was determined from the separation of the companion in the SPHERE/IFS data and the host star's proper motion difference. \citet{kuzuharaDirectImagingDiscovery_2022} conducted a joint orbit fit of relative astrometry, radial velocities, and HGCA proper motions, finding values of $M_{\mathrm{comp}} = 28^{+9}_{-5} \, \mathrm{M_{Jup}}$, $a = 17^{+7}_{-4} \, \mathrm{au}$, and $i = 105^{+18}_{-7} {}^{\circ}$ that are consistent with our fit. Our uncertainties on the dynamical mass and inclination measurements are 29\% and 76\% lower than the uncertainties from \citet{kuzuharaDirectImagingDiscovery_2022}, respectively. This is likely due to our orbit fit's relative astrometry covering a longer time-baseline. Eccentricity is poorly constrained in both orbit fits. Continued orbital monitoring with high-contrast imaging and RVs, particularly as the companion approaches periastron, will be essential in improving the constraint on the companion's eccentricity to facilitate comparisons with observed brown dwarf and giant planet eccentricity distributions \citep{bowlerPopulationlevelEccentricityDistributions_2020}.

\section{Discussion}

\subsection{Evolutionary Model Comparison \label{sec:ev_model_comp}}
Our model-independent mass of HIP~21152~B and age constraint from the host star's Hyades membership enable a direct comparison against the mass predictions of different substellar evolutionary models. We select a variety of hot-start model grids that cover the lumifnosity and age of the companion: \citet{burrowsNongrayTheoryExtrasolar_1997}, Cond \citep{baraffeEvolutionaryModelsCool_2003}, \texttt{ATMO-2020} \citep{phillipsNewSetAtmosphere_2020}, and the \citet{saumonEvolutionDwarfsColor_2008} models with three cloud prescriptions (no clouds, hybrid, and cloudy).

To determine the luminosity of HIP~21152~B, we synthesize $K_s$-band photometry from the companion's CHARIS spectrum. As an estimate of the uncertainty from the absolute flux calibration of the CHARIS spectrum, we scale the CHARIS spectrum to the $K_s$-band magnitude of $K_s = \SI{16.57 \pm 0.17}{mag}$ reported in \citet{kuzuharaDirectImagingDiscovery_2022}. The uncertainty of the resulting scale factor is then incorporated into the measurement of the photometry, in addition to the individual uncertainties on each flux density measurement in the spectrum. This produces $K_s = \SI{16.35 \pm 0.17}{mag}$, which corresponds to an absolute magnitude of $M_{K_s} = \SI{13.13 \pm 0.18}{mag}$ and a bolometric luminosity of $\log(L_{\mathrm{bol}}/\mathrm{L_\odot}) = \SI{-4.57 \pm 0.07}{dex}$, applying the $M_{K_s}{-}\log(L_{\mathrm{bol}} / L_\odot)$ relation from \citet{dupuyIndividualDynamicalMasses_2017}.

\begin{figure}
    \centering
    \includegraphics[width=\linewidth]{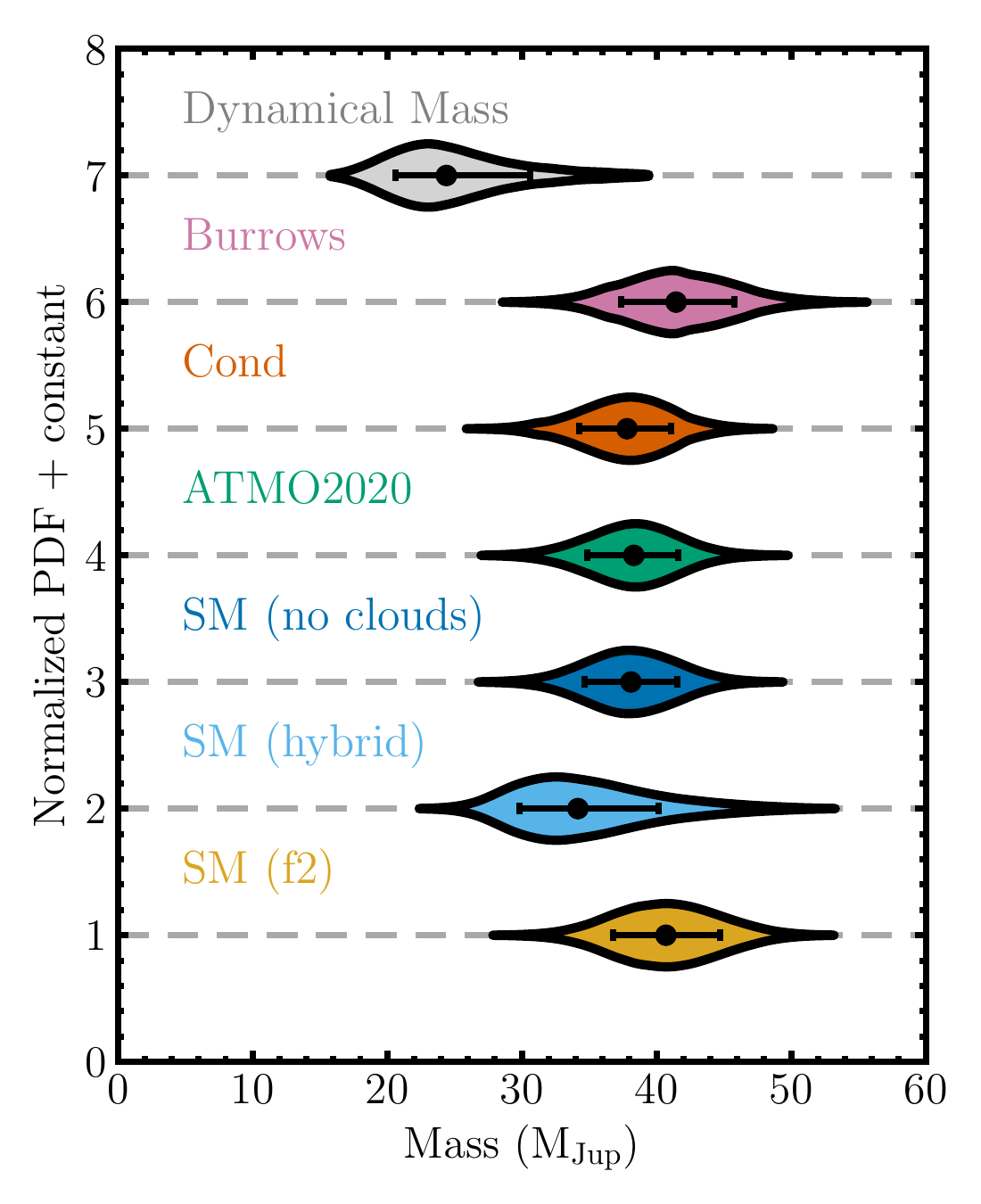}
    \caption{Comparison between the mass predictions of hot-start evolutionary models and the dynamical mass (top distribution) of HIP~21152~B. The median and 68.3\% confidence interval are highlighted for each distribution. The dynamical mass is lower than most model predictions by $1{-}2\sigma$. The hybrid \citet{saumonEvolutionDwarfsColor_2008} grid is most consistent with the dynamical mass, agreeing to within $1.2\sigma$.
    \label{fig:ev_model_comp_mass}}
\end{figure}

To generate predicted masses for each evolutionary model, we draw from the luminosity and age distributions $10^6$ times following \citet{fransonDynamicalMassYoung_2022}. For each of these trials, we then linearly interpolate the cooling curves for a given model to obtain a corresponding inferred mass. Figure \ref{fig:ev_model_comp_mass} compares the inferred masses for the array of hot-start models with our model-independent, dynamical mass of $24^{+6}_{-4} \, \mathrm{M_{Jup}}$. The model predictions are summarized in Table \ref{tab:model_comp}. We quantify the consistency between the model-inferred masses and the companion's dynamical mass by computing $P(M_{\mathrm{Inferred}}{>}M_{\mathrm{Dynamical}})$, the probability that a randomly drawn value from a given inferred mass distribution is higher than a random value drawn from the dynamical mass prior. If the two distributions are consistent with one another, $P(M_{\mathrm{Inferred}} {>} M_{\mathrm{Dynamical}})$ should be ${\approx}50\%$. Values below 50\% signify that the inferred mass is lower than the dynamical mass, while values above 50\% indicate that the inferred mass is higher than the dynamical mass. We determine $P(M_{\mathrm{Inferred}} {>} M_{\mathrm{Dynamical}})$ by drawing $10^6$ pairs of masses from each inferred mass distribution and the dynamical mass posterior; this probability is simply the fraction of draws where the inferred mass is higher than the dynamical mass.

We find that the model predictions are higher than the dynamical mass by $1{-}2\sigma$. The model that produces the most consistent inferred mass with the companion's true mass is the hybrid \citet{saumonEvolutionDwarfsColor_2008} model, which agrees to within $1.2\sigma$. This grid aims to represent the progression of brown dwarfs through their spectral sequence and the L/T transition by using cloudy ($f_{\mathrm{sed}} {=} 2$) model atmospheres for $T_{\mathrm{eff}} > \SI{1400}{K}$ (L dwarfs) and clear model atmospheres for $T_{\mathrm{eff}} < \SI{1200}{K}$ (T dwarfs). Between 1200 K and 1400 K, the atmosphere grid is linearly interpolated between the cloudy and clear models to emulate the breakup of iron and silicate clouds at the L/T transition. This makes these models better-suited for L/T transition objects and T dwarfs than the other models we consider here. The \texttt{ATMO-2020} \citep{phillipsNewSetAtmosphere_2020} grids are designed to model cloudless T and Y dwarfs, adopting the same assumption in Cond and the cloud-free \citet{saumonEvolutionDwarfsColor_2008} model that dust grains settle below the photosphere and minimally impact the radiative transfer. The \citet{burrowsNongrayTheoryExtrasolar_1997} grid uses gray cloudless (grain-free) atmospheres from \citet{saumonTheoryExtrasolarGiant_1996} for $T_{\mathrm{eff}} {\gtrsim} \SI{1300}{K}$ and non-gray model atmospheres computed in a similar manner to \citet{marleyAtmosphericEvolutionarySpectral_1996} for lower effective temperatures. This approach is best suited for lower temperature T dwarfs. The slightly better agreement between the dynamical mass and the hybrid \citet{saumonEvolutionDwarfsColor_2008} grid than the other evolutionary models lends support for the potential presence of clouds.

\begin{figure}
    \centering
    \includegraphics[width=\linewidth]{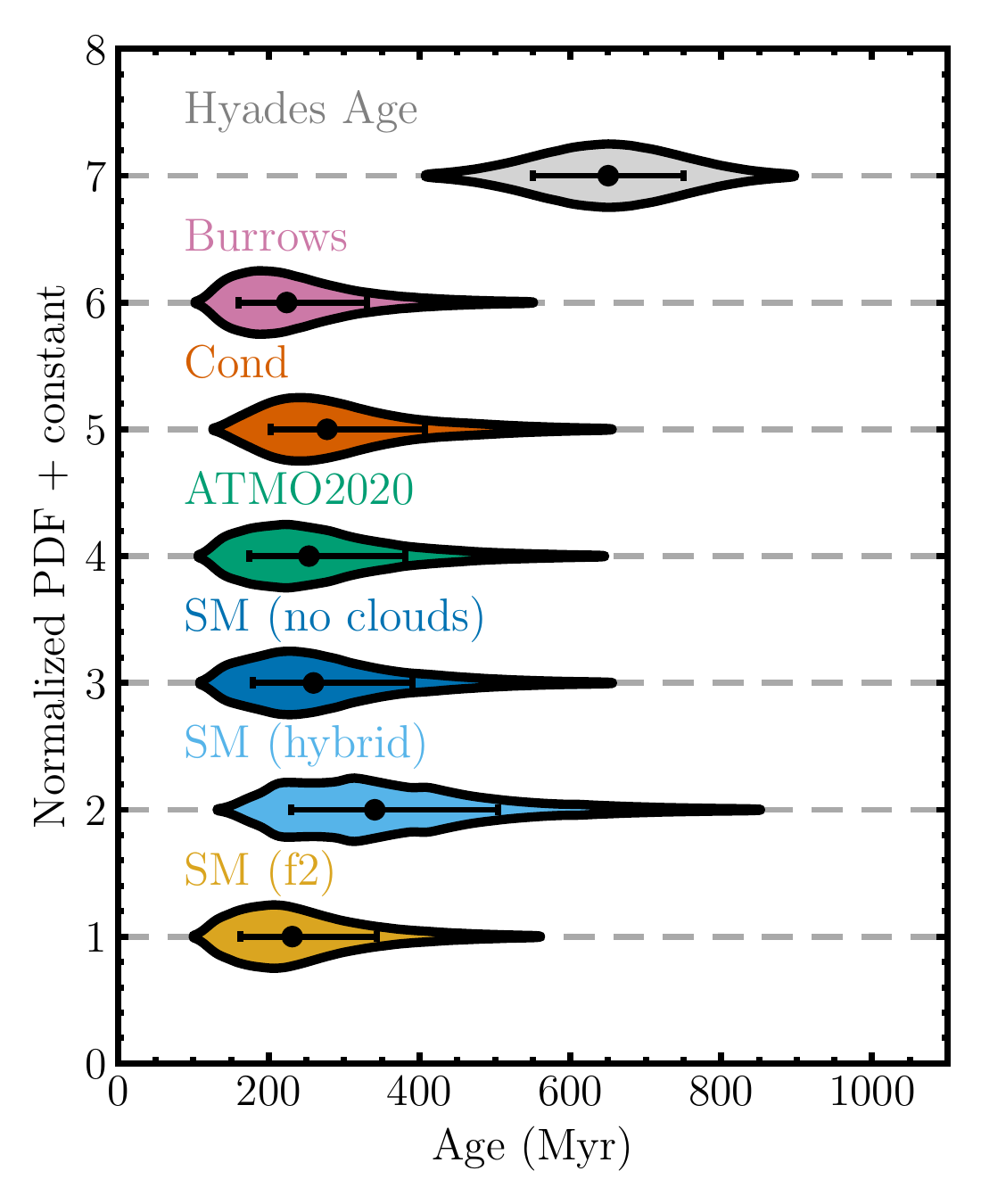}
    \caption{Age predictions from the luminosity and mass of HIP~21152~B for several hot-start evolutionary models. The median and 68.3\% confidence interval are highlighted for each distribution. In general, the inferred ages are about $2\sigma$ lower than the age of the Hyades (\SI{650 \pm 100}{Myr}).\label{fig:ev_model_comp_age}}
\end{figure}

We can also investigate the ages that evolutionary models predict given the companion's luminosity and dynamical mass. Figure \ref{fig:ev_model_comp_age} and Table \ref{tab:model_comp} present the predicted ages for the same suite of hot-start models previously examined. The model-inferred ages are determined in a similar manner as the model-inferred masses: we draw from the luminosity and dynamical mass distributions $10^6$ times and linearly interpolate the corresponding ages for each model. The probability $P(\mathrm{Age}_{\mathrm{Inferred}} {>} \mathrm{Age}_{\mathrm{Hyades}})$ is determined by drawing $10^6$ samples from the inferred age distribution and our adopted Hyades age of \SI{650 \pm 100}{Myr}. The inferred ages are generally 200--\SI{300}{Myr} lower than the nominal age of the Hyades. The model that produces the closest age of $330^{+190}_{-120}\, \mathrm{Myr}$ is the hybrid \citet{saumonEvolutionDwarfsColor_2008} grid.

\begin{deluxetable*}{lcccccc} 
\tablecaption{\label{tab:model_comp}Model-Inferred Parameters}
\tablehead{\colhead{Model} & \colhead{Predicted Mass\tablenotemark{a}} & \colhead{$P(M_{\mathrm{Inferred}} {>} M_{\mathrm{Dynamical}})$\tablenotemark{b}} & \colhead{$\Delta_{\mathrm{mass_pri}}$\tablenotemark{b}} & \colhead{Predicted Age} & \colhead{$P(\mathrm{Age}_{\mathrm{Inferred}} {>} \mathrm{Age}_{\mathrm{Hyades}})$\tablenotemark{b}} & \colhead{$\Delta_{\mathrm{age}}$\tablenotemark{c}}\\
\colhead{} & \colhead{($\mathrm{M_{Jup}}$)} & \colhead{(\%)} & \colhead{} & \colhead{(Myr)} & \colhead{(\%)} & \colhead{}}
\startdata
Burrows & $41^{+4}_{-4}$ & 96.1 & $1.8\sigma$ & $220^{+110}_{-60}$ & 0.7 & $-2.4\sigma$\\
Cond & $37.8^{+3.3}_{-3.5}$ & 94.0 & $1.6\sigma$ & $280^{+130}_{-80}$ & 2.2 & $-2.0\sigma$\\
ATMO2020 & $38.3^{+3.3}_{-3.5}$ & 94.4 & $1.6\sigma$ & $250^{+130}_{-80}$ & 1.9 & $-2.1\sigma$\\
SM (no clouds) & $38.1^{+3.5}_{-3.4}$ & 94.3 & $1.6\sigma$ & $260^{+130}_{-80}$ & 1.9 & $-2.1\sigma$\\
SM (hybrid) & $34^{+6}_{-4}$ & 89.1 & $1.2\sigma$ & $340^{+160}_{-110}$ & 7.3 & $-1.5\sigma$\\
SM (f2) & $41^{+4}_{-4}$ & 95.8 & $1.7\sigma$ & $230^{+110}_{-70}$ & 0.9 & $-2.4\sigma$
\enddata
\tablenotetext{a}{Predicted mass and age entries are $\mathrm{median} \pm 1\sigma$.}
\tablenotetext{b}{A probability of 50\% signifies that the model-inferred quantities and measurement are equivalent. Probabilities below 50\% indicate that the model-inferred values are lower than the dynamical mass or system age. Probabilities above 50\% indicate that the model-inferred values are higher.}
\tablenotetext{c}{One-sided Gaussian equivalent $\sigma$. Calculated via $\sigma = \sqrt{2}\, \mathrm{erf}^{-1} (1 - 2P)$, where $P$ is $P(M_{\mathrm{Inferred}} {>} M_{\mathrm{Dynamical}})$ or $P(\mathrm{Age}_{\mathrm{Inferred}} {>} \mathrm{Age}_{\mathrm{Hyades}})$.}
\end{deluxetable*}

\subsection{Spectral Analysis \label{sec:spec_analysis}}
\subsubsection{Spectral Type \label{sec:spt}}
\begin{figure*}
    \centering
    \includegraphics[width=0.8\textwidth]{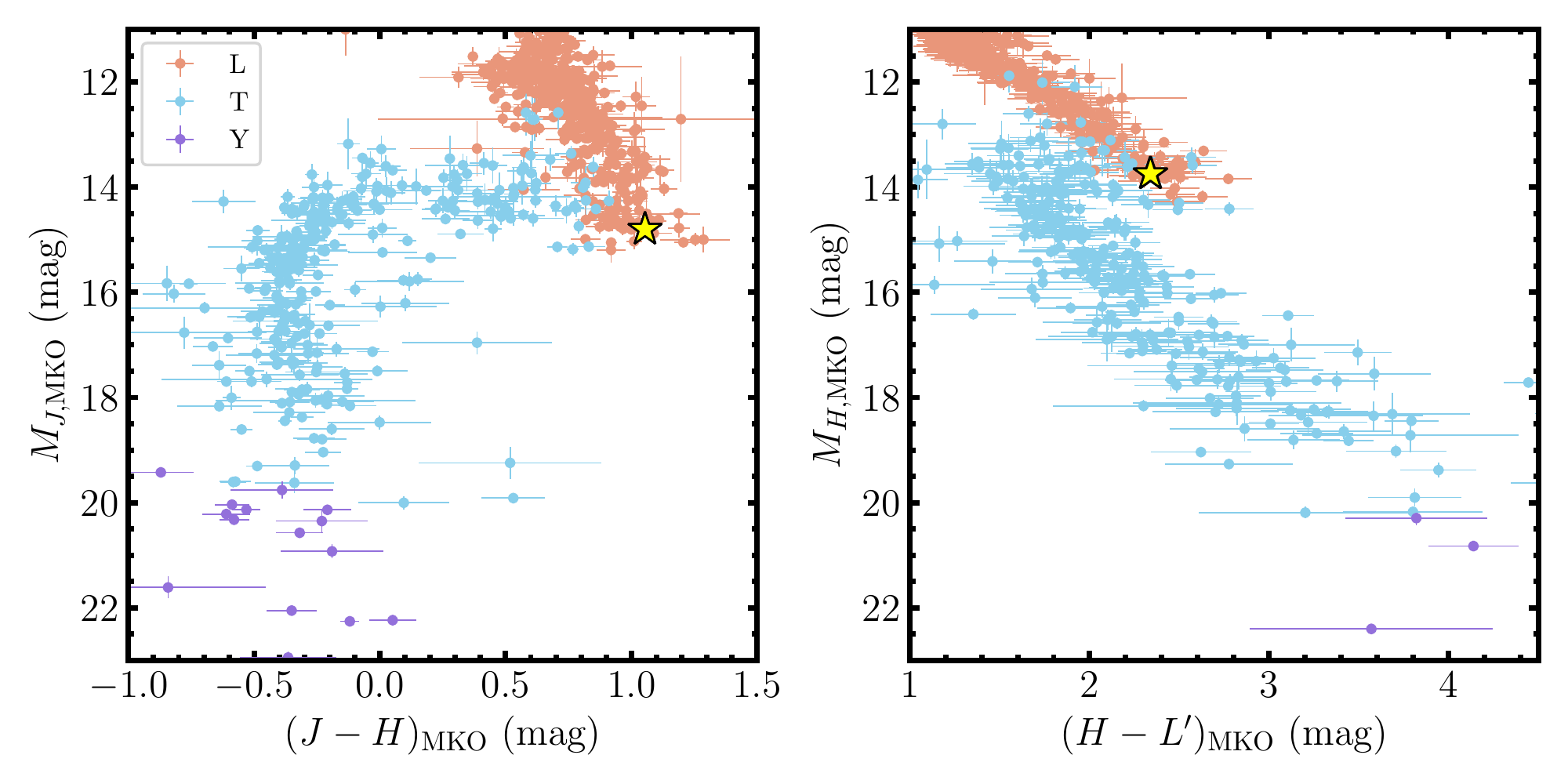}
    \caption{Near-infrared color-magnitude diagrams of HIP~21152~B. The companion's position in each plot is denoted by the yellow star. The uncertainties on the companion's magnitudes are smaller than the symbol size. The background points are L (red), T (blue) and Y (purple) dwarfs from The UltracoolSheet. The companion's positions on these two CMDs are consistent with a late-L or early-T spectral type.\label{fig:cmd}}
\end{figure*}

Our photometry and spectroscopy together with observations from \citet{bonavitaResultsCopainsPilot_2022} are used to determine the spectral type of HIP~21152~B. $J$ and $H$-band photometry are synthesized from our 2022 CHARIS spectrum to compare the position of HIP~21152~B in near-infrared color-magnitude diagrams (CMDs) with the empirical cooling sequence of brown dwarfs. Figure \ref{fig:cmd} shows two CMDs as a function of the companion's $J-H$ and $H - L^\prime$ colors. The sequence of substellar objects is from The UltracoolSheet \citep{bestwilliamm.j.UltracoolsheetPhotometryAstrometry_2020}. $W1$ magnitudes of field brown dwarfs are converted to $L'$ magnitudes using the relation derived in Appendix \ref{sec:w1_lp}. We select all L, T, and Y-dwarfs in the compilation with parallax measurements. Subdwarfs, young brown dwarfs, and unresolved close binaries are excluded. The companion's positions on these CMDs point to a late-L or early-T spectral type.

\begin{figure*}
    \centering
    \includegraphics[width=0.8\textwidth]{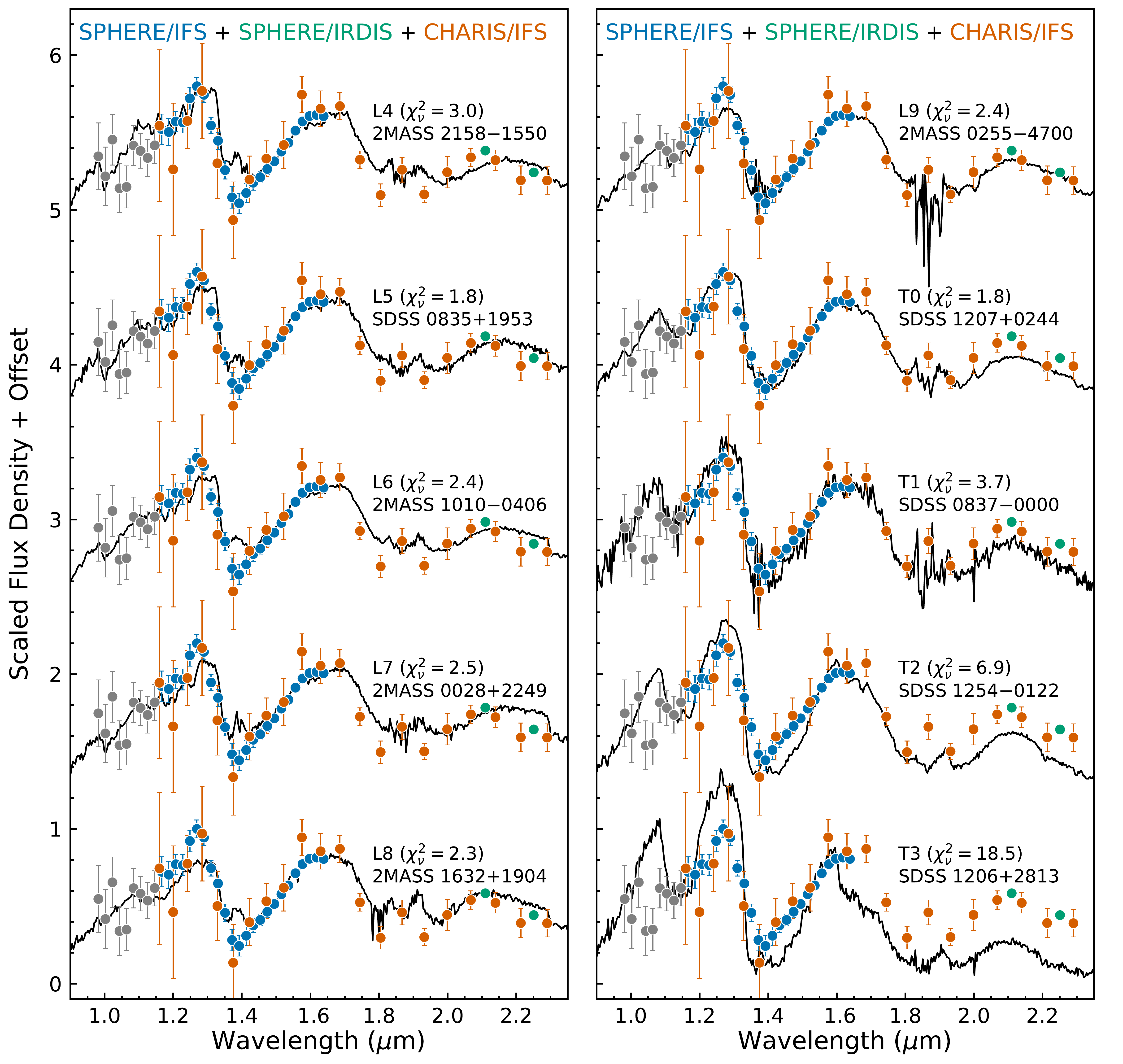}
    \caption{Comparison of the SPHERE/IFS spectrum (red), CHARIS spectrum (blue), and SPHERE/IRDIS photometry (red) of HIP~21152~B to L4--T3 spectral standards. We find that the T0 spectrum provides the best match to the data. We thus assign a spectral type of $\mathrm{T0} \pm 1$ for the brown dwarf companion.
    \label{fig:spec_std_comp}}
\end{figure*}

Figure \ref{fig:spec_std_comp} compares the SPHERE/IFS spectrum, CHARIS spectrum, and SPHERE/IRDIS photometry of HIP~21152~B with the sequence of L and early-T spectral standards. The L-type standards are from Table 4 of \citet{kirkpatrickDiscoveriesNearinfraredProper_2010}. The T-type standards are from \citet{burgasserUnifiedNearinfraredSpectral_2006}, with the exception of the T3 standard SDSS J120602.51+281328.7, which was proposed by \citet{liuDiscoveryHighlyUnequalmass_2010} as a replacement for the original standard which was later found to be a binary. Each template spectrum is optimally scaled to the SPHERE and CHARIS data using the scale factor that minimizes the resulting $\chi^2$ value. Reduced chi-square values ($\chi^2_\nu$) are computed between the template spectra and HIP 21152 B's spectra and photometry. The L5 (SDSS J083506.16+195304.3) and T0 (SDSS J120747.17+024424.8) standards yield the lowest $\chi^2_\nu$ values. Between those two spectral types, T0 is more consistent with the companion's position on the CMD (Figure \ref{fig:cmd}). We therefore assign HIP~21152~B a spectral type of $\mathrm{T0} \pm 1$.

\subsubsection{Grid-Based Model Comparison \label{sec:atm_model_comp}}
To determine the physical properties of HIP~21152~B, we compare the SPHERE/IFS spectrum, SPHERE/IRDIS photometry, CHARIS spectrum, and Keck/NIRC2 photometry against model spectra from \citet{saumonEvolutionDwarfsColor_2008}. Here, we adopt Reduction 2 for the SPHERE/IFS data (see Appendix \ref{sec:atm_model_comp_red1} for the same model comparison using Reduction 1). The \citet{saumonEvolutionDwarfsColor_2008} model grid is a set of one-dimensional, hydrostatic, nongray radiative-convective atmosphere models in chemical equilibrium (see \citealt{marleyCoolSideModeling_2015} for a detailed review). It is part of a lineage of grids originally developed to model the atmosphere of Titan \citep{mckayThermalStructureTitan_1989}. They have since been extended to model the atmospheres of giant planets and brown dwarfs \citep[e.g.,][]{marleyAtmosphericEvolutionarySpectral_1996,marleyCloudsChemistryUltracool_2002,burrowsNongrayTheoryExtrasolar_1997,marleyThermalStructureUranus_1999,saumonAmmoniaTracerChemical_2006,cushingAtmosphericParametersField_2008}. 

The models we consider here were computed using the two-stream source function approach \citep{toonRapidCalculationRadiative_1989} to solve the 1D plane-parallel radiative transfer equation. For convective portions of the profile, the temperature gradient was iteratively fixed to the adiabatic gradient, since super-adiabicity is negligible for cool, $\mathrm{H_2}$-dominated atmospheres \citep{baraffeEvolutionaryModelsLowmass_2002}. Chemical equilibrium calculations were carried out following \cite{fegleyChemicalModelsDeep_1994,fegleyAtmosphericChemistryBrown_1996}, \cite{loddersAtmosphericChemistryGiant_2002,loddersChemistryLowMass_2006}, and \citet{loddersAlkaliElementChemistry_1999,loddersTitaniumVanadiumChemistry_2002}, with elemental abundances from \citet{loddersSolarSystemAbundances_2003}. These abundances confer a C/O ratio of 0.5. The $k$-distribution method with the correlated-$k$ approximation \citep{goodyCorrelatedkMethodRadiation_1989} was used to incorporate opacities. \citet{freedmanLineMeanOpacities_2008} outlines the opacity data used for these models. The \citet{ackermanPrecipitatingCondensationClouds_2001} cloud prescription was used to account for the effect of condensates on the emergent spectrum. This is parameterized by a sedimentation efficiency factor $f_{\mathrm{sed}}$, where larger values of the parameter imply larger particle sizes and more efficient settling. Additionally, a set of models were computed where the effects of condensation and rainout were included in the chemical equilibrium calculation but the opacity from the condensates is ignored.

The model atmospheres we consider have effective temperatures that range from $\SI{800}{K} \leq T_{\mathrm{eff}} \leq \SI{2400}{K}$, surface gravities that range from $4.5 \leq \log g \leq 5.5$, and sedimentation efficiencies $f_{\mathrm{sed}} {=} 1, 2, 3, 4$, alongside the clear version. All synthetic spectra have solar metallicities. To compare our data against the model spectra, we first smooth and resample the models to the wavelength grid of the SPHERE/IFS spectrum, CHARIS spectrum, and photometry. Using a Gaussian kernel, we smooth the spectrum to a spectral resolution of $R{=}25$, based on the resolutions of CHARIS ($R \sim 20$) and the IFS ($R \sim 30$ for \texttt{IRDIFS-EXT} mode). Each spectrum is then anchored to our data using the scale factor that minimizes the $\chi^2$ value between the data and model. For each model, we then compute a reduced chi-square value $\chi^2_{\mathrm{\nu}}$ and a goodness-of-fit statistic $G$, using Equations 1 and 2 of \citet{cushingAtmosphericParametersField_2008}. The $G$ statistic weights spectral and photometric points by their wavelength coverage. This causes photometric points, whose bandpasses span a relatively wide range of wavelengths, to be weighted higher than individual spectral points, whose spacing covers a smaller wavelength coverage. On the other hand, the $\chi^2_\nu$ metric weights each photometric and spectral monochromatic flux density equally and is more statistically meaningful (assuming independent, Gaussian-distributed measurement uncertainties and a model free of systematic errors), although photometry inevitably provides little additional information compared to a spectrum.

\begin{figure*}
    \centering
    \includegraphics[width=\linewidth]{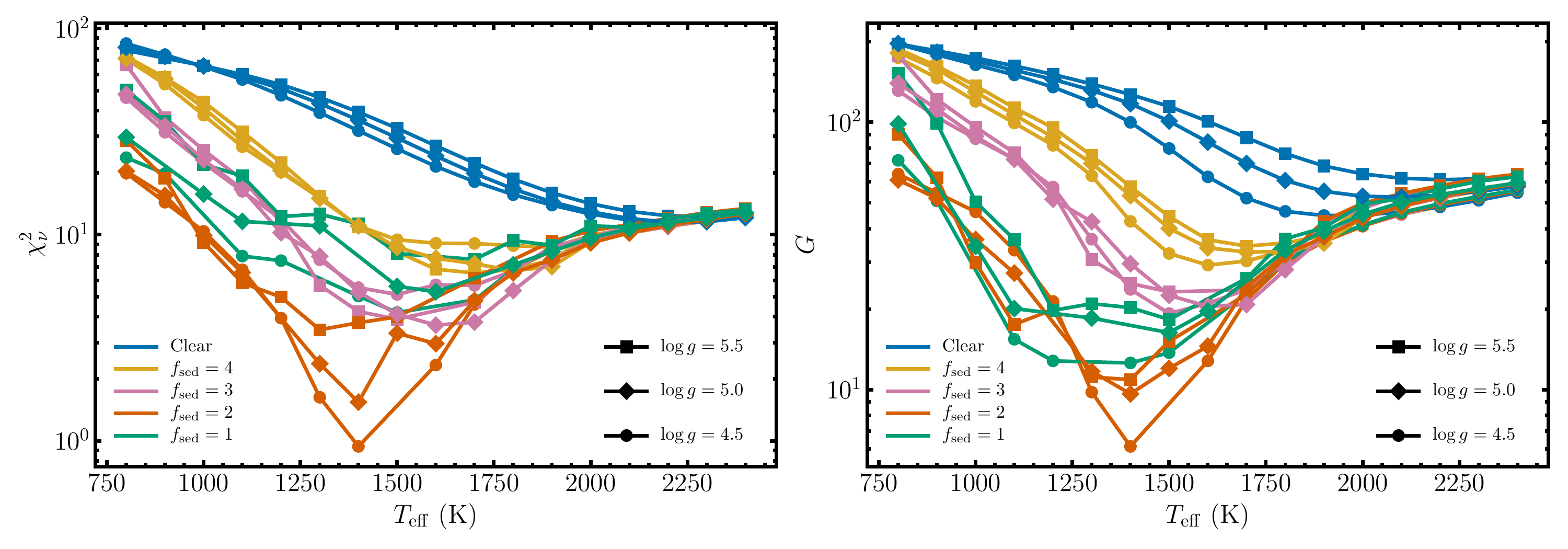}
    \caption{Reduced chi-square ($\chi^2_\nu$; left) and goodness-of-fit ($G$; right) values for model spectra from \citet{saumonEvolutionDwarfsColor_2008} compared with HIP~21152~B's spectra and photometry. Lower values of $\chi^2_\nu$ or $G$ signify a better fit to the data. Colors indicate different values of $f_{\mathrm{sed}}$, while symbols correspond to different values of $\log g$. Reduction 2 is used for the SPHERE/IFS spectrum. The two metrics yield the same best-fitting spectrum, with $T_{\mathrm{eff}} = \SI{1400}{K}$, $\log g = 4.5$, and $f_{\mathrm{sed}} = 2$. 
    \label{fig:model_spec_comp}}
\end{figure*}

\begin{figure}
    \centering
    \includegraphics[width=\linewidth]{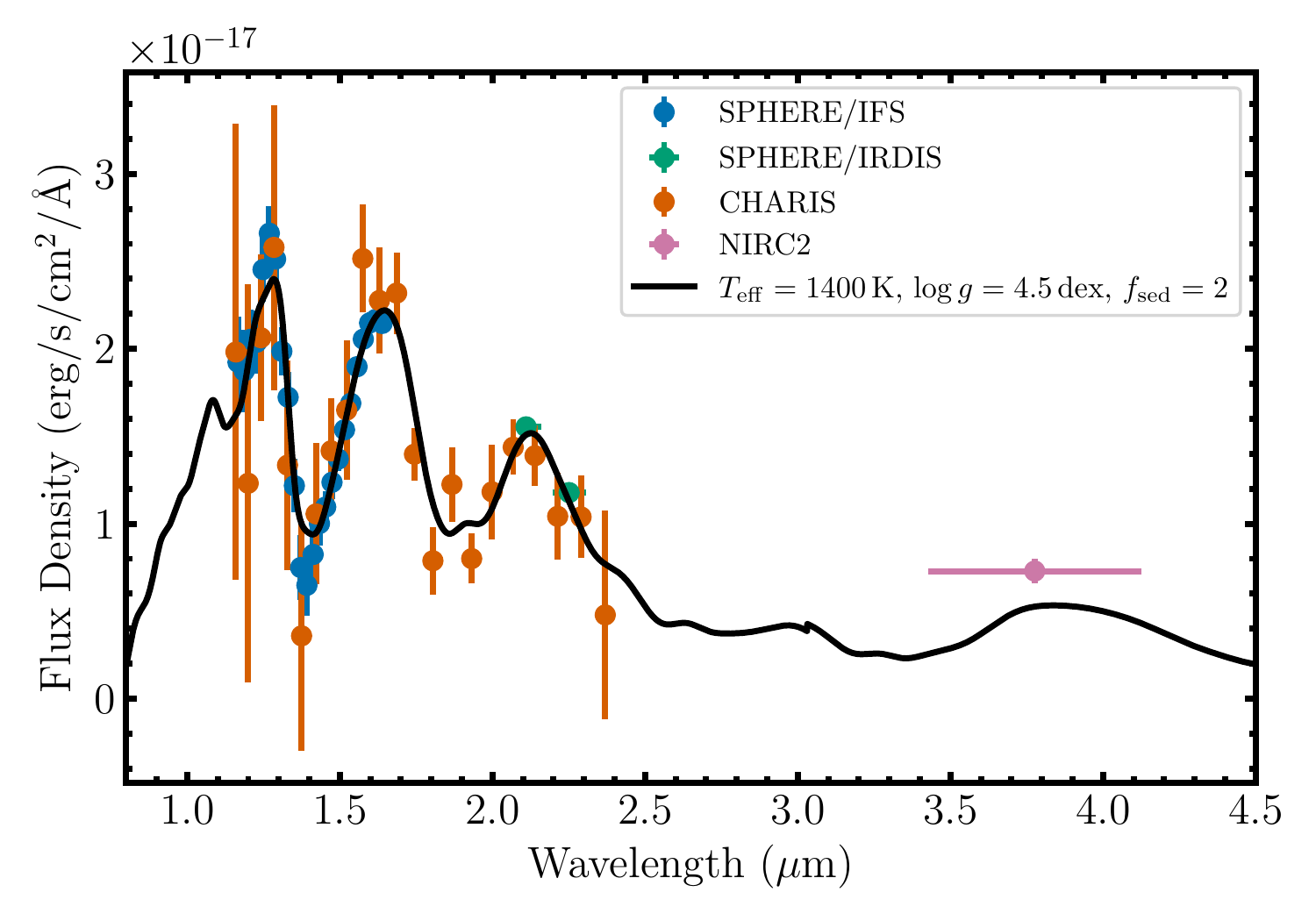}
    \caption{Spectra and photometry of HIP~21152~B compared with the best-fitting \citet{saumonEvolutionDwarfsColor_2008} model spectrum (black). The model represents $T_{\mathrm{eff}} = \SI{1400}{K}$, $\log{g} = \SI{4.5}{dex}$, and $f_{\mathrm{sed}} = 2$, and has been smoothed to a resolving power of $R=25$. Reduction 2 is used for the SPHERE/IFS spectrum. The atmospheric model reproduces the companion's spectra well, although we find a slight excess in $L^\prime$ that might hint at disequilibrium chemistry from 3--\SI{4}{\mu m}.
    \label{fig:spectrum_model_plot}}
\end{figure}

Figure \ref{fig:model_spec_comp} shows these goodness-of-fit values as a function of $T_{\mathrm{eff}}$, $\log g$, and $f_{\mathrm{sed}}$. Generally, the two metrics yield similar results, with a clear minimum at $T_{\mathrm{eff}} = \SI{1400}{K}$ and best-fitting spectrum of $(T_{\mathrm{eff}},\log g,f_{\mathrm{sed}}) = (\SI{1400}{K}, \SI{4.5}{dex}, 2)$ produced by both statistics. The best-fit model corresponds to a radius of $0.72 \, \mathrm{R_{Jup}}$. Here, the radius is determined from the flux-calibration scale factor $C$ via $R = d \sqrt{C}$, where $d$ is the distance to the system \citep[see e.g.,][]{bowlerBenchmarkUltracoolSubdwarf_2009}. This model spectrum is plotted against HIP~21152~B's spectra and photometry in Figure \ref{fig:spectrum_model_plot}. Models with $f_{\mathrm{sed}} = 2$ tend to provide the best fit compared to other $f_{\mathrm{sed}}$ values for $T_{\mathrm{eff}} < \SI{1600}{K}$. The clear models yield the poorest fit to the data.

Another approach to determine an effective temperature is by using the Stefan-Boltzmann law:
\begin{align}
    T_{\mathrm{eff}} = \bigg(\frac{L_{\mathrm{bol}}}{4 \pi R^2 \sigma}\bigg)^{1/4}
\end{align}
Here, $L_{\mathrm{bol}}$ is the companion's bolometric luminosity and $R$ is its radius. In Section \ref{sec:ev_model_comp}, we determined the bolometric luminosity to be $\log(L_\mathrm{bol}/\mathrm{L_\odot}) = \SI{-4.57 \pm 0.07}{dex}$. For the radius, we interpolate evolutionary model tracks from \citet{burrowsNongrayTheoryExtrasolar_1997} in a similar fashion as the mass and age distributions in Section \ref{sec:ev_model_comp}. We draw values from an age distribution of \SI{650 \pm 100}{Myr} and the dynamical mass posterior of $24^{+6}_{-4} \, \mathrm{M_{Jup}}$ and interpolate the corresponding radius. This yields $R=\SI{0.997 \pm 0.023}{R_{Jup}}$. The Stefan-Boltzmann law then produces an effective temperature of $T_{\mathrm{eff}} = \SI{1300 \pm 50}{K}$, which is consistent with the best-fit \citet{saumonEvolutionDwarfsColor_2008} model of \SI{1400}{K}.

\begin{figure}
    \centering
    \includegraphics[width=\linewidth]{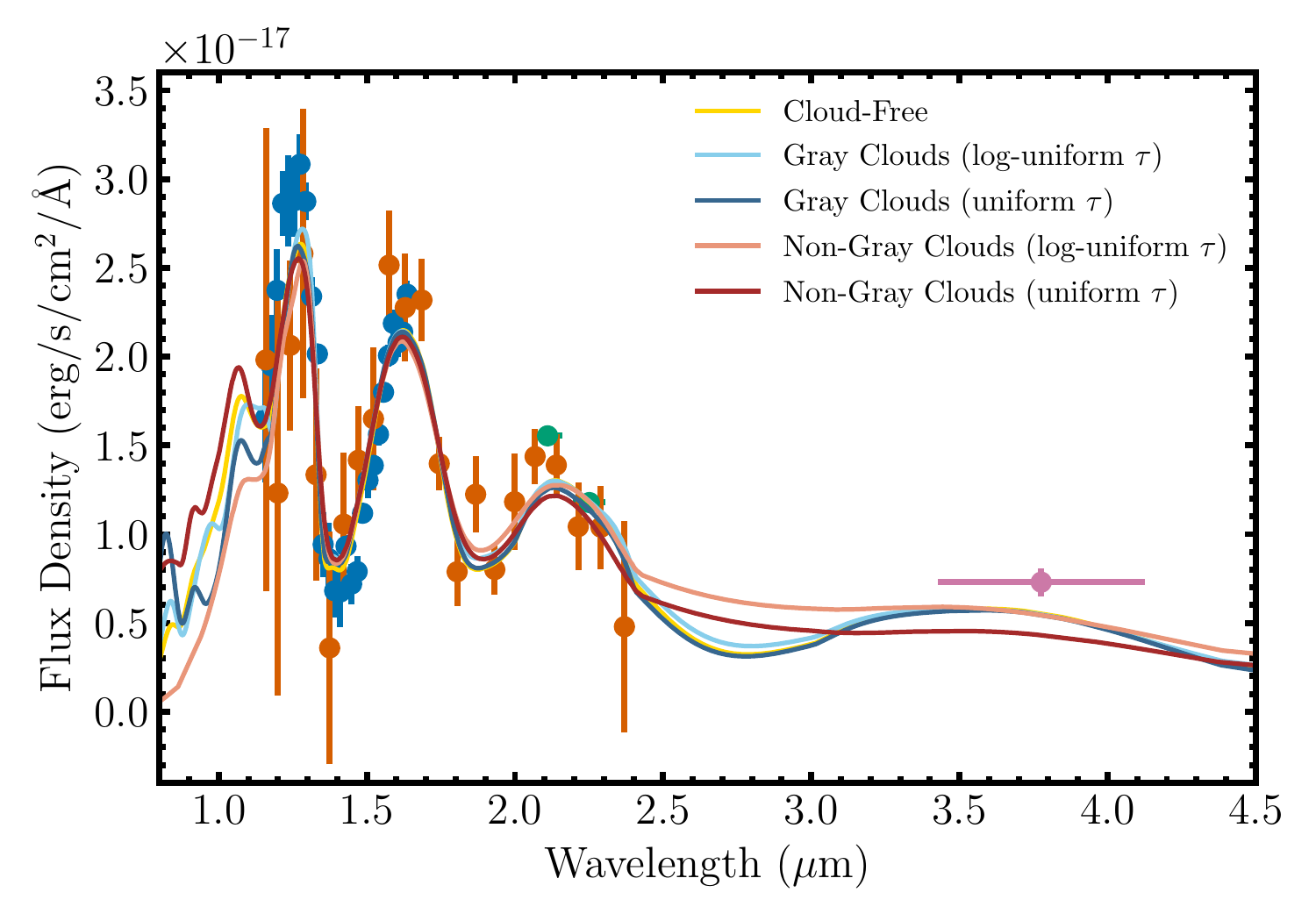}
    \caption{Spectra and photometry of HIP~21152~B compared with the best-fit retrievals for each of the five prescriptions we consider. Here we show the retrievals performed with the complete dataset (IFS and CHARIS spectra, and photometry) using Reduction 2 for the SPHERE/IFS spectrum. The colors of the spectra and photometry are the same as in Figures \ref{fig:spectrum} and \ref{fig:spectrum_model_plot}. The retrieval models have been smoothed to a resolving power of $R = 25$.
    \label{fig:retrieval_comp_plot}}
\end{figure}

\subsubsection{Atmospheric Retrieval \label{sec:retrieval}}
We also perform a suite of atmospheric retrievals on HIP~21152~B's SPHERE/IFS spectrum, CHARIS spectrum, SPHERE/IRDIS $K_1/K_2$ photometry, and Keck/NIRC2 $L^\prime$ photometry. Here, we adopt Reduction 2 for the SPHERE/IFS data (see Appendix \ref{sec:retrieval_red1} for retrievals using Reduction 1). The atmospheric retrievals are performed with the open-source \texttt{Helios-r2} code \citep{kitzmannHeliosr2NewBayesian_2020}, which implements the \texttt{MULTINEST} \citep{ferozMultinestEfficientRobust_2009} multi-modal nested sampling \citep{skillingNestedSamplingGeneral_2006} algorithm for Bayesian exploration of parameter space. The forward model, which computes the outgoing radiation flux as a function of wavelength, uses the method of short characteristics \citep{olsonShortCharacteristicSolution_1987}. The temperature-pressure profile is described using seven free parameters and 70 levels (i.e. 69 layers) via a finite element approach \citep{kitzmannHeliosr2NewBayesian_2020}. The opacities of atoms and molecules are computed using the open-source \texttt{HELIOS-K} calculator \citep{grimmHelioskUltrafastOpensource_2015,grimmHelioskOpacityCalculator_2021}. We include $\mathrm{H_2 O}$, $\mathrm{CH_4}$, $\mathrm{NH_3}$, $\mathrm{CO_2}$, CO, $\mathrm{H_2 S}$, CrH, FeH, CaH, TiH, as well as the alkali metals Na and K. Corresponding line lists are taken from the ExoMol database \citep{barberHighaccuracyComputedWater_2006,yurchenkoVariationallyComputedLine_2011,yurchenkoExomolLineLists_2014,azzamExomolMolecularLine_2016} and the HITEMP database \citep{rothmanHitempHightemperatureMolecular_2010}. Collision-induced absorption coefficients for $\mathrm{H_2 {-} H_2}$ and $\mathrm{H_2{-}He}$ are based on \citet{abelCollisioninducedAbsorptionH2pairs_2011} and \citet{abelInfraredAbsorptionCollisional_2012}, respectively. For the resonance lines of Na and K, we use the line profile descriptions from \citet{allardLineShapesSpectra_2016} and \citet{allardNewStudyLine_2019}, as described in \citet{kitzmannHeliosr2NewBayesian_2020}. \texttt{Helios-r2} was previously applied to a curated sample of 19 L and T dwarfs \citep{lueberRetrievalStudyBrown_2022}. We adopt the same criterion for model comparison using the ratio of Bayesian evidences (i.e., the Bayes factor; \citealt{trottaBayesSkyBayesian_2008}), which are natural outcomes of the nested sampling algorithm \citep{skillingNestedSampling_2004}.

\begin{deluxetable*}{lcc}[htb!]
    \tablecaption{Summary of Free Parameters for Retrievals\label{tab:retrievalpriors}}
    %. and prior distributions for the free chemistry approach used in the cloud-free, gray cloud, and non-gray cloud models, was well as the permutations with a new prior distribution for the cloud optical depth $\tau$.
    \tablehead{
    \colhead{Parameter} & \colhead{Description} & \colhead{Prior}
    % \colhead{Parameter} & \multicolumn{2}{c}{Prior}\\ %\cmidrule{1-3}
    % & \colhead{Type} & \colhead{Value}
    }
    \startdata
    \emph{All Retrievals} & & \\
    $\log{g}$    & Surface Gravity  & $3.5-6.0$ (cgs; Uniform) \\
    $d$             & Distance  & \SI{43.27 \pm 0.052}{pc} (Gaussian)  \\
    $f$                   & Flux Scaling Factor  &  $0.1-5.0$ (Uniform) \\
    $T_1$              & Temperature at Base of Modeled Atmosphere &   $1000-\SI{5000}{K}$ (Uniform)\\
    $b_{i=1\dots6}$    & P/T Profile Coefficients  & $0.1-0.95$ (Uniform)  \\
    $\ln{\delta}$         & Error Inflation Term &   $-10-1.0$ (Uniform) \\
    $x_i$           & Mixing Ratio of Species $i$ &   $10^{-12}-0.1$ (log-uniform) \\
    \hline
    \textit{Gray Clouds} &   &   \\
    $p_{\mathrm{t}}$  & Pressure at Top of Cloud  & $10^{-2}-\SI{50}{bar}$ (log-uniform)  \\
    $b_{\mathrm{c}}$     & Cloud Base Pressure Scale Factor  &  $1-10$ (log-uniform) \\
    $\tau$ & Optical Depth  &  $10^{-5}-20$ (log-uniform) \\
    \hline
    \textit{Non-Gray Clouds}  &  &   \\
    $p_{\mathrm{t}}$  & Pressure at Top of Cloud  & $10^{-2}-\SI{50}{bar}$ (Uniform)\\
    $b_{\mathrm{c}}$  & Cloud Base Pressure Scale Factor  &  $1-10$ (log-uniform) \\
    $\tau_{\mathrm{ref}}$  & Optical Depth at Reference Wavelength  & $10^{-5}-20$ (log-uniform) \\
    $Q_0$                 & Particle Size (dimensionless) with Highest Extinction & $1-100$ (log-uniform)  \\
    $a_0$                 & Power-Law Index for Small Particles  &  $3-7$ (Uniform) \\
    $a$ & Particle Size & $0.1-\SI{50}{\mu m}$ (log-uniform) \\
    \hline
    \textit{Additional $\tau$ Prior}   &  &   \\
    $\tau$             & Optical Depth & $-10-20$ (Uniform) \\
    \enddata
    % \tablenotetext{i}{Note that all the other cloud retrieval parameters are similar for "newtau" retrievals.}
    \tablecomments{We consider five different retrieval versions: one with no clouds, two with gray clouds, and two with non-gray clouds. The four cloudy retrievals consist of two with log-uniform priors on $\tau$ and two with uniform priors on $\tau$.}
\end{deluxetable*}

The priors for our suite of retrievals are listed in Table \ref{tab:retrievalpriors}. A general description of the brown dwarf forward model used here can be found in \citet{kitzmannHeliosr2NewBayesian_2020} and \citet{lueberRetrievalStudyBrown_2022}. We consider both cloud-free and cloudy atmospheres. For the cloudy cases, we either use gray or non-gray clouds. Additionally, for each retrieval with clouds, both a log-uniform and a uniform prior on the cloud optical depth $\tau$ is adopted. This follows \citet{lueberRetrievalStudyBrown_2022}, who found that when a uniform (instead of a log-uniform) prior is used for the $\tau$, it becomes constrained. In total, we run the companion's spectrum and photometry through five different versions of the retrieval: one with no clouds, one with gray clouds and a log-uniform prior on $\tau$% (GC log-uniform)
, one with gray clouds and a uniform prior on $\tau$% (GC uniform)
, one with non-gray clouds and a log-uniform prior on $\tau$% (NGC log-uniform)
, and one with non-gray clouds and a uniform prior on $\tau$% (NGC uniform)
. In total, we have 21 free parameters for the cloud-free %(NC) 
retrieval, 24 free parameters for the gray cloud %(GC) 
retrievals, and 27 for the non-gray cloud %(NGC) 
retrievals.

\begin{deluxetable*}{lccccccc}%[b!]
    % \tabletypesize{\scriptsize}
    \tablecolumns{12}
    \tablewidth{\columnwidth}
    \tablecaption{Summary of Bayesian Statistics for Atmospheric Retrievals}
    \label{tab:bayes_vals}
    \tablehead{\colhead{Model} & \colhead{Parameter} & \colhead{IFS} & \colhead{CHARIS} & \colhead{IFS + CHARIS + Photometry}}
    \startdata 
    Cloud-Free & $\log$ evidence & 275.83 & 275.26 & 558.35 \\
    Gray Clouds (log-uniform $\tau$) & $\log$ evidence & 274.18 & 274.18 & 557.41 \\
    Gray Clouds (uniform $\tau$) & $\log$ evidence & 274.08 & 273.78 & 556.62 \\
    Non-Gray Clouds (log-uniform $\tau$) & $\log$ evidence & 273.71 & 273.26 & 556.28 \\
    Non-Gray Clouds (uniform $\tau$) & $\log$ evidence & 272.84 & 272.97 & 555.63 \\
    \hline
    Gray Clouds (log-uniform $\tau$) vs. Cloud-Free & $\ln B_{ij}$\tablenotemark{a} & 1.66 & 1.08 & 0.94 \\
    Gray Clouds (uniform $\tau$) vs. Cloud-Free & $\ln B_{ij}$\tablenotemark{a}& 1.75 & 1.48 & 1.74 \\
    Non-Gray Clouds (log-uniform $\tau$) vs. Cloud-Free & $\ln B_{ij}$\tablenotemark{a} & 2.12 & 1.99 & 2.08 \\
    Non-Gray Clouds (uniform $\tau$) vs. Cloud-Free & $\ln B_{ij}$\tablenotemark{a} & 2.99 & 2.29 & 2.73 \\
    \hline
    \enddata
    \tablenotetext{a}{The Bayes factor $B_{\mathrm{ij}}$ is the ratio of the Bayesian evidence values for two models \citep{trottaBayesSkyBayesian_2008}. $\ln B_{\mathrm{ij}} \sim 1$ signifies weak evidence for a statistical preference between the two models, while $\ln B_{\mathrm{ij}} \gtrsim 5$ indicates strong evidence.}
\end{deluxetable*}
We consider retrievals for three combinations of HIP~21152~B's spectra and photometry: the SPHERE/IFS spectrum only; the CHARIS spectrum only; and the combined dataset of the IFS spectrum, CHARIS spectrum, SPHERE/IRDIS $K_{12}$ photometry, and Keck/NIRC2 $L^{\prime}$ photometry. Due to the discrepancy between the two reductions of the SPHERE/IFS data at short wavelengths (see Section \ref{sec:second_reduction} and Figure \ref{fig:model_spec_comp}), we exclude points with $\lambda < \SI{1.15}{\mu m}$. This also circumvents unresolved issues with the shapes of the alkali metal resonance line wings \citep{oreshenkoSupervisedMachineLearning_2020}. The posteriors for the suite of retrievals are shown in Table \ref{tab:retrieval_posteriors_red2}, while Table \ref{tab:bayes_vals} summarizes the Bayesian evidences for each retrieval and the relative Bayes factors $B_{\mathrm{ij}}$ between each cloudy retrieval and the cloud-free retrieval. The posteriors of the five permutations of different retrieval models (cloud-free and cloudy scenarios) for each dataset yield very similar values. This is consistent with the Bayes factors being close to unity for all cases, indicating that there is not a significant statistical preference between the retrieval models. Though there are slight differences in the posteriors between retrievals with subsets of the data (i.e., a slightly lower median $T_{\mathrm{eff}}$ for the retrieval with only the CHARIS spectrum), the retrieval results remain generally consistent. Depending on the retrieval and the subset of the data used, we obtain a surface gravity of $\log g \sim 4.9{-}5.4$ and effective temperature of $T_{\mathrm{eff}} \sim 1400{-}\SI{1600}{K}$. These effective temperatures are roughly consistent with the value from the companion's bolometric luminosity and radius (\SI{1300 \pm 50}{K}) and the best-fit \citet{saumonEvolutionDwarfsColor_2008} model spectrum (\SI{1400}{K}). The small retrieved radii of 0.5--\SI{0.8}{R_{\mathrm{Jup}}} is a recurrent issue with retrievals, possibly indicating missing physics or chemistry \citep{kitzmannHeliosr2NewBayesian_2020,lueberRetrievalStudyBrown_2022}.

Water is formally detected in all retrievals with a volume mixing ratio $x_{\mathrm{H_2 O}} \sim 10^{-3}{-}10^{-2}$, while potassium, chromium hydride, iron hydride, and titanium hydride are only partially detected, depending on the wavelength coverage and spectral resolution of the dataset. For these chemical species, the mixing ratios are roughly constant with values of about $10^{-10}{-}10^{-8}$. Similar to \citet{lueberRetrievalStudyBrown_2022}, $\tau$ is unconstrained for the cloudy retrievals that utilize a log-uniform prior for $\tau$, but it becomes constrained when a uniform $\tau$ prior is adopted. Besides the cloud top pressures, other cloud properties --- especially the non-gray cloud properties --- remain unconstrained.

\subsection{Transit Search in TESS}
The Transiting Exoplanet Survey Satellite \citep[TESS;][]{rickerTransitingExoplanetSurvey_2014} observed HIP~21152 at 2-minute cadence during Sectors 4 and 32 for a total of 62 days. We downloaded the Science Processing Operations Center (SPOC) reduced Pre-search Data Conditioning Simple Aperture Photometry (PDCSAP) light curve \citep{smithKeplerPresearchData_2012, stumpeKeplerPresearchData_2012, stumpeMultiscaleSystematicError_2014} from the Mikulski Archive for Space Telescopes (MAST) data archive\footnote{\href{https://archive.stsci.edu/missions-and-data/tess}{https://archive.stsci.edu/missions-and-data/tess/}} using the \texttt{lightkurve} \citep{lightkurvecollaborationLightkurveKeplerTess_2018} software package. All photometric measurements flagged as poor quality by the SPOC pipeline (\texttt{DQUALITY} $>$ 0) or listed as \texttt{NaN} are removed. Positive outliers at 3$\sigma$ and negative outliers at 10$\sigma$ were further rejected to allow for possible transit events. We de-trend the light curve using a 1D box smoothing kernel with a one-hour width. The de-trended light curve has an RMS of 0.2 ppt (in units of relative flux). We searched for signals of a transiting planet in the de-trended light curve using a Box Least Squares search \citep{kovacsBoxfittingAlgorithmSearch_2002} between 0.5 and 30 day orbits but did not identify any significant periodic transit-like events.

\begin{figure*}
    \centering
    \includegraphics[width=0.8\textwidth]{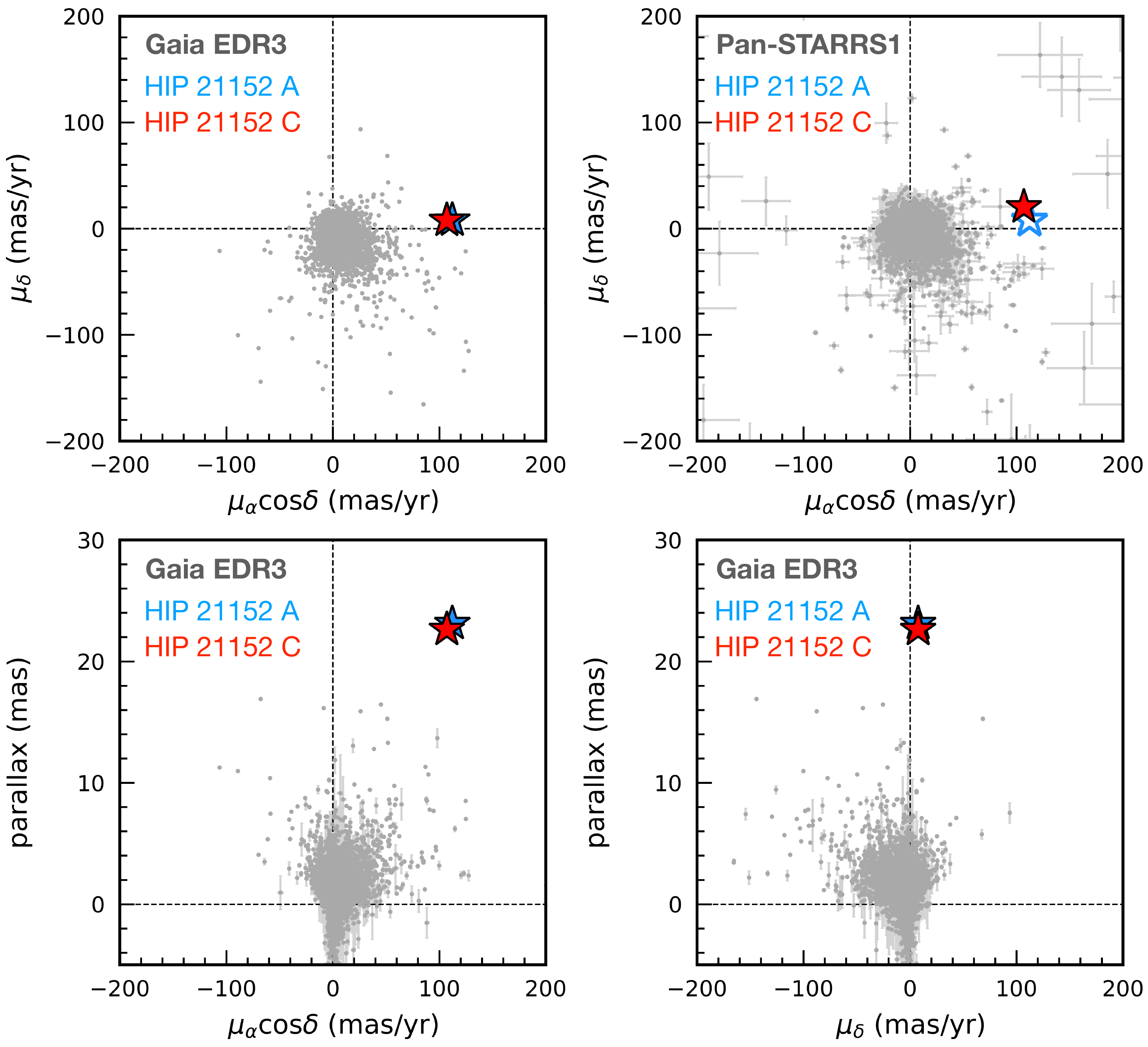}
    \caption{{\it Top}: Proper motions of HIP~21152~A (blue) and HIP~21152~C (red) from Gaia EDR3 (top left) and PS1 (top right), overlaid with nearby sources within a radius of 1$^{\circ}$ (grey). We only show PS1 neighbors if their proper motions have S/N $>5$ or have errors $<10$ mas/yr. The host star does not have a PS1 proper motion given that it is saturated, so we plot its Gaia EDR3 proper motion (blue open symbol) in the top right panel. {\it Bottom}: Gaia EDR3 proper motions and parallaxes of HIP~21152~AC and nearby sources shown in the top left panel. The potential association between the A and C components is indicated by their common proper motions and parallaxes.
    \label{fig:cpm_compan}}
\end{figure*}

\subsection{A Wide Common-Proper-Motion Companion}
To investigate the outer architecture of the HIP~21152 system, we search for any wide-separation comoving companions using Gaia EDR3 \citep[][]{gaiacollaborationGaiaMission_2016,gaiacollaborationGaiaEarlyData_2021} and Pan-STARRS1 \citep[PS1;][]{chambersPanstarrs1Surveys_2016,magnierPanstarrsPhotometricAstrometric_2020}. For each survey, we follow \citet{zhangSecondDiscoveryCool_2021} and apply the following conditions to identify wide companions: (1) the projected separation between the two components is $<10^{5}$~au; (2) the companion's proper motion has a signal-to-noise ratio $>5$; (3) the vector difference between the proper motions of HIP~21152~A ($\overrightarrow{\mu_{P}}$; from Gaia EDR3) and the companion ($\overrightarrow{\mu_{C}}$; from Gaia or PS1) is within $30\%$ of HIP~21152~A's total proper motion (i.e., $|\overrightarrow{\mu_{C}} - \overrightarrow{\mu_{P}}| \leqslant 0.3 |\mu_{P}|$); (4) the parallax difference between HIP~21152~A and the companion is within $30\%$ of HIP~21152~A's parallax; (5) if present, the radial velocities of two components are consistent to within $3\sigma$. We only applied the last two criteria when searching for companions in Gaia. This process produced one candidate common proper motion companion: 2MASS~J04335658+0537235 (HIP~21152~C hereinafter). The properties of HIP~21152~C are summarized in Table \ref{tab:prop_c}. The PS1 and Gaia EDR3 proper motions and parallax values for HIP~21152~A and HIP~21152~C are plotted against nearby sources in Figure \ref{fig:cpm_compan}.

HIP~21152~C has an angular separation of $1837\arcsec$ from HIP~21152~A, corresponding to a projected physical separation of \SI{7.9e4}{au} at the primary star's distance. The Gaia astrometry of HIP~21152~C is determined from 20 visibility periods, with a RUWE of 1.05, suggesting this wide-orbit companion is likely single. HIP~21152~C was previously identified as a Hyades member by \citet{gagneBanyanXiiiFirst_2018} using BANYAN~$\Sigma$ \citep[][]{gagneBanyanXiBanyan_2018} and Gaia~DR2 astrometry. HIP~21152~C was also independently found by \citet{reyleNewUltracoolBrown_2018} in their Gaia-based search for ultracool dwarfs. Based on the object's $G$-band absolute magnitude, \citet{gagneBanyanXiiiFirst_2018} and \citet{reyleNewUltracoolBrown_2018} assigned a spectral type of L1 and L0 to this companion, respectively.

To determine the bolometric luminosity and infer the mass of the object, we first convert its $H$-band magnitude from 2MASS to an absolute magnitude of $M_{H_{\mathrm{2MASS}}} = \SI{11.17 \pm 0.08}{mag}$ using HIP~21152~C's Gaia EDR3 parallax. The $M_{H_{\mathrm{2MASS}}} - \log{(L_{\mathrm{bol}} / L_{\odot})}$ relation from \citet{dupuyIndividualDynamicalMasses_2017} then yields a luminosity of $\log{(L_{\mathrm{bol}} / L_{\odot})} = \SI{-3.62 \pm 0.04}{dex}$. We determine a mass for HIP~21152~C by drawing $10^6$ samples from its luminosity and the age distribution of the Hyades (\SI{650 \pm 100}{Myr}; Section \ref{sec:props}) and interpolating the BHAC15 evolutionary model grid \citep{baraffeNewEvolutionaryModels_2015}. This yields a model-inferred mass of \SI{73 \pm 7}{M_{Jup}}. The hydrogen burning limit (HBL) is defined as the mass at which half of an object's luminosity is generated by fusing hydrogen \citep{reidNewLightDark_2005} and is primarily a function of opacity (metallicity) and helium fraction \citep{burrowsTheoryBrownDwarfs_2001}. Rotation rate can also influence the HBL \citep{chowdhuryStableHydrogenBurning_2022}. Stars with higher opacities, metallicities, helium fractions, and slower rotation rates can sustain hydrogen burning at lower masses. Generally, model and empirically determined HBLs range from $\approx70{-}\SI{80}{M_{Jup}}$ \citep{scuflaireClesCodeLiegeois_2008,saumonEvolutionDwarfsColor_2008,baraffeNewEvolutionaryModels_2015,dupuyIndividualDynamicalMasses_2017,fernandesEvolutionaryModelsUltracool_2019}. HIP~21152~C's mass of \SI{73\pm7}{M_{Jup}} places it at the HBL, so it is unclear whether it is a low-mass star or a brown dwarf. This corresponds to a mass-ratio of $q = 0.049 \pm 0.005$. The binding energy of the HIP 21152 AC pair is $(5 \pm 4) \times 10^{39} \, \mathrm{erg}$. This is similar to other wide common proper motion low-mass companions with very low binding energies of $\sim 10^{40} \, \mathrm{erg}$ \citep[see e.g.,][]{fahertyBrownDwarfKinematics_2010,dhitalSloanLowmassWide_2010}.

We determine the odds that HIP~21152~AC is a chance alignment with unrelated Hyades members by calculating the probability of there being at least two cluster members within a $10^5  \times 10^5  \times (2 \times 10^5) \, \mathrm{au}$ ellipsoid. The first two dimensions are based on the search criterion that the projected separation between two sources is ${<}10^5 \, \mathrm{au}$. The last dimension is from the parallax difference between HIP~21152~A and HIP~21152~C, which corresponds to a line-of-sight distance of $\SI{0.94}{pc} \approx \SI{2e5}{au}$. We estimate the number density of Hyades stars at HIP~21152's position by selecting the Hyades sources in \citet{lodieu3dViewHyades_2019} with physical distances in a volumetric shell from the cluster center within $\SI{1}{pc}$ of HIP~21152's distance of \SI{9.75}{pc}. After dividing by the volume of the shell, this yields a density of $n = \SI{0.023}{stars/pc^3}$. The probability of two or more sources occurring in a region of volume $V$ and number density $n$ is given by the complement of a Poisson distribution:
\begin{align}
    P(N \geq 2) = 1 - P(N=1) = 1 - \lambda e^{-\lambda},
\end{align}
where the rate is $\lambda = nV$. We find that the probability of chance alignment $P(N \geq 2) = 0.7\%$, so this companion is likely physical. Although HIP~21152~A and C have consistent distances (\SI{43.27 \pm 0.05}{pc} and \SI{44.2 \pm 0.9}{pc}), we note that there is a modest uncertainty on the line-of-sight distance between the objects of $\pm \SI{0.9}{pc}$. If we extend the ellipsoid to have a depth of $\SI{2.74}{pc} \approx \SI{5.7e5}{au}$, representing a $2\sigma$ increase in the distance between the sources, the resulting probability of chance alignment $P(N \geq 2) = 2\%$, which remains low. We therefore conclude that HIP~21152~C is likely a bound companion and not a chance alignment of an isolated Hyades member.

One can compare the proper motion of the barycenter of the system from the joint orbit fit of HIP~21152~B (Section \ref{sec:orbitfit}) and the Gaia EDR3 proper motion of HIP~21152~C to assess whether the proper motions are compatabile with the object being bound. We first convert the barycenter proper motions in Table (\ref{tab:elements}) to the position of HIP~21152~C using the Gaia EDR3 coordinates, parallaxes of the A and C components, and our radial velocity of \SI{40.6 \pm 1.5}{km.s^{-1}} (Section \ref{sec:tull}). The resulting barycenter proper motions at the position of HIP~21152~C are $\mu_\alpha = \SI{108.4 \pm 2.5}{mas.yr^{-1}}$ and $\mu_\delta = \SI{6.8 \pm 0.7}{mas.yr^{-1}}$. Compared with the EDR3 proper motions of $\mu_\alpha = \SI{107.17 \pm 0.49}{mas.yr^{-1}}$ and $\mu_\delta = \SI{7.60 \pm 0.39}{mas.yr^{-1}}$, the barycenter proper motion differs by $\Delta_\mu = \SI{2.6 \pm 1.6}{mas.yr^{-1}}$, or $\Delta_\mu = \SI{0.55 \pm .32}{km.s^{-1}}$. The 3D separation between HIP~21152~A and C is $(2.5 \pm 1.4) \times 10^5 \, \mathrm{au}$. At that distance, the escape velocity from the host-star is $v_{\mathrm{esc}} = \SI{0.11 \pm 0.04}{km.s^{-1}}$. $\Delta_\mu$ is thus compatabile with the companion being bound to within $2 \sigma$. A future RV measurement of HIP~21152~C would enable the determination of the 3D relative velocity between the barycenter and the source as an additional assessment of whether the companion is bound.

\begin{deluxetable}{lcc}
\tablecaption{\label{tab:prop_c}Properties of HIP 21152 C}
\tablehead{\colhead{Property} & \colhead{Value} & \colhead{Refs}}
\startdata
$\alpha_{2000.0}$ & 04:33:56.60 & 1\\
$\delta_{2000.0}$ & $+05$:37:23.54 & 1\\
$\pi$ (mas) & $22.62 \pm 0.45$ & 1\\
$\mu_{\alpha}$ (mas/yr) & $107.17 \pm 0.49$ & 1\\
$\mu_{\delta}$ (mas/yr) & $7.60 \pm 0.39$ & 1\\
Distance (pc) & $44.2 \pm 0.9$ & 1\\
Projected Separation ($\prime\prime$) & 1837 & 2\\
Projected Separation ($10^4 \, \mathrm{au}$) & 7.9 & 2\\
3D Separation ($10^5 \, \mathrm{au}$) & $2.5 \pm 1.4$ & 2\\
SpT & L0/1 & 3, 4\\
RUWE & 1.05 & 1\\
$\log(L_{\mathrm{bol}} / L_\odot)$ (dex) & $-3.62 \pm 0.04$ & 2\\
Mass ($\mathrm{M_{Jup}}$)\tablenotemark{a} & $73 \pm 7$ & 2\\
Binding Energy ($10^{39} \, \mathrm{erg}$) & $-5 \pm 4$ & 2\\
$r$ (mag) & $21.64\pm0.07$ & 5\\
$i$ (mag) & $19.538 \pm 0.009$ & 5\\
$z$ (mag) & $18.039 \pm 0.008$ & 5\\
$y$ (mag) & $17.087 \pm 0.009$ & 5\\
Gaia $G$ (mag) & $19.969 \pm 0.004$ & 1\\
$J$ (mag) & $15.28 \pm 0.06$ & 6\\
$H$ (mag) & $14.39 \pm 0.06$ & 6\\
$K$ (mag) & $13.91 \pm 0.06$ & 6\\
\enddata
\tablenotetext{a}{Inferred via the BHAC15 evolutionary model \citep{baraffeNewEvolutionaryModels_2015}.}\tablerefs{(1) \citet{gaiacollaborationGaiaEarlyData_2021}; (2) This work; (3) \citet{reyleNewUltracoolBrown_2018}; (4) \citet{gagneBanyanXiiiFirst_2018}; (5) \citet{chambersPanstarrs1Surveys_2016}; (6) \citet{skrutskieTwoMicronAll_2006}}
\end{deluxetable}

\section{Conclusions}
In this work, we reported the independent discovery of HIP~21152~B, the first imaged brown dwarf companion in the Hyades, along with a comprehensive characterization of its orbital and atmospheric properties. HIP~21152 was targeted in our ongoing Astrometric Accelerations as Dynamical Beacons program due to its small but significant Hipparcos-Gaia astrometric acceleration of \SI{8.4 \pm 0.8}{m s^{-1} yr^{-1}}. A joint orbit fit of its relative astrometry including new Keck/NIRC2 and CHARIS imaging, radial velocities, and HGCA acceleration yields a dynamical mass of $24^{+6}_{-4} \, \mathrm{M_{Jup}}$. This measurement is slightly lower ($1{-}2\sigma$) than model-inferred masses from substellar evolutionary models, with the \citet{saumonEvolutionDwarfsColor_2008} hybrid grid that incorporates both the presence and dissipation of clouds producing the most consistent inferred mass. HIP~21152~B is the first benchmark brown dwarf with a model-independent mass in the Hyades, the lowest mass brown dwarf companion with a dynamical mass, and the fourth lowest mass benchmark substellar companion after $\beta$ Pic b \citep{brandtPreciseDynamicalMasses_2021}, $\beta$ Pic c \citep{nowakDirectConfirmationRadialvelocity_2020,brandtPreciseDynamicalMasses_2021}, and HR 8799 e \citep{brandtFirstDynamicalMass_2021}. We also identified a wide-separation (1837\arcsec) comoving low-mass star or high-mass brown dwarf, HIP 21152 C, that is likely bound with a model-inferred mass of \SI{73 \pm 7}{M_{Jup}}. 

We investigated the atmospheric properties of HIP~21152~B using new reductions of the SPHERE/IFS spectrum from \citet{bonavitaResultsCopainsPilot_2022}, a new 2022 CHARIS spectrum, and both SPHERE/IRDIS $K_{12}$ and Keck/NIRC2 $L^{\prime}$ photometry. Comparing the observed spectra against spectral standards yields a spectral type of $\mathrm{T0} \pm 1$. The best-fit model atmosphere from \citet{saumonEvolutionDwarfsColor_2008} has $T_{\mathrm{eff}} = \SI{1400}{K}$, $\log g = 4.5$, and $f_{\mathrm{sed}} = 2$. This is consistent with the companion residing at the L/T transition. We additionally performed a suite of retrievals with the \texttt{Helios-r2} models. The retrievals produce a similar temperature to the grid-based comparison of $1400{-}\SI{1600}{K}$, while the retrieved surface gravity is slightly higher ($\log g \sim 4.9{-}5.4$).

This companion joins a small but growing list of benchmark companions with well-constrained dynamical masses and independent age determinations (see \citealt{fransonDynamicalMassYoung_2022} for a recent compilation). The modest difference between the model-predicted mass and the dynamical mass of this companion echoes other cases where benchmark companions are somewhat less massive than model-inferred masses \citep[e.g.,][]{dupuyDynamicalMassSubstellar_2009,beattySignificantOverluminosityTransiting_2018, rickmanSpectralAtmosphericCharacterisation_2020}. Additionally, it follows the trend suggested in \citet{brandtImprovedDynamicalMasses_2021} of younger, lower-mass benchmark brown dwarfs being systematically under-massive. These discrepancies underscore the need for additional benchmark companions to empirically test and calibrate substellar evolutionary models. As demonstrated by the discovery of HIP~21152~B, astrometric accelerations offer a promising avenue forward.

\begin{acknowledgements}
    We thank Garreth Ruane, Bin Ren, Nicole Wallack, and Dimitri Mawet for helpful discussions regarding the reduction of VVC ADI sequences. We thank Masayuki Kuzuhara and Thayne Currie for helpful discussions about this object and for sharing their astrometry and radial velocities. K.F. acknowledges support from the National Science Foundation Graduate Research Fellowship Program under Grant No. DGE-1610403. B.P.B. acknowledges support from the National Science Foundation grant AST-1909209, NASA Exoplanet Research Program grant 20-XRP20$\_$2-0119, and the Alfred P. Sloan Foundation. KM acknowledges funding by the Science and Technology Foundation of Portugal (FCT), grants No. PTDC/FIS-AST/28731/2017 and UIDB/00099/2020. This work was supported by a NASA Keck PI Data Award, administered by the NASA Exoplanet Science Institute.

    This work has made use of data fram the European Space Agency (ESA) space mission Gaia. Gaia data are being processed by the Gaia Data Processing and Analysis Consortium (DPAC). Funding for the DPAC is provided by national institutions, in particular the institutions participating in the Gaia MultiLateral Agreement (MLA). The Gaia mission website is \href{https://www.cosmos.esa.int/gaia}{https://www.cosmos.esa.int/gaia}. The Gaia archive website is \href{https://archives.esac.esa.int/gaia}{https://archives.esac.esa.int/gaia}. This publication makes use of data products from the Two Micron All Sky Survey, which is a joint project of the University of Massachusetts and the Infrared Processing and Analysis Center/California Institute of Technology, funded by the National Aeronautics and Space Administration and the National Science Foundation. This publication makes use of data products from the Wide-field Infrared Survey Explorer, which is a joint project of the University of California, Los Angeles, and the Jet Propulsion Laboratory/California Institute of Technology, funded by the National Aeronautics and Space Administration. The Pan-STARRS1 Surveys (PS1) have been made possible through contributions of the Institute for Astronomy, the University of Hawaii, the Pan-STARRS Project Office, the Max-Planck Society and its participating institutes, the Max Planck Institute for Astronomy, Heidelberg and the Max Planck Institute for Extraterrestrial Physics, Garching, The Johns Hopkins University, Durham University, the University of Edinburgh, Queen's University Belfast, the Harvard-Smithsonian Center for Astrophysics, the Las Cumbres Observatory Global Telescope Network Incorporated, the National Central University of Taiwan, the Space Telescope Science Institute, the National Aeronautics and Space Administration under Grant No. NNX08AR22G issued through the Planetary Science Division of the NASA Science Mission Directorate, the National Science Foundation under Grant No. AST-1238877, the University of Maryland, and Eotvos Lorand University (ELTE). This research has made use of the SIMBAD database and the VizieR catalogue access tool, CDS, Strasbourg, France. This work has made use of the SPHERE Data Centre, jointly operated by OSUG/IPAG (Grenoble), PYTHEAS/LAM/CeSAM (Marseille), OCA/Lagrange (Nice) and Observatoire de Paris/LESIA (Paris) and supported by a grant from Labex OSUG@2020 (Investissements d'avenir - ANR10 LABX56). This work has benefited from The UltracoolSheet at \href{http://bit.ly/UltracoolSheet}{http://bit.ly/UltracoolSheet}, maintained by Will Best, Trent Dupuy, Michael Liu, Rob Siverd, and Zhoujian Zhang, and developed from compilations by \citet{dupuyHawaiiInfraredParallax_2012}, \citet{dupuyDistancesLuminositiesTemperatures_2013}, \citet{liuHawaiiInfraredParallax_2016}, \citet{bestPhotometryProperMotions_2018}, \citet{bestVolumelimitedSampleL0t8_2021}.
    
    The development of SCExAO is supported by the Japan Society for the Promotion of Science (Grant-in-Aid for Research \#23340051, \#26220704, \#23103002, \#19H00703, \#19H00695 and \#21H04998), the Subaru Telescope, the National Astronomical Observatory of Japan, the Astrobiology Center of the National Institutes of Natural Sciences, Japan, the Mt Cuba Foundation and the Heising-Simons Foundation. CHARIS was built at Princeton University under a Grant-in-Aid for Scientific Research on Innovative Areas from MEXT of the Japanese government (\#23103002). Data presented herein were obtained at the W. M. Keck Observatory from telescope time allocated to the National Aeronautics and Space Administration through the agency's scientific partnership with the California Institute of Technology and the University of California. The Observatory was made possible by the generous financial support of the W. M. Keck Foundation.
    
    The authors wish to recognize and acknowledge the very significant cultural role and reverence that the summit of Maunakea has always had within the indigenous Hawaiian community. We are most fortunate to have the opportunity to conduct observations from this mountain.
\end{acknowledgements}

\facilities{Keck:II (NIRC2), VLT:Melipal (SPHERE), Subaru (SCExAO/CHARIS), Smith (Tull Coud\'e spectrograph)}
\software{\texttt{VIP} \citep{gomezgonzalezVipVortexImage_2017}, \texttt{pyKLIP} \citep{wangPyklipPsfSubtraction_2015}, \texttt{orvara} \citep{brandtOrvaraEfficientCode_2021}, \texttt{ccdproc} \citep{craigAstropyCcdprocV1_2017}, \texttt{photutils} \citep{bradleyAstropyPhotutilsV0_2019}, \texttt{astropy} \citep{astropycollaborationAstropyCommunityPython_2013,astropycollaborationAstropyProjectBuilding_2018}, \texttt{pandas} \citep{mckinneyDataStructuresStatistical_2010}, \texttt{matplotlib} \citep{hunterMatplotlib2dGraphics_2007}, \texttt{numpy} \citep{harrisArrayProgrammingNumpy_2020}, \texttt{scipy} \citep{virtanenScipyFundamentalAlgorithms_2020}, \texttt{emcee} \citep{foreman-mackeyEmceeMcmcHammer_2013}, \texttt{corner} \citep{foreman-mackeyCornerPyScatterplot_2016}, \texttt{lightkurve} \citep{lightkurvecollaborationLightkurveKeplerTess_2018}, \texttt{Helios-r2} \citep{kitzmannHeliosr2NewBayesian_2020}, \texttt{scikit-learn} \citep{pedregosaScikitlearnMachineLearning_2011}}

\begin{deluxetable*}{cccccc}
\tablecaption{\label{tab:rel_astrometry_hip55507}Astrometry of HIP 55507 B}
\tablehead{\colhead{Filter} & \colhead{Date} & \colhead{Epoch} & \colhead{Separation} & \colhead{PA} & \colhead{Instrument}\\ \colhead{ } & \colhead{(UT)} & \colhead{(UT)} & \colhead{(mas)} & \colhead{(\si{\degree})} & \colhead{ }}
\startdata
$H$ & 2012 Jan 07 & 2012.016 & $475.8 \pm 1.4$ & $292.40 \pm 0.12$ & NIRC2 \\
$K^{\prime}$ & 2012 Jan 07 & 2012.016 & $475.3 \pm 1.4$ & $292.65 \pm 0.12$ & NIRC2 \\
$K_{\mathrm{cont}}$ & 2015 May 29 & 2015.406 & $549.7 \pm 1.4$ & $274.79 \pm 0.11$ & NIRC2 \\
$J_{\mathrm{cont}}$ & 2015 May 29 & 2015.406 & $548.6 \pm 1.4$ & $274.67 \pm 0.11$ & NIRC2 \\
$K_s$ & 2021 Dec 21 & 2021.970 & $748 \pm 5$ & $254.04 \pm 0.20$ & NIRC2
\enddata
\end{deluxetable*}

\begin{deluxetable*}{lccc} 
\tablecaption{\label{tab:elements_hip55507}HIP 55507 B Orbit Fit Results}
\tablehead{\colhead{Parameter} & \colhead{Median $\pm 1\sigma$} & \colhead{95.4\% C.I.} & \colhead{Prior}}
\startdata
\multicolumn{4}{c}{Fitted Parameters} \\
\hline
$M_{\mathrm{comp}}$ $(\mathrm{M_{Jup}})$ & ${98}_{-11}^{+13}$ & (78, 126) & $1/M_{\mathrm{comp}}$ (log-flat)\\
$M_{\mathrm{host}}$ $(\mathrm{M_\odot})$ & ${0.79}_{-0.10}^{+0.12}$ & (0.61, 1.05) & $\SI{0.7 \pm 0.2}{M_\odot}$ (Gaussian)\\
$a$ $(\mathrm{AU})$ & ${34}_{-4}^{+7}$ & (26, 51) & $1/a$ (log-flat)\\
$i$ $(\si{\degree})$ & $117.5 \pm 1.5$ & (114.6, 120.5) & $\sin (i)$, $\SI{0}{\degree} < i < \SI{180}{\degree}$\\
$\sqrt{e} \sin{\omega}$ & ${-0.43}_{-0.14}^{+0.17}$ & (-0.66, -0.03) & Uniform\\
$\sqrt{e} \cos{\omega}$ & ${-0.35}_{-0.08}^{+0.10}$ & (-0.50, -0.13) & Uniform\\
$\Omega$ $(\si{\degree})$ & $219.9 \pm 2.2$ & (215.9, 224.3) & Uniform\\
$\lambda_{\mathrm{ref}}$ $(\si{\degree})$\tablenotemark{a} & $253 \pm 7$ & (240, 266) & Uniform\\
Parallax $(\si{mas})$ & $39.319 \pm 0.015$ & (39.288, 39.349) & $\SI{39.3187 \pm 0.0147}{mas}$ (Gaussian)\\
$\mu_\alpha$ ($\si{mas.yr^{-1}}$) & ${-200.60}_{-0.37}^{+0.32}$ & (-201.38, -200.01) & Uniform\\
$\mu_\delta$ ($\si{mas.yr^{-1}}$) & ${-139.0}_{-0.5}^{+0.4}$ & (-140.0, -138.2) & Uniform\\
RV Jitter $\sigma_{\mathrm{RV}}$ ($\si{m.s^{-1}}$) & ${4.4}_{-0.7}^{+0.9}$ & (3.2, 6.5) & $1/\sigma_{\mathrm{RV}}$ (log-flat), $\sigma_{\mathrm{RV}} \in (0, 1000\si{m.s^{-1}}]$\\
\hline
\multicolumn{4}{c}{Derived Parameters} \\
\hline
$P$ (yr) & ${210}_{-50}^{+80}$ & (130, 420) & . . .\\
$e$ & ${0.31}_{-0.10}^{+0.12}$ & (0.13, 0.54) & . . .\\
$\omega$ $(\si{\degree})$ & ${231}_{-19}^{+14}$ & (185, 257) & . . .\\
$T_0$ $(\mathrm{JD})$ & ${2450600}_{-2000}^{+1400}$ & (2446000, 2453300) & . . .\\
$q$ $(=M_{\mathrm{comp}}/M_{\mathrm{host}})$ & ${0.118}_{-0.013}^{+0.015}$ & (0.095, 0.150) & . . .\\
$\rho$ on 2022.159 $(\si{mas})$ & $759 \pm 4$ & (750, 767) & . . .\\
$\theta$ on 2022.159 $(\si{\degree})$ & $253.08 \pm 0.15$ & (252.78, 253.39) & . . .
\enddata
\tablenotetext{a}{Mean longitude at the reference epoch of 2010.0.}

\end{deluxetable*}
%file: ../../../results/HIP 55507_chain004.fits
%burnin: 5000

\appendix
\section{CHARIS Astrometric Calibration with HIP 55507 \label{sec:hip55507}}
We calibrate our February 2022 CHARIS astrometry of HIP~21152~AB using the 0.75\arcsec visual binary HIP~55507. We targeted HIP~55507 with CHARIS on UT 2022 February 27 during the same run we observed HIP~21152. HIP~55507 was previously imaged with Keck/NIRC2 on UT 2012 January 07 and 2015 May 29 (PI: Justin Crepp) as part of the TRENDS survey \citep{gonzalesTrendsHighcontrastImaging_2020}. We also obtained coronagraphic adaptive optics imaging of this system with Keck/NIRC2 on UT 2021 December 21. HIP~55507 additionally has Keck/HIRES radial velocities from 2009--2013 reported in \citet{butlerLcesHiresKeck_2017}. To correct for small but potentially significant orbital motion between the December 2021 imaging and the February 2022 CHARIS dataset, we perform a joint orbit fit with \texttt{orvara} of the relative astrometry, RVs, and HGCA acceleration ($\chi^2 = 674$, or $26\sigma$ with 2 degrees of freedom).

Basic image reduction is carried out following the description in Section \ref{sec:nirc2_obs}. To correct for geometric distortions, we apply the solution from \citet{serviceNewDistortionSolution_2016} for imaging after the NIRC2 camera and AO system were realigned on UT 2015 April 13 and the solution from \citet{yeldaImprovingGalacticCenter_2010} for imaging prior to that date. We fit 2D Gaussians to the positions of the host star and the binary companion in each exposure to measure astrometry from the calibrated images. The individual exposures within a sequence are weighted by their total integration times. We also incorporate the uncertainty in the distortion solution, north alignment, and plate scale from \citet{serviceNewDistortionSolution_2016} or \citet{yeldaImprovingGalacticCenter_2010}, following \citet{fransonDynamicalMassYoung_2022}. For the December 2021 imaging with the 600-mas Lyot coronagraph, we adopt a noise floor of $\pm 5$ mas in separation and $\pm 0.2^{\circ}$ in position angle to account for potential systematic offsets in the host star position behind the partially transparent coronagraph (see \citealt{bowlerOrbitDynamicalMass_2018}). The resulting astrometry is listed in Table \ref{tab:rel_astrometry_hip55507}.

We perform a joint orbit fit of HIP~55507~A with \texttt{orvara} to obtain the predicted companion position on UT 2022 February 28 for anchoring the CHARIS astrometry. Our fit uses 100 walkers, 20 temperatures, and $5 \times 10^5$ total steps. We discard the first $50\%$ of each chain as burn-in. The priors for the orbit fit of HIP~21152~B (see Table \ref{tab:elements}) are adopted except for host star mass, for which we use a prior of $0.7 \pm 0.2$ $\mathrm{M_\odot}$ based on the estimate in \citet{andersPhotoastrometricDistancesExtinctions_2022}, and parallax, which was measured to be \SI{39.3187 \pm 0.0147}{mas} from Gaia EDR3. The orbit elements from the fit are shown in Table \ref{tab:elements_hip55507}. Upon propogating the position to the UT 2022 February 28 epoch, the orbit fit predicts the companion to have a separation of $\rho = \SI{759 \pm 4}{mas}$ and a position angle of $\theta = 253\fdg08 \pm 0\fdg15$.

The CHARIS sequence of HIP~55507 that we obtained consists of 30 frames in low-resolution ($R \sim 20$), broadband ($1.15{-}2.39$ $\mathrm{\mu m}$) mode with integration times of \SI{30}{s} and the \SI{113}{mas} Lyot coronagraph. The CHARIS plate scale has been found to be stable over long periods, so we adopt the value of \SI{16.15 \pm 0.1}{mas.spaxel^{-1}} from Chen et al. (in prep.). We measure astrometry using the same procedure from Section \ref{sec:charis_imaging} and find a separation of \SI{758 \pm 6}{mas} and a position angle of $252\fdg2 \pm 0\fdg2$. While the separation value is consistent with the prediction, the position angle is slightly lower, with a difference of $0\fdg9 \pm 0\fdg3$. To account for this slight offset in north alignment, we add this value to the position angle of the HIP~21152~B CHARIS measurement (Section \ref{sec:charis_imaging}).

\begin{figure*}
    \centering
    \includegraphics[width=0.5\linewidth]{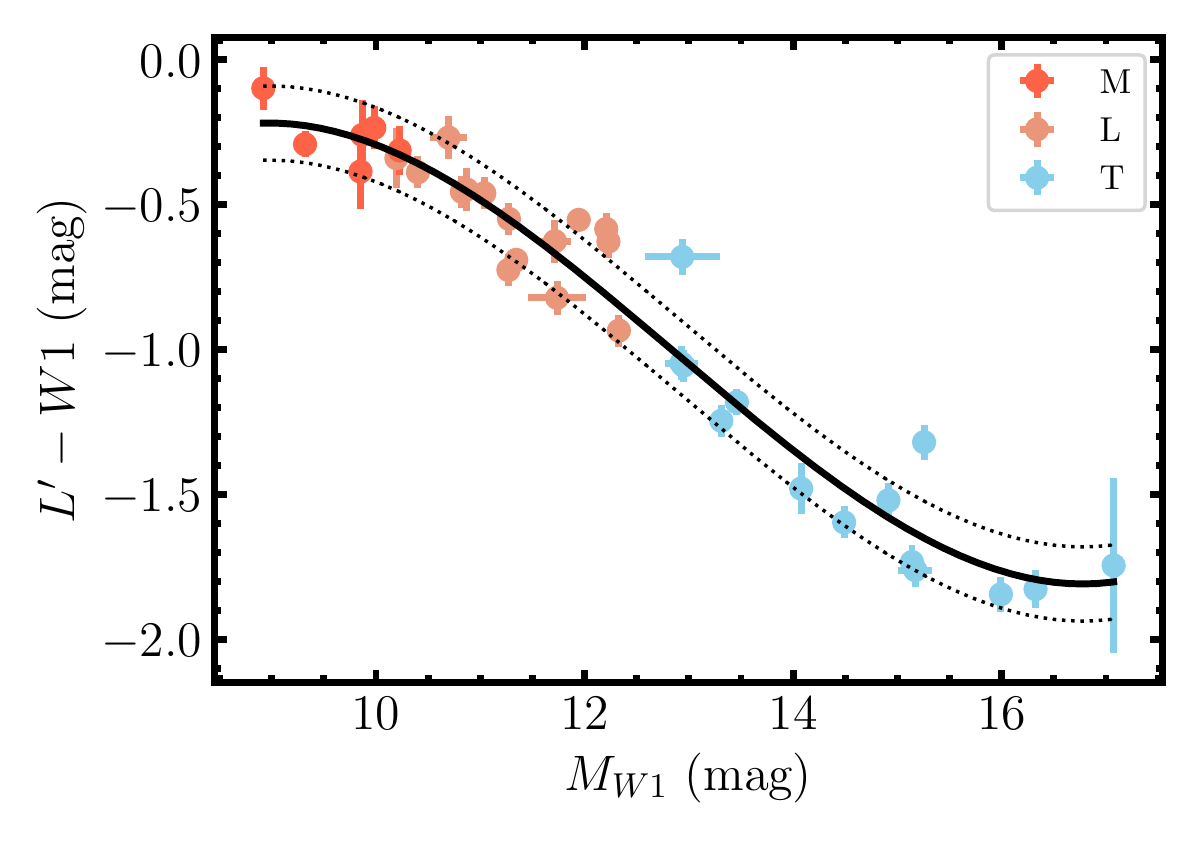}
    \caption{$L' - W1$ as a function of $M_{W1}$ for field-age M dwarfs (orange), L dwarfs (light orange), and T dwarfs (blue) from \citet{filippazzoFundamentalParametersSpectral_2015}. The black line denotes a 4th-order polynomial fit to the objects' magnitudes and the dotted lines show the rms of the relation.
    \label{fig:w1_lp_relation}}
\end{figure*}
\section{Converting $W1$ to $L'$ Magnitudes \label{sec:w1_lp}}
To convert $W1$ magnitudes to $L'$ magnitudes in constructing the CMD (Section \ref{sec:spt}), we develop a relation between the absolute $W1$ magnitude $M_{W1}$ and $L' - W1$ using the 35 field-age substellar objects in \citet{filippazzoFundamentalParametersSpectral_2015} with parallaxes and both $L'$ and $W1$ photometry. We fit a 4th-order polynomial of the form $P(x) = \sum_{i=0}^4 c_i x^i$, where $x = M_{W1}$ and $P(x) = L' - W1$. The fit produced coefficients of $c_0 = -11.449$, $c_1 = 3.0429$, $c_2 = 0.26004$, and $c_3 = 6.7306 \times 10^{-3}$, and has an rms of 0.128. The absolute $W1$ magnitudes of the objects used in the fit range from $8.9 \ \mathrm{mag} \ \leq M_{W1} \leq 17.1 \ \mathrm{mag}$. The relation is shown in Figure \ref{fig:w1_lp_relation}.

\begin{figure*}
    \centering
    \includegraphics[width=\linewidth]{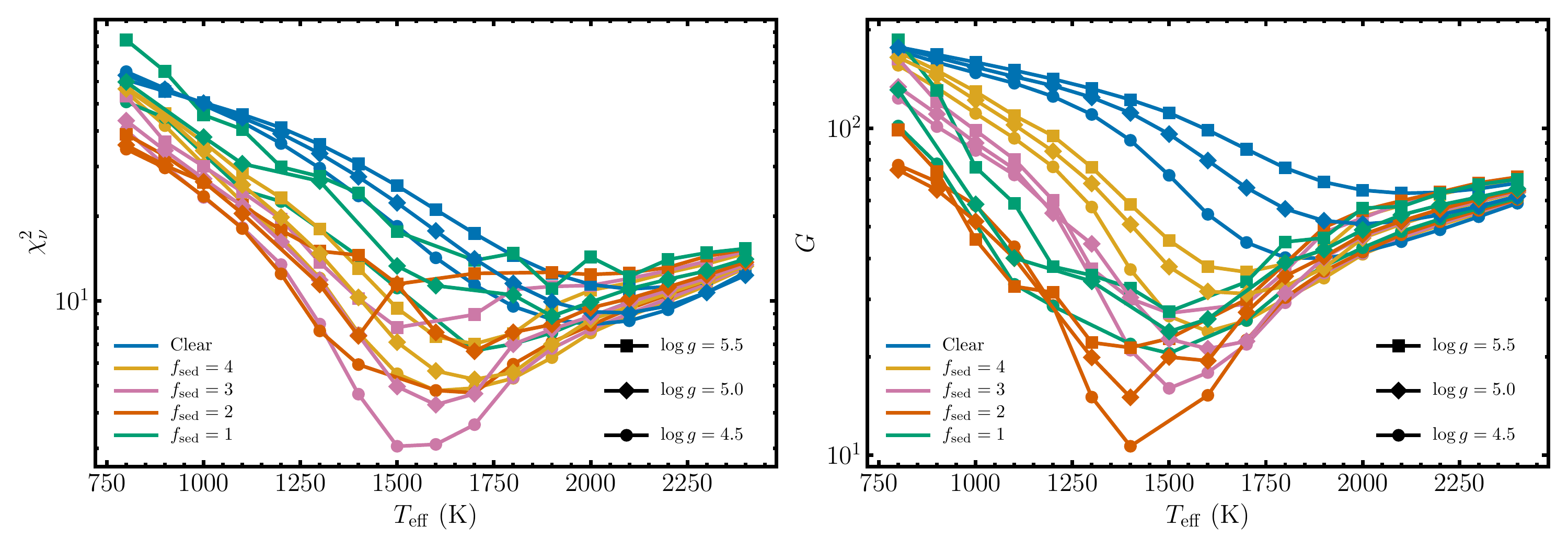}
    \caption{Reduced chi-square ($\chi^2_\nu$; left) and goodness-of-fit ($G$; right) values for model spectra from \citet{saumonEvolutionDwarfsColor_2008} compared with HIP~21152~B's spectra and photometry. Here, Reduction 1 is used for the SPHERE/IFS spectrum. The two metrics produce different best-fitting spectra. $\chi^2_\nu$ yields a best-fit spectrum with $T_{\mathrm{eff}} = \SI{1500}{K}$, $\log{g} = 4.5$, and $f_{\mathrm{sed}} = 3$, while $G$ gives a best-fitting spectrum with $T_{\mathrm{eff}} = \SI{1400}{K}$, $\log g = 4.5$, and $f_{\mathrm{sed}} = 2$.
    \label{fig:model_spec_comp_red1}}
\end{figure*}

\begin{figure*}
    \centering
    \includegraphics[width=0.5\linewidth]{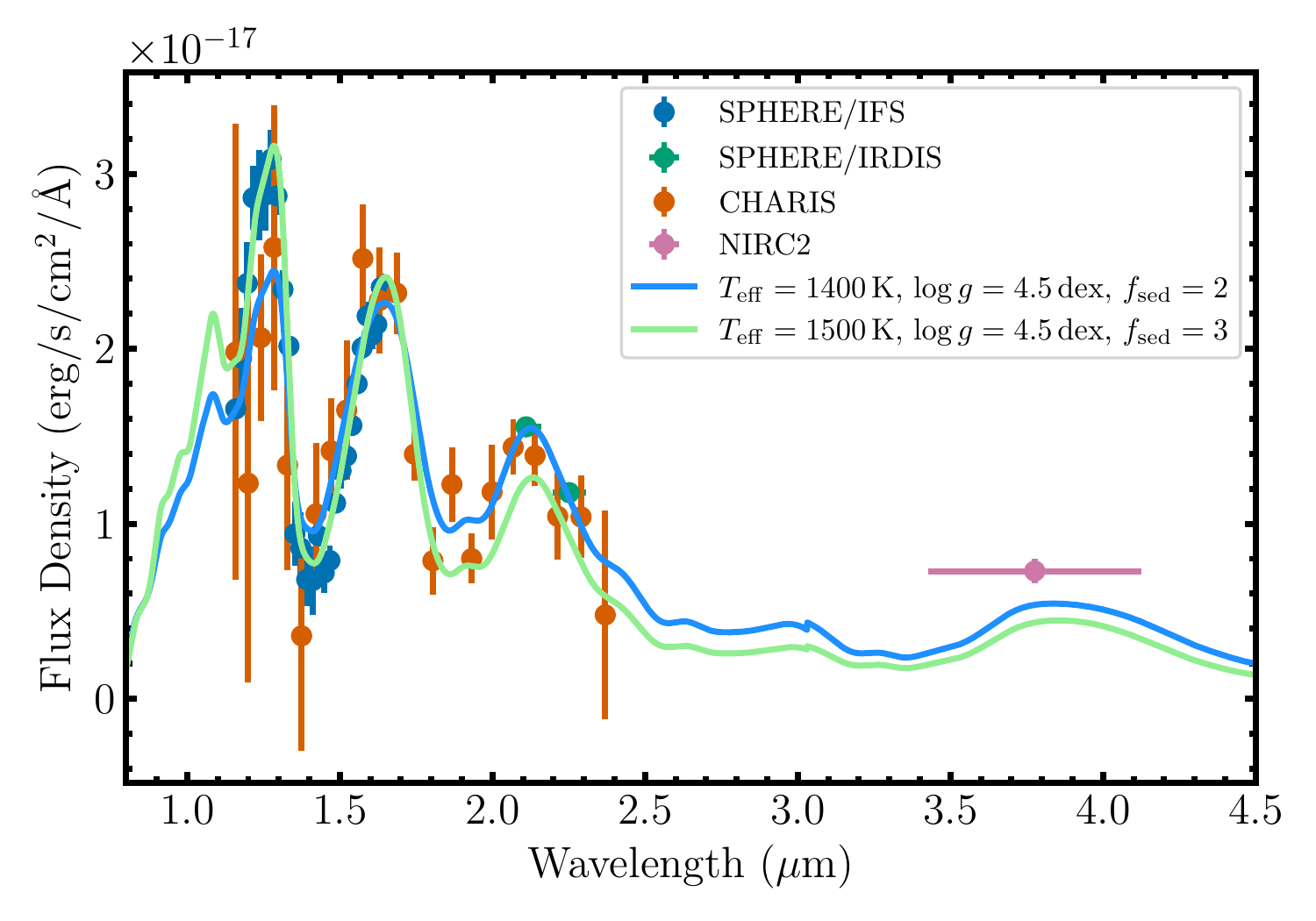}
    \caption{Comparison of best-fitting model spectra with HIP~21152~B's spectra and photometry. The two curves show the best-fitting spectra based on $\chi^2_\nu$ (light green) and $G$ (blue). Here, Reduction 1 is used for the SPHERE/IFS spectrum. The model spectra have been smoothed to a resolving power of $R = 25$.
    \label{fig:spectrum_model_plot_red1}}
\end{figure*}

\begin{figure}
    \centering
    \includegraphics[width=0.5\linewidth]{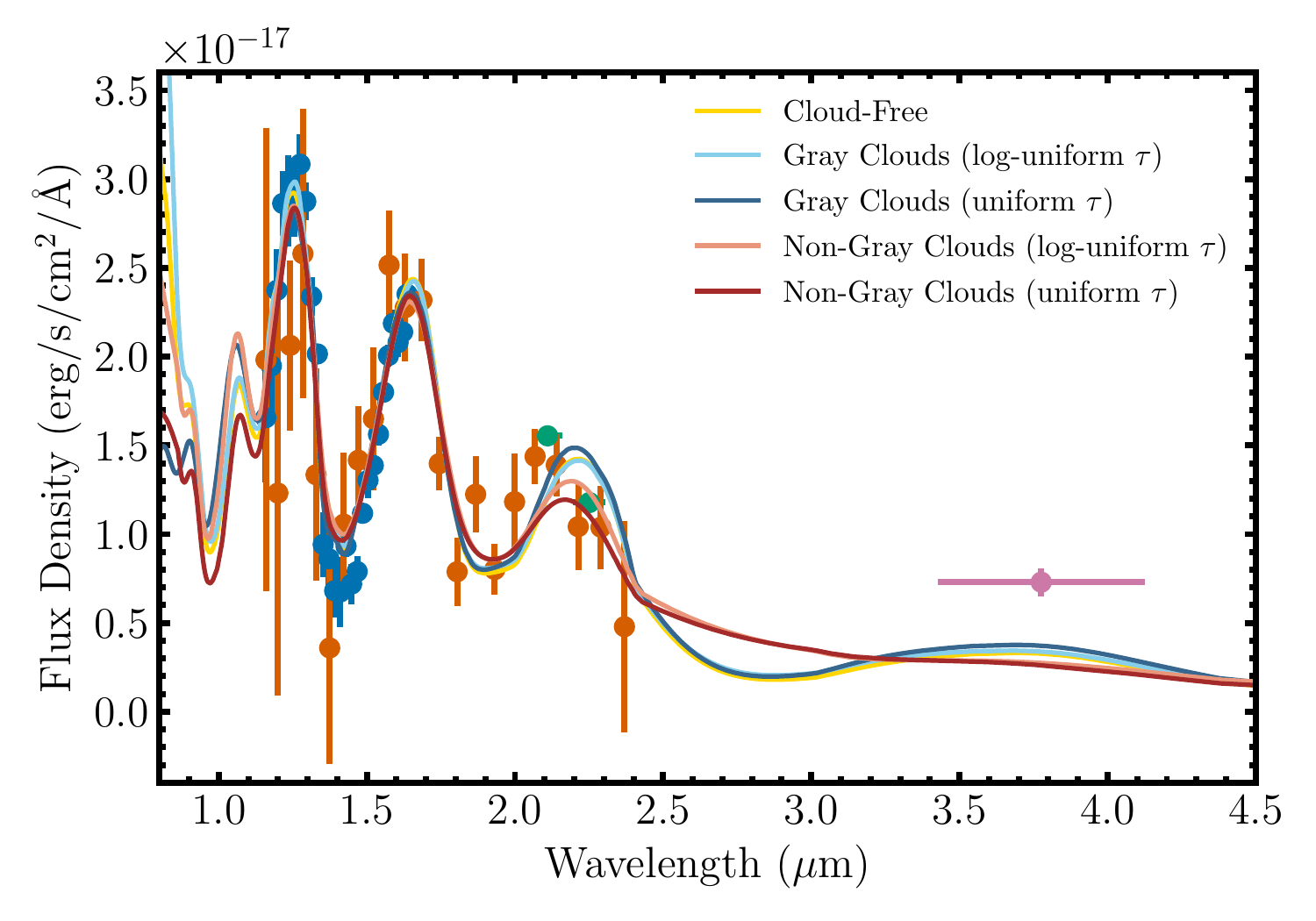}
    \caption{Spectra and photometry of HIP~21152~B compared with the best-fit retrievals from the five retrieval versions we consider. These retrievals incorporate all available data (IFS + CHARIS + Photometry) and adopt Reduction 1 for the SPHERE/IFS spectrum. The retrieved spectra are smoothed to a resolving power of $R=25$. 
    \label{fig:retrieval_comp_plot_red1}}
\end{figure}
\section{Grid-Based Model Comparison (Reduction 1) \label{sec:atm_model_comp_red1}}
Here, we show the results of performing the model comparison with \citet{saumonEvolutionDwarfsColor_2008} atmospheric models using Reduction 1 for the SPHERE/IFS spectrum (see Section \ref{sec:atm_model_comp} for details on the model comparison). Figure \ref{fig:model_spec_comp_red1} displays reduced chi-square ($\chi^2_\nu$) and goodness-of-fit ($G$) statistics (analogous to Figure \ref{fig:model_spec_comp}). Unlike with Reduction 2, the best-fitting atmospheric model is dependent on the metric used. $\chi^2_\nu$ yields a minimum around $\SI{1600}{K}$, with the best-fit spectrum having $T_{\mathrm{eff}} = \SI{1500}{K}$, $\log{g}=4.5$, and $f_{\mathrm{sed}} = 3$. The best-fit model produced by the $G$ statistic has $T_{\mathrm{eff}} = \SI{1400}{K}$, $\log{g} = 4.5$, and $f_{\mathrm{sed}} = 2$, which matches the best-fit model with Reduction 2 of the IFS spectrum (Section \ref{sec:atm_model_comp}). The $\chi^2_\nu$ metric preferring models with higher $T_{\mathrm{eff}}$ and $f_{\mathrm{sed}}$ indicates that Reduction 1 drives effective temperature and sedimentation efficiency upward. As $G$ weights the companion's photometry more heavily, this effect is minimized for that metric. This can be seen on Figure \ref{fig:spectrum_model_plot_red1}, which shows the two best-fit model spectra. The $\chi^2_\nu$ metric (green curve) better captures the enhanced flux in the $J$-band from Reduction 1, but it departs from the compared to the \SI{1400}{K} model. 

\section{Atmospheric Retrieval (Reduction 1) \label{sec:retrieval_red1}}
Here, we show the retrieval results using Reduction 1 for the SPHERE/IFS spectrum instead of Reduction 2 (see Section \ref{sec:retrieval} for details on the retrievals). Table \ref{tab:retrieval_posteriors_red1} shows the posteriors on the retrieved parameters. The CHARIS-only retrieval is shown for completeness; it is identical to the version in Table \ref{tab:retrieval_posteriors_red2}. These retrievals produce significantly higher effective temperatures of $T_{\mathrm{eff}}\sim\SI{1800}{K}$ for the IFS-only version and $T_{\mathrm{eff}}\sim\SI{1900}{K}$ for the IFS + CHARIS + Photometry retrievals. These values are inconsistent with the effective temperature derived from the Stefan-Boltzmann relation (\SI{1300 \pm 50}{K}) and the effective temperature from the CHARIS-only retrieval of ${\sim}\SI{1400}{K}$. We detect water in all retrievals, with mixing ratios $x_{\mathrm{H_2 O}} {\sim} 10^{-3}{-}10^{-2}$. FeH and TiH are detected in some versions of the retrieval models or included datasets, with small mixing ratios ${\sim} 10^{-9}{-}10^{-6}$. Figure \ref{fig:retrieval_comp_plot_red1} shows the best-fit spectra for the five retrieval versions against the companion's spectra and photometry (analogous to Figure \ref{fig:retrieval_comp_plot}).

\begin{figure}
    \centering
    \includegraphics[width=0.4\linewidth]{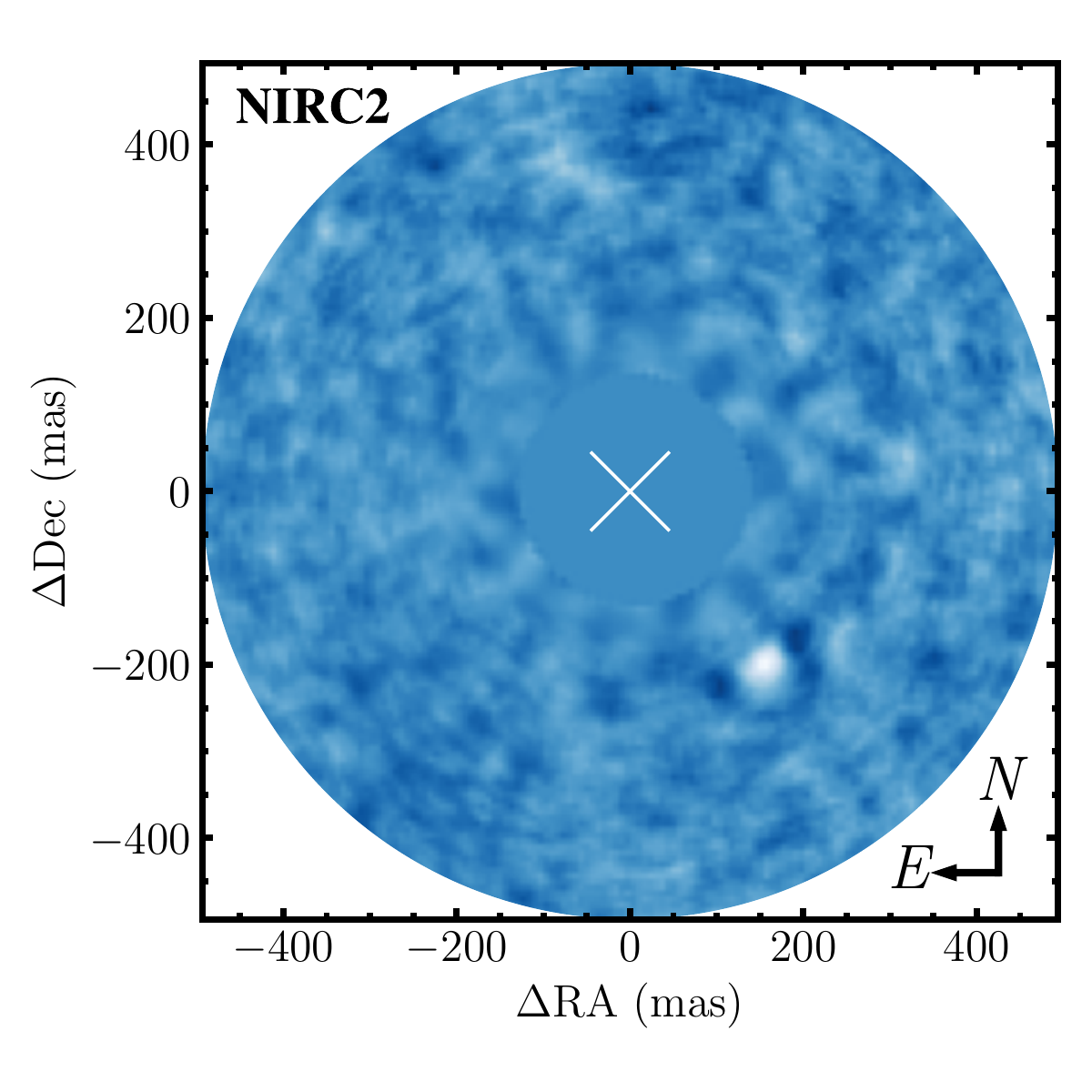}
    \caption{Keck/NIRC2 $L'$ imaging of HIP~21152~B, reduced using the PACO algorithm. The frame is oriented so that north is up and east is to the left. We detect HIP~21152~B with a S/N of 7.8 using PACO.
    \label{fig:nirc2_paco}}
\end{figure}

\section{Additional Reduction of Keck/NIRC2 Imaging \label{sec:nirc2_paco}}
Here, we apply an independent reduction of the 2021 December $L'$-band NIRC2 observation of HIP 21152 using an alternative framework to further assess the significance of the detection from this ADI sequence. We developed a custom implementation of the PAtch COvariances (PACO) algorithm \citep{flasseurExoplanetDetectionAngular_2018}, which uses spatially local models of the expected speckle noise statistics in a maximum-likelihood estimator to find the best-fit flux estimate of a potential off-axis point source. Fixed spatial patches centered at a pixel $\phi_t$ in the $t^{\mathrm{th}}$ frame, and spanning the time axis of the ADI image cube, are used to construct a mean vector and covariance matrix, $ \bm m_{\phi_t}$ and $\hat{C}_{\phi_t}$. These parameterize a multivariate Gaussian distribution that describes the speckle noise local to $\phi_t$. The set of spatial patches centered at pixels $\phi_{\{1:T\}}$, determined by the parallactic rotation of the ADI sequence, each contain an off-axis point source if it were located at $\phi_1$ in the first frame. The maximum likelihood estimate of a potential point source is found by minimizing the quadratic cost function 

\begin{equation}
        \hat{\alpha}_{\phi_1} = \argminA_{\alpha} \sum_{t = 1}^{T}(\bm p_{\phi_t} - \alpha \bm h - \bm m_{\phi_t})^\top \hat{C}_{\phi_t}^{-1} (\bm p_{\phi_t} - \alpha \bm h  -  \bm m_{\phi_t})\mbox{,} \label{PACO_cost}
\end{equation}
\noindent
where $T$ is the number of frames in the ADI sequence, and $\bm h$ is a unit-normalized template of the expected signal profile of a point source in the data patch, $\bm p_{\phi_i}$.  Equation \ref{PACO_cost} has a closed form solution, and can be solved with linear least-squares. Under this framework, the S/N of a potential source is

\begin{equation}
    \mbox{S/N} = \frac{\hat{\alpha}_{\phi_1}}{ \sqrt{\frac{1}{T}\sum _{t = 1}^{T} \bm h^\top \hat{C}_{\phi_t}^{-1} \bm h}} \mbox{.} \label{PACO_snr}
\end{equation}

Our implementation of the PACO algorithm deviates slightly from the original paper. We model the local speckle noise as a multivariate double exponential distribution, which we find is more robust at close separations. This model results in finding the minimum mean absolute error estimate of $\hat{\alpha}_{\phi_1}$, but does not permit an analytical solution. We used the \texttt{SciPy} implementation of the Levenberg-Marquardt nonlinear least-squares algorithm \citep{virtanenScipyFundamentalAlgorithms_2020} instead. Additionally, we used the \texttt{scikit-learn} implementation of the Ledoit-Wolf covariance shrinkage method \citep{pedregosaScikitlearnMachineLearning_2011} to estimate each $\hat{C}_{\phi_t}$ in lieu of the covariance shrinkage approach proposed by \citet{flasseurExoplanetDetectionAngular_2018}. To improve the run-time of our implimentation, we process the ADI sequence serially over pixel-width annuli, and parallelize the algorithm for each pixel in the annulus. The choice of $\bm h$ is instrument specific. For the $L'$-band filter with NIRC2, we use a symmetric 2D-Guassian profile with a 3.5-pixel standard deviation. The size of the local spatial patch centered at each pixel $\phi_t$ must also be tuned. We find that using a square patch with a width of 8-pixels works well for most data sets. 

Using our implementation of PACO, we were able to formally recover HIP 21152 B at an S/N of 24.7 as determined using Equation \ref{PACO_snr}. However, we note that the S/N values across the image shows correlated noise features with S/N levels up to $\pm$10. To be consistent with the reduction produced with the PCA-based algorithm in Section \ref{sec:nirc2_obs}, we also calculate the S/N with the standard empirical approach, using the RMS flux value in an annular section at the separation of HIP 21152 B, which gives an S/N value of 7.8, which is comparable to the companion's S/N from the PCA reduction (Section \ref{sec:nirc2_obs}). Figure \ref{fig:nirc2_paco} shows the PACO reduction of the Keck/NIRC2 imaging.

\bibliography{references}{}
\bibliographystyle{aasjournal}
\newpage
\
\begin{deluxetable*}{lccccr} 
\tablecaption{\label{tab:ifs_spectrum}SPHERE/IFS Spectrum}
\tablehead{\multicolumn{3}{c}{Reduction 1} & \multicolumn{3}{c}{Reduction 2}\\
\colhead{$\lambda$} & \colhead{$f_\lambda \times 10^{-17}$} & \colhead{$\sigma_{f_\lambda} \times 10^{-17}$} & \colhead{$\lambda$} & \colhead{$f_\lambda \times 10^{-17}$} & \colhead{$\sigma_{f_\lambda} \times 10^{-17}$}\\
\colhead{($\mathrm{\mu m}$)} & \colhead{($\mathrm{erg / s / cm^2 / \text{\AA}}$)} & \colhead{($\mathrm{erg / s / cm^2 / \text{\AA}}$)} & \colhead{($\mathrm{\mu m}$)} & \colhead{($\mathrm{erg / s / cm^2 / \text{\AA}}$)} & \colhead{($\mathrm{erg / s / cm^2 / \text{\AA}}$)}}
\startdata
0.957 & $1.0$ & $0.5$ & 0.982 & $1.5$ & $0.6$\\
0.972 & $1.4$ & $0.3$ & 1.002 & $1.1$ & $0.5$\\
0.987 & $2.2$ & $0.4$ & 1.023 & $1.7$ & $0.4$\\
1.002 & $2.2$ & $0.4$ & 1.043 & $0.9$ & $0.4$\\
1.018 & $1.88$ & $0.10$ & 1.064 & $0.9$ & $0.4$\\
1.034 & $1.9$ & $0.4$ & 1.084 & $1.6$ & $0.3$\\
1.051 & $2.0$ & $0.5$ & 1.105 & $1.5$ & $0.3$\\
1.068 & $1.7$ & $0.4$ & 1.125 & $1.4$ & $0.3$\\
1.086 & $1.2$ & $0.5$ & 1.146 & $1.6$ & $0.3$\\
1.104 & $1.2$ & $0.3$ & 1.166 & $1.9$ & $0.3$\\
1.122 & $1.4$ & $0.3$ & 1.187 & $1.87$ & $0.23$\\
1.140 & $1.6$ & $0.4$ & 1.207 & $2.05$ & $0.17$\\
1.159 & $1.7$ & $0.4$ & 1.228 & $2.04$ & $0.18$\\
1.178 & $1.9$ & $0.3$ & 1.248 & $2.45$ & $0.19$\\
1.197 & $2.37$ & $0.23$ & 1.269 & $2.66$ & $0.15$\\
1.216 & $2.86$ & $0.18$ & 1.289 & $2.51$ & $0.14$\\
1.235 & $2.9$ & $0.3$ & 1.310 & $1.99$ & $0.14$\\
1.255 & $2.88$ & $0.20$ & 1.331 & $1.72$ & $0.15$\\
1.274 & $3.08$ & $0.17$ & 1.351 & $1.22$ & $0.15$\\
1.294 & $2.87$ & $0.11$ & 1.372 & $0.75$ & $0.19$\\
1.314 & $2.34$ & $0.11$ & 1.392 & $0.65$ & $0.17$\\
1.333 & $2.02$ & $0.03$ & 1.413 & $0.82$ & $0.17$\\
1.353 & $0.94$ & $0.18$ & 1.433 & $1.00$ & $0.13$\\
1.372 & $0.86$ & $0.21$ & 1.454 & $1.10$ & $0.09$\\
1.391 & $0.68$ & $0.15$ & 1.474 & $1.24$ & $0.10$\\
1.411 & $0.67$ & $0.20$ & 1.495 & $1.37$ & $0.07$\\
1.430 & $0.93$ & $0.06$ & 1.515 & $1.54$ & $0.05$\\
1.449 & $0.72$ & $0.11$ & 1.536 & $1.69$ & $0.06$\\
1.467 & $0.79$ & $0.09$ & 1.556 & $1.90$ & $0.05$\\
1.486 & $1.12$ & $0.06$ & 1.577 & $2.06$ & $0.05$\\
1.504 & $1.30$ & $0.10$ & 1.597 & $2.15$ & $0.06$\\
1.522 & $1.39$ & $0.05$ & 1.618 & $2.17$ & $0.05$\\
1.539 & $1.56$ & $0.06$ & 1.638 & $2.15$ & $0.06$\\
1.556 & $1.80$ & $0.03$ & $\cdots$ & $\cdots$ & $\cdots$\\
1.573 & $2.01$ & $0.06$ & $\cdots$ & $\cdots$ & $\cdots$\\
1.589 & $2.19$ & $0.08$ & $\cdots$ & $\cdots$ & $\cdots$\\
1.605 & $2.08$ & $0.08$ & $\cdots$ & $\cdots$ & $\cdots$\\
1.621 & $2.14$ & $0.12$ & $\cdots$ & $\cdots$ & $\cdots$\\
1.636 & $2.35$ & $0.08$ & $\cdots$ & $\cdots$ & $\cdots$
\enddata
\end{deluxetable*}
\startlongtable
\begin{deluxetable*}{lcccc} 
\tablecaption{\label{tab:retrieval_posteriors_red2}Retrieved Parameters of HIP~21152~B (Reduction 2)}
\tablehead{\colhead{Model} & \colhead{Parameter} & \colhead{IFS} & \colhead{CHARIS} & \colhead{IFS + CHARIS + Photometry}}
\startdata
Cloud-Free & $\log g$ & $5.1_{-0.7}^{+0.5}$ & $5.0_{-0.7}^{+0.6}$ & $5.3_{-0.6}^{+0.4}$\\
Gray Clouds (log-uniform $\tau$) & $\log g$ & $5.0_{-0.7}^{+0.6}$ & $4.9_{-0.7}^{+0.6}$ & $5.3_{-0.5}^{+0.4}$\\
Gray Clouds (uniform $\tau$) & $\log g$ & $4.9_{-0.7}^{+0.6}$ & $5.0_{-0.7}^{+0.6}$ & $5.3_{-0.4}^{+0.4}$\\
Non-Gray Clouds (log-uniform $\tau$) & $\log g$ & $4.9_{-0.7}^{+0.6}$ & $4.9_{-0.7}^{+0.7}$ & $5.2_{-0.5}^{+0.4}$\\
Non-Gray Clouds (uniform $\tau$) & $\log g$ & $4.9_{-0.7}^{+0.6}$ & $4.9_{-0.7}^{+0.6}$ & $5.2_{-0.5}^{+0.5}$\\
\hline
Cloud-Free & $R$ & $0.74_{-0.14}^{+0.21}$ & $0.8_{-0.3}^{+0.5}$ & $0.68_{-0.09}^{+0.18}$\\
Gray Clouds (log-uniform $\tau$) & $R$ & $0.72_{-0.17}^{+0.31}$ & $0.8_{-0.3}^{+0.4}$ & $0.66_{-0.09}^{+0.13}$\\
Gray Clouds (uniform $\tau$) & $R$ & $0.70_{-0.16}^{+0.32}$ & $0.66_{-0.20}^{+0.39}$ & $0.54_{-0.09}^{+0.08}$\\
Non-Gray Clouds (log-uniform $\tau$) & $R$ & $0.77_{-0.19}^{+0.28}$ & $0.76_{-0.25}^{+0.43}$ & $0.67_{-0.11}^{+0.15}$\\
Non-Gray Clouds (uniform $\tau$) & $R$ & $0.71_{-0.18}^{+0.37}$ & $0.73_{-0.23}^{+0.41}$ & $0.65_{-0.10}^{+0.12}$\\
\hline
Cloud-Free & $d$ & $43.27_{-0.04}^{+0.04}$ & $43.27_{-0.04}^{+0.04}$ & $43.27_{-0.04}^{+0.04}$\\
Gray Clouds (log-uniform $\tau$) & $d$ & $43.27_{-0.04}^{+0.04}$ & $43.27_{-0.04}^{+0.04}$ & $43.27_{-0.04}^{+0.04}$\\
Gray Clouds (uniform $\tau$) & $d$ & $43.27_{-0.04}^{+0.04}$ & $43.27_{-0.04}^{+0.04}$ & $43.27_{-0.04}^{+0.04}$\\
Non-Gray Clouds (log-uniform $\tau$) & $d$ & $43.27_{-0.04}^{+0.04}$ & $43.27_{-0.04}^{+0.04}$ & $43.27_{-0.04}^{+0.04}$\\
Non-Gray Clouds (uniform $\tau$) & $d$ & $43.27_{-0.04}^{+0.04}$ & $43.27_{-0.04}^{+0.04}$ & $43.27_{-0.04}^{+0.04}$\\
\hline
Cloud-Free & $T_\mathrm{{eff}}$ & $1460_{-140}^{+180}$ & $1360_{-210}^{+290}$ & $1480_{-140}^{+100}$\\
Gray Clouds (log-uniform $\tau$) & $T_\mathrm{{eff}}$ & $1500_{-200}^{+250}$ & $1370_{-220}^{+320}$ & $1500_{-110}^{+100}$\\
Gray Clouds (uniform $\tau$) & $T_\mathrm{{eff}}$ & $1500_{-190}^{+200}$ & $1480_{-270}^{+300}$ & $1630_{-100}^{+130}$\\
Non-Gray Clouds (log-uniform $\tau$) & $T_\mathrm{{eff}}$ & $1460_{-170}^{+240}$ & $1400_{-230}^{+300}$ & $1490_{-120}^{+110}$\\
Non-Gray Clouds (uniform $\tau$) & $T_\mathrm{{eff}}$ & $1500_{-220}^{+220}$ & $1420_{-240}^{+280}$ & $1520_{-110}^{+120}$\\
\hline
Cloud-Free & $\log$ \ch{H2O} & $-3.4_{-0.5}^{+1.5}$ & $-3.6_{-0.6}^{+0.8}$ & $-3.61_{-0.31}^{+0.24}$\\
Gray Clouds (log-uniform $\tau$) & $\log$ \ch{H2O} & $-2.4_{-1.3}^{+0.8}$ & $-3.6_{-0.7}^{+0.9}$ & $-3.58_{-0.27}^{+0.24}$\\
Gray Clouds (uniform $\tau$) & $\log$ \ch{H2O} & $-2.4_{-1.3}^{+0.8}$ & $-3.4_{-0.8}^{+1.0}$ & $-3.4_{-0.3}^{+0.5}$\\
Non-Gray Clouds (log-uniform $\tau$) & $\log$ \ch{H2O} & $-2.8_{-1.2}^{+1.1}$ & $-3.6_{-0.6}^{+0.9}$ & $-3.61_{-0.29}^{+0.29}$\\
Non-Gray Clouds (uniform $\tau$) & $\log$ \ch{H2O} & $-2.2_{-1.3}^{+0.7}$ & $-3.5_{-0.7}^{+1.0}$ & $-3.59_{-0.29}^{+0.33}$\\
\hline
Cloud-Free & $\log$ \ch{K} & $\cdots$ & $\cdots$ & $-8.9_{-1.8}^{+1.9}$\\
Gray Clouds (log-uniform $\tau$) & $\log$ \ch{K} & $\cdots$ & $\cdots$ & $-8.7_{-1.9}^{+1.9}$\\
Gray Clouds (uniform $\tau$) & $\log$ \ch{K} & $\cdots$ & $\cdots$ & $-8.6_{-1.9}^{+2.0}$\\
Non-Gray Clouds (log-uniform $\tau$) & $\log$ \ch{K} & $\cdots$ & $\cdots$ & $-8.7_{-1.9}^{+1.8}$\\
Non-Gray Clouds (uniform $\tau$) & $\log$ \ch{K} & $\cdots$ & $\cdots$ & $-8.8_{-1.8}^{+1.9}$\\
\hline
Cloud-Free & $\log$ \ch{CrH} & $-9.5_{-1.4}^{+1.5}$ & $\cdots$ & $-9.8_{-1.3}^{+1.3}$\\
Gray Clouds (log-uniform $\tau$) & $\log$ \ch{CrH} & $-9.1_{-1.6}^{+1.8}$ & $\cdots$ & $-9.6_{-1.3}^{+1.2}$\\
Gray Clouds (uniform $\tau$) & $\log$ \ch{CrH} & $-9.1_{-1.6}^{+1.9}$ & $\cdots$ & $-9.5_{-1.4}^{+1.4}$\\
Non-Gray Clouds (log-uniform $\tau$) & $\log$ \ch{CrH} & $-9.4_{-1.5}^{+1.8}$ & $\cdots$ & $-9.7_{-1.3}^{+1.3}$\\
Non-Gray Clouds (uniform $\tau$) & $\log$ \ch{CrH} & $-9.0_{-1.7}^{+1.9}$ & $\cdots$ & $-9.7_{-1.3}^{+1.3}$\\
\tablebreak
Cloud-Free & $\log$ \ch{FeH} & $-9.5_{-1.4}^{+1.7}$ & $-8.7_{-1.8}^{+1.6}$ & $-9.8_{-1.3}^{+1.3}$\\
Gray Clouds (log-uniform $\tau$) & $\log$ \ch{FeH} & $-9.0_{-1.7}^{+2.0}$ & $-8.6_{-1.8}^{+1.6}$ & $-9.7_{-1.3}^{+1.3}$\\
Gray Clouds (uniform $\tau$) & $\log$ \ch{FeH} & $-9.1_{-1.7}^{+2.0}$ & $-8.6_{-1.9}^{+1.8}$ & $-9.6_{-1.4}^{+1.4}$\\
Non-Gray Clouds (log-uniform $\tau$) & $\log$ \ch{FeH} & $-9.3_{-1.5}^{+2.0}$ & $-8.5_{-1.9}^{+1.6}$ & $-9.6_{-1.3}^{+1.3}$\\
Non-Gray Clouds (uniform $\tau$) & $\log$ \ch{FeH} & $-8.9_{-1.7}^{+2.0}$ & $-8.6_{-1.9}^{+1.7}$ & $-9.7_{-1.3}^{+1.3}$\\
\hline
Cloud-Free & $\log$ \ch{TiH} & $\cdots$ & $-8.9_{-1.7}^{+1.6}$ & $-10.0_{-1.2}^{+1.2}$\\
Gray Clouds (log-uniform $\tau$) & $\log$ \ch{TiH} & $\cdots$ & $-8.9_{-1.7}^{+1.7}$ & $-9.9_{-1.2}^{+1.2}$\\
Gray Clouds (uniform $\tau$) & $\log$ \ch{TiH} & $\cdots$ & $-8.8_{-1.7}^{+1.7}$ & $-9.6_{-1.3}^{+1.4}$\\
Non-Gray Clouds (log-uniform $\tau$) & $\log$ \ch{TiH} & $\cdots$ & $-8.9_{-1.7}^{+1.7}$ & $-9.9_{-1.2}^{+1.2}$\\
Non-Gray Clouds (uniform $\tau$) & $\log$ \ch{TiH} & $\cdots$ & $-9.0_{-1.7}^{+1.7}$ & $-9.9_{-1.2}^{+1.3}$\\
\hline
Cloud-Free & $\log p_{\mathrm{t}}$ & $\cdots$ & $\cdots$ & $\cdots$\\
Gray Clouds (log-uniform $\tau$) & $\log p_{\mathrm{t}}$ & $-0.1_{-1.0}^{+1.0}$ & $-0.1_{-1.1}^{+1.1}$ & $-0.0_{-1.1}^{+1.0}$\\
Gray Clouds (uniform $\tau$) & $\log p_{\mathrm{t}}$ & $0.2_{-0.9}^{+0.8}$ & $0.7_{-1.0}^{+0.6}$ & $1.0_{-0.7}^{+0.4}$\\
Non-Gray Clouds (log-uniform $\tau$) & $\log p_{\mathrm{t}}$ & $-0.1_{-1.0}^{+1.0}$ & $-0.1_{-1.1}^{+1.0}$ & $0.1_{-1.1}^{+1.0}$\\
Non-Gray Clouds (uniform $\tau$) & $\log p_{\mathrm{t}}$ & $0.3_{-0.7}^{+0.7}$ & $0.7_{-0.9}^{+0.6}$ & $0.0_{-1.1}^{+1.0}$\\
\hline
Cloud-Free & $\log p_{\mathrm{b}}$ & $\cdots$ & $\cdots$ & $\cdots$\\
Gray Clouds (log-uniform $\tau$) & $\log p_{\mathrm{b}}$ & $0.52_{-0.28}^{+0.27}$ & $0.50_{-0.28}^{+0.29}$ & $0.52_{-0.29}^{+0.28}$\\
Gray Clouds (uniform $\tau$) & $\log p_{\mathrm{b}}$ & $0.49_{-0.28}^{+0.28}$ & $0.51_{-0.29}^{+0.28}$ & $0.52_{-0.29}^{+0.28}$\\
Non-Gray Clouds (log-uniform $\tau$) & $\log p_{\mathrm{b}}$ & $0.50_{-0.28}^{+0.27}$ & $0.52_{-0.29}^{+0.28}$ & $0.49_{-0.28}^{+0.28}$\\
Non-Gray Clouds (uniform $\tau$) & $\log p_{\mathrm{b}}$ & $0.51_{-0.28}^{+0.27}$ & $0.50_{-0.28}^{+0.28}$ & $0.51_{-0.28}^{+0.28}$\\
\hline
Cloud-Free & $\log \tau$ & $\cdots$ & $\cdots$ & $\cdots$\\
Gray Clouds (log-uniform $\tau$) & $\log \tau$ & $-2.2_{-1.6}^{+1.7}$ & $-2.0_{-1.6}^{+1.5}$ & $-2.7_{-1.3}^{+1.4}$\\
Gray Clouds (uniform $\tau$) & $\tau$ & $2_{-8}^{+10}$ & $11_{-6}^{+5}$ & $5_{-10}^{+9}$\\
Non-Gray Clouds (log-uniform $\tau$) & $\log \tau$ & $-2.2_{-1.6}^{+1.6}$ & $-2.1_{-1.6}^{+1.6}$ & $-2.7_{-1.3}^{+1.4}$\\
Non-Gray Clouds (uniform $\tau$) & $\tau$ & $5_{-8}^{+9}$ & $3_{-8}^{+10}$ & $-4_{-3}^{+4}$\\
\hline
Cloud-Free & $\log Q_0$ & $\cdots$ & $\cdots$ & $\cdots$\\
Gray Clouds (log-uniform $\tau$) & $\log Q_0$ & $\cdots$ & $\cdots$ & $\cdots$\\
Gray Clouds (uniform $\tau$) & $\log Q_0$ & $\cdots$ & $\cdots$ & $\cdots$\\
Non-Gray Clouds (log-uniform $\tau$) & $\log Q_0$ & $1.0_{-0.6}^{+0.6}$ & $1.0_{-0.6}^{+0.6}$ & $1.0_{-0.6}^{+0.6}$\\
Non-Gray Clouds (uniform $\tau$) & $\log Q_0$ & $1.0_{-0.6}^{+0.6}$ & $1.0_{-0.6}^{+0.6}$ & $1.0_{-0.6}^{+0.6}$\\
\hline
Cloud-Free & $a_0$ & $\cdots$ & $\cdots$ & $\cdots$\\
Gray Clouds (log-uniform $\tau$) & $a_0$ & $\cdots$ & $\cdots$ & $\cdots$\\
Gray Clouds (uniform $\tau$) & $a_0$ & $\cdots$ & $\cdots$ & $\cdots$\\
Non-Gray Clouds (log-uniform $\tau$) & $a_0$ & $5.0_{-1.1}^{+1.1}$ & $5.0_{-1.1}^{+1.1}$ & $5.0_{-1.1}^{+1.1}$\\
Non-Gray Clouds (uniform $\tau$) & $a_0$ & $5.0_{-1.1}^{+1.1}$ & $5.0_{-1.1}^{+1.1}$ & $4.9_{-1.1}^{+1.1}$\\
\hline
Cloud-Free & $\log a$ & $\cdots$ & $\cdots$ & $\cdots$\\
Gray Clouds (log-uniform $\tau$) & $\log a$ & $\cdots$ & $\cdots$ & $\cdots$\\
Gray Clouds (uniform $\tau$) & $\log a$ & $\cdots$ & $\cdots$ & $\cdots$\\
Non-Gray Clouds (log-uniform $\tau$) & $\log a$ & $0.4_{-0.8}^{+0.7}$ & $0.3_{-0.8}^{+0.8}$ & $0.3_{-0.7}^{+0.8}$\\
Non-Gray Clouds (uniform $\tau$) & $\log a$ & $0.3_{-0.7}^{+0.7}$ & $0.3_{-0.8}^{+0.8}$ & $0.2_{-0.7}^{+0.8}$
\enddata
\end{deluxetable*}

\startlongtable
\begin{deluxetable*}{lcccc} 
\tablecaption{\label{tab:retrieval_posteriors_red1}Retrieved Parameters of HIP~21152~B (Reduction 1)}
\tablehead{\colhead{Model} & \colhead{Parameter} & \colhead{IFS} & \colhead{CHARIS} & \colhead{IFS + CHARIS + Photometry}}
\startdata
Cloud-Free & $\log g$ & $5.3_{-0.7}^{+0.4}$ & $5.0_{-0.7}^{+0.6}$ & $5.4_{-0.6}^{+0.4}$\\
Gray Clouds (log-uniform $\tau$) & $\log g$ & $4.8_{-0.6}^{+0.6}$ & $4.9_{-0.7}^{+0.6}$ & $5.4_{-0.6}^{+0.4}$\\
Gray Clouds (uniform $\tau$) & $\log g$ & $5.1_{-0.7}^{+0.6}$ & $5.0_{-0.7}^{+0.6}$ & $5.4_{-0.5}^{+0.4}$\\
Non-Gray Clouds (log-uniform $\tau$) & $\log g$ & $4.8_{-0.6}^{+0.6}$ & $4.9_{-0.7}^{+0.7}$ & $5.4_{-0.6}^{+0.4}$\\
Non-Gray Clouds (uniform $\tau$) & $\log g$ & $5.2_{-0.7}^{+0.5}$ & $4.9_{-0.7}^{+0.6}$ & $5.4_{-0.5}^{+0.4}$\\
\hline
Cloud-Free & $R$ & $0.52_{-0.06}^{+0.10}$ & $0.8_{-0.3}^{+0.5}$ & $0.43_{-0.04}^{+0.05}$\\
Gray Clouds (log-uniform $\tau$) & $R$ & $0.54_{-0.06}^{+0.09}$ & $0.8_{-0.3}^{+0.4}$ & $0.43_{-0.04}^{+0.05}$\\
Gray Clouds (uniform $\tau$) & $R$ & $0.52_{-0.06}^{+0.10}$ & $0.66_{-0.20}^{+0.39}$ & $0.43_{-0.03}^{+0.05}$\\
Non-Gray Clouds (log-uniform $\tau$) & $R$ & $0.54_{-0.06}^{+0.10}$ & $0.76_{-0.25}^{+0.43}$ & $0.43_{-0.04}^{+0.05}$\\
Non-Gray Clouds (uniform $\tau$) & $R$ & $0.50_{-0.06}^{+0.09}$ & $0.73_{-0.23}^{+0.41}$ & $0.42_{-0.04}^{+0.05}$\\
\hline
Cloud-Free & $d$ & $43.27_{-0.04}^{+0.04}$ & $43.27_{-0.04}^{+0.04}$ & $43.27_{-0.04}^{+0.04}$\\
Gray Clouds (log-uniform $\tau$) & $d$ & $43.27_{-0.04}^{+0.04}$ & $43.27_{-0.04}^{+0.04}$ & $43.27_{-0.04}^{+0.04}$\\
Gray Clouds (uniform $\tau$) & $d$ & $43.27_{-0.04}^{+0.04}$ & $43.27_{-0.04}^{+0.04}$ & $43.27_{-0.04}^{+0.04}$\\
Non-Gray Clouds (log-uniform $\tau$) & $d$ & $43.27_{-0.04}^{+0.04}$ & $43.27_{-0.04}^{+0.04}$ & $43.27_{-0.04}^{+0.04}$\\
Non-Gray Clouds (uniform $\tau$) & $d$ & $43.27_{-0.04}^{+0.04}$ & $43.27_{-0.04}^{+0.04}$ & $43.27_{-0.04}^{+0.04}$\\
\hline
Cloud-Free & $T_\mathrm{{eff}}$ & $1790_{-130}^{+230}$ & $1360_{-210}^{+290}$ & $1890_{-110}^{+270}$\\
Gray Clouds (log-uniform $\tau$) & $T_\mathrm{{eff}}$ & $1780_{-130}^{+190}$ & $1370_{-220}^{+320}$ & $1890_{-110}^{+260}$\\
Gray Clouds (uniform $\tau$) & $T_\mathrm{{eff}}$ & $1780_{-130}^{+160}$ & $1480_{-270}^{+300}$ & $1870_{-90}^{+130}$\\
Non-Gray Clouds (log-uniform $\tau$) & $T_\mathrm{{eff}}$ & $1770_{-130}^{+200}$ & $1400_{-230}^{+300}$ & $1890_{-110}^{+250}$\\
Non-Gray Clouds (uniform $\tau$) & $T_\mathrm{{eff}}$ & $1780_{-120}^{+120}$ & $1420_{-240}^{+280}$ & $1880_{-100}^{+150}$\\
\hline
Cloud-Free & $\log$ \ch{H2O} & $-1.8_{-0.6}^{+0.5}$ & $-3.6_{-0.6}^{+0.8}$ & $-1.9_{-0.7}^{+0.5}$\\
Gray Clouds (log-uniform $\tau$) & $\log$ \ch{H2O} & $-1.8_{-0.6}^{+0.5}$ & $-3.6_{-0.7}^{+0.9}$ & $-1.9_{-0.7}^{+0.5}$\\
Gray Clouds (uniform $\tau$) & $\log$ \ch{H2O} & $-1.8_{-0.6}^{+0.4}$ & $-3.4_{-0.8}^{+1.0}$ & $-1.9_{-0.7}^{+0.5}$\\
Non-Gray Clouds (log-uniform $\tau$) & $\log$ \ch{H2O} & $-1.7_{-0.6}^{+0.4}$ & $-3.6_{-0.6}^{+0.9}$ & $-1.9_{-0.7}^{+0.5}$\\
Non-Gray Clouds (uniform $\tau$) & $\log$ \ch{H2O} & $-1.8_{-0.6}^{+0.4}$ & $-3.5_{-0.7}^{+1.0}$ & $-1.9_{-0.7}^{+0.5}$\\
\hline
Cloud-Free & $\log$ \ch{K} & $-7.7_{-2.4}^{+2.3}$ & $\cdots$ & $-7.3_{-2.6}^{+2.3}$\\
Gray Clouds (log-uniform $\tau$) & $\log$ \ch{K} & $-7.5_{-2.5}^{+2.4}$ & $\cdots$ & $-7.5_{-2.5}^{+2.3}$\\
Gray Clouds (uniform $\tau$) & $\log$ \ch{K} & $-7.5_{-2.5}^{+2.4}$ & $\cdots$ & $-7.5_{-2.5}^{+2.3}$\\
Non-Gray Clouds (log-uniform $\tau$) & $\log$ \ch{K} & $-7.5_{-2.5}^{+2.4}$ & $\cdots$ & $-7.5_{-2.5}^{+2.3}$\\
Non-Gray Clouds (uniform $\tau$) & $\log$ \ch{K} & $-7.5_{-2.5}^{+2.3}$ & $\cdots$ & $-7.5_{-2.5}^{+2.3}$\\
\hline
Cloud-Free & $\log$ \ch{CrH} & $-9.3_{-1.5}^{+1.6}$ & $\cdots$ & $-9.3_{-1.5}^{+1.6}$\\
Gray Clouds (log-uniform $\tau$) & $\log$ \ch{CrH} & $-9.3_{-1.5}^{+1.6}$ & $\cdots$ & $-9.3_{-1.5}^{+1.6}$\\
Gray Clouds (uniform $\tau$) & $\log$ \ch{CrH} & $-9.3_{-1.5}^{+1.6}$ & $\cdots$ & $-9.2_{-1.5}^{+1.6}$\\
Non-Gray Clouds (log-uniform $\tau$) & $\log$ \ch{CrH} & $-9.3_{-1.5}^{+1.5}$ & $\cdots$ & $-9.3_{-1.5}^{+1.6}$\\
Non-Gray Clouds (uniform $\tau$) & $\log$ \ch{CrH} & $-9.2_{-1.5}^{+1.5}$ & $\cdots$ & $-9.2_{-1.6}^{+1.6}$\\
\tablebreak
Cloud-Free & $\log$ \ch{FeH} & $-8.8_{-1.8}^{+1.8}$ & $-8.7_{-1.8}^{+1.6}$ & $-8.1_{-2.2}^{+1.9}$\\
Gray Clouds (log-uniform $\tau$) & $\log$ \ch{FeH} & $-8.6_{-1.9}^{+1.9}$ & $-8.6_{-1.8}^{+1.6}$ & $-8.0_{-2.1}^{+1.8}$\\
Gray Clouds (uniform $\tau$) & $\log$ \ch{FeH} & $-8.6_{-1.9}^{+1.9}$ & $-8.6_{-1.9}^{+1.8}$ & $-8.3_{-2.0}^{+1.9}$\\
Non-Gray Clouds (log-uniform $\tau$) & $\log$ \ch{FeH} & $-9.3_{-1.5}^{+2.0}$ & $-8.5_{-1.9}^{+1.6}$ & $-8.2_{-2.1}^{+1.8}$\\
Non-Gray Clouds (uniform $\tau$) & $\log$ \ch{FeH} & $-8.7_{-1.8}^{+1.8}$ & $-8.6_{-1.9}^{+1.7}$ & $-8.0_{-2.2}^{+1.8}$\\
\hline
Cloud-Free & $\log$ \ch{TiH} & $-6.7_{-2.1}^{+0.9}$ & $-8.9_{-1.7}^{+1.6}$ & $-6.5_{-1.5}^{+0.8}$\\
Gray Clouds (log-uniform $\tau$) & $\log$ \ch{TiH} & $-6.5_{-2.0}^{+0.9}$ & $-8.9_{-1.7}^{+1.7}$ & $-6.4_{-1.4}^{+0.7}$\\
Gray Clouds (uniform $\tau$) & $\log$ \ch{TiH} & $-6.6_{-2.1}^{+0.9}$ & $\cdots$ & $-6.5_{-1.9}^{+0.8}$\\
Non-Gray Clouds (log-uniform $\tau$) & $\log$ \ch{TiH} & $-6.6_{-2.2}^{+0.9}$ & $-8.9_{-1.7}^{+1.7}$ & $-6.5_{-1.4}^{+0.7}$\\
Non-Gray Clouds (uniform $\tau$) & $\log$ \ch{TiH} & $-6.5_{-1.9}^{+0.8}$ & $-9.0_{-1.7}^{+1.7}$ & $-6.5_{-1.9}^{+0.8}$\\
\hline
Cloud-Free & $\log p_{\mathrm{t}}$ & $\cdots$ & $\cdots$ & $\cdots$\\
Gray Clouds (log-uniform $\tau$) & $\log p_{\mathrm{t}}$ & $-0.0_{-1.0}^{+1.0}$ & $-0.1_{-1.1}^{+1.1}$ & $0.1_{-1.0}^{+0.9}$\\
Gray Clouds (uniform $\tau$) & $\log p_{\mathrm{t}}$ & $0.2_{-0.9}^{+0.8}$ & $0.7_{-1.0}^{+0.6}$ & $0.4_{-0.6}^{+0.7}$\\
Non-Gray Clouds (log-uniform $\tau$) & $\log p_{\mathrm{t}}$ & $0.0_{-1.0}^{+1.0}$ & $-0.1_{-1.1}^{+1.0}$ & $0.1_{-1.0}^{+0.9}$\\
Non-Gray Clouds (uniform $\tau$) & $\log p_{\mathrm{t}}$ & $0.4_{-0.7}^{+0.7}$ & $0.7_{-0.9}^{+0.6}$ & $0.2_{-0.7}^{+0.7}$\\
\hline
Cloud-Free & $\log p_{\mathrm{b}}$ & $\cdots$ & $\cdots$ & $\cdots$\\
Gray Clouds (log-uniform $\tau$) & $\log p_{\mathrm{b}}$ & $0.50_{-0.28}^{+0.28}$ & $0.50_{-0.28}^{+0.29}$ & $0.52_{-0.29}^{+0.27}$\\
Gray Clouds (uniform $\tau$) & $\log p_{\mathrm{b}}$ & $0.51_{-0.29}^{+0.28}$ & $0.51_{-0.29}^{+0.28}$ & $0.49_{-0.28}^{+0.29}$\\
Non-Gray Clouds (log-uniform $\tau$) & $\log p_{\mathrm{b}}$ & $0.50_{-0.28}^{+0.28}$ & $0.52_{-0.29}^{+0.28}$ & $0.50_{-0.28}^{+0.28}$\\
Non-Gray Clouds (uniform $\tau$) & $\log p_{\mathrm{b}}$ & $0.50_{-0.28}^{+0.28}$ & $0.50_{-0.28}^{+0.28}$ & $0.49_{-0.28}^{+0.28}$\\
\hline
Cloud-Free & $\log \tau$ & $\cdots$ & $\cdots$ & $\cdots$\\
Gray Clouds (log-uniform $\tau$) & $\log \tau$ & $-2.2_{-1.6}^{+1.7}$ & $-2.0_{-1.6}^{+1.5}$ & $-2.2_{-1.6}^{+1.8}$\\
Gray Clouds (uniform $\tau$) & $\tau$ & $1_{-7}^{+10}$ & $11_{-6}^{+5}$ & $4_{-8}^{+8}$\\
Non-Gray Clouds (log-uniform $\tau$) & $\log \tau$ & $-2.0_{-1.7}^{+1.7}$ & $-2.1_{-1.6}^{+1.6}$ & $-2.2_{-1.6}^{+1.8}$\\
Non-Gray Clouds (uniform $\tau$) & $\tau$ & $4_{-8}^{+9}$ & $3_{-8}^{+10}$ & $3_{-8}^{+10}$\\
\hline
Cloud-Free & $\log Q_0$ & $\cdots$ & $\cdots$ & $\cdots$\\
Gray Clouds (log-uniform $\tau$) & $\log Q_0$ & $\cdots$ & $\cdots$ & $\cdots$\\
Gray Clouds (uniform $\tau$) & $\log Q_0$ & $\cdots$ & $\cdots$ & $\cdots$\\
Non-Gray Clouds (log-uniform $\tau$) & $\log Q_0$ & $1.0_{-0.6}^{+0.6}$ & $1.0_{-0.6}^{+0.6}$ & $1.0_{-0.6}^{+0.6}$\\
Non-Gray Clouds (uniform $\tau$) & $\log Q_0$ & $1.0_{-0.6}^{+0.6}$ & $1.0_{-0.6}^{+0.6}$ & $1.0_{-0.6}^{+0.6}$\\
\hline
Cloud-Free & $a_0$ & $\cdots$ & $\cdots$ & $\cdots$\\
Gray Clouds (log-uniform $\tau$) & $a_0$ & $\cdots$ & $\cdots$ & $\cdots$\\
Gray Clouds (uniform $\tau$) & $a_0$ & $\cdots$ & $\cdots$ & $\cdots$\\
Non-Gray Clouds (log-uniform $\tau$) & $a_0$ & $5.0_{-1.1}^{+1.1}$ & $5.0_{-1.1}^{+1.1}$ & $5.0_{-1.1}^{+1.1}$\\
Non-Gray Clouds (uniform $\tau$) & $a_0$ & $5.0_{-1.1}^{+1.1}$ & $5.0_{-1.1}^{+1.1}$ & $5.0_{-1.1}^{+1.1}$\\
\hline
Cloud-Free & $\log a$ & $\cdots$ & $\cdots$ & $\cdots$\\
Gray Clouds (log-uniform $\tau$) & $\log a$ & $\cdots$ & $\cdots$ & $\cdots$\\
Gray Clouds (uniform $\tau$) & $\log a$ & $\cdots$ & $\cdots$ & $\cdots$\\
Non-Gray Clouds (log-uniform $\tau$) & $\log a$ & $0.3_{-0.8}^{+0.8}$ & $0.3_{-0.8}^{+0.8}$ & $0.3_{-0.8}^{+0.8}$\\
Non-Gray Clouds (uniform $\tau$) & $\log a$ & $0.4_{-0.8}^{+0.7}$ & $0.3_{-0.8}^{+0.8}$ & $0.4_{-0.8}^{+0.8}$
\enddata
\end{deluxetable*}
% need space to render second page of table for some reason
\ 
\end{document}